\documentclass[11pt, a4paper]{article}
\usepackage{jheparxiv}
\usepackage{wasysym}
\usepackage[utf8]{inputenc}
\usepackage{amsmath}
\usepackage{amsfonts}
\usepackage{amssymb}
\usepackage{latexsym}
\usepackage{mathrsfs}
\usepackage{mathtools}
\usepackage{braket}		%Bra Ket Notation
\usepackage{graphicx}
\usepackage{color}
\usepackage{xcolor}
\usepackage{slashed}
\usepackage{twistor}
\usepackage[all]{xy}
\usepackage{tikz-cd}
\usepackage{bbm}
\usetikzlibrary{arrows.meta,decorations.pathmorphing,calc,shadings,shapes.misc}

\newcommand{\vphi}{\varphi}
\renewcommand{\d}{\mathrm{d}}
\newcommand{\br}[1]{\overline{#1}}

\renewcommand{\bz}{{\bar z}}
\renewcommand{\by}{{\bar y}}

\newcommand{\eps}{\epsilon}
\newcommand{\veps}{\varepsilon}

\newcommand{\bi}{{\bar\imath}}

\newcommand{\AdS}{{\text{AdS}}}
\newcommand{\PV}{{\mathrm{PV}}}
\newcommand{\ip}{\lrcorner\,}
\newcommand{\til}[1]{\widetilde{#1}}

\newcommand{\bX}{{\mathbb{X}}}
\newcommand{\bY}{{\mathbb{Y}}}
\newcommand{\gl}{{\mathfrak{gl}}}
\newcommand{\da}{{\dot a}}

\newcommand{\bA}{{\mathbf{A}}}
\newcommand{\bB}{{\mathbf{B}}}
\newcommand{\bC}{{\mathbf{C}}}

\newcommand{\sa}{{\mathsf{a}}}
\newcommand{\msf}[1]{\mathsf{#1}}
\newcommand{\W}{{\mathrm{W}}}

\renewcommand{\bx}{{\bar x}}
\renewcommand{\by}{{\bar y}}

\newcommand{\sA}{{\mathsf{A}}}
\newcommand{\sM}{{\mathsf{M}}}
\newcommand{\sN}{{\mathsf{N}}}
\newcommand{\sP}{{\mathsf{P}}}
\newcommand{\sQ}{{\mathsf{Q}}}
\newcommand{\sR}{{\mathsf{R}}}
\newcommand{\sB}{{\mathsf{B}}}

\newcommand{\fu}{{\mathfrak{u}}}
\newcommand{\YM}{{\text{YM}}}
\newcommand{\sdYM}{{\text{sdYM}}}
\newcommand{\Tr}{{\mathrm{Tr}}}
\newcommand{\pt}{{\mathbf{PT}}}

\newcommand{\bs}[1]{\boldsymbol{#1}}
\newcommand{\Bsi}{{\boldsymbol{\Psi}}}
\renewcommand{\sl}{{\mathfrak{sl}}}
\newcommand{\rO}{{\mathrm{O}}}
\newcommand{\rA}{{\mathscr{A}}}
\newcommand{\sX}{{\mathsf{X}}}
\newcommand{\sZ}{{\mathsf{Z}}}
\newcommand{\psl}{{\mathfrak{psl}}}

\title{Chiral holography}

\author[a]{Atul Sharma,}
\author[b]{David Skinner}

\affiliation[a]{Center for the Fundamental Laws of Nature, Harvard University,\\
Cambridge, MA, 02138, USA \vspace{0.1cm}}
\emailAdd{atulsharma@fas.harvard.edu}

\affiliation[b]{Department of Applied Maths \& Theoretical Physics, University of Cambridge,\\
Wilberforce Road, Cambridge CB3 0WA, UK}
\emailAdd{d.b.skinner@damtp.cam.ac.uk}

\abstract{Using twistor string theory, we engineer a gravitational holographic dual of the self-dual sector of $\cN=4$ super Yang-Mills theory. This provides a top-down example of holography at zero 't Hooft coupling. Our holographic dictionary operates at the level of D-branes wrapping twistor lines and generalizes the D-instanton prescription used to compute scattering amplitudes. As an application, we study the correlators of determinant operators that act as generating functions of half-BPS operators in the self-dual theory. We show that these correlators are dual to configurations of giant gravitons in the bulk.}

\begin{document}

\maketitle

\section{Introduction}
\label{sec:intro}

Holographic dualities have revolutionized our conception of quantum gravity. The various AdS/CFT correspondences introduced in \cite{Maldacena:1997re} provide a sharp tool to translate gravitational questions into large $N$ gauge theory questions and vice versa. Often, a question that is hard to answer on one side of the duality becomes amenable to conventional methods on the other side. Over the years, this technique has been applied to analyze many physical observables ranging from CFT correlation functions to black hole entropy \cite{Strominger:1996sh,Gubser:1998bc,Witten:1998qj,Aharony:1999ti,Ammon:2015wua}.

Most precision tests of the AdS/CFT correspondence rely on matching protected quantities in the boundary gauge theory to perturbative observables in the bulk. When working with non-protected quantities, one needs to deal with the additional complication of a highly nontrivial 't Hooft coupling dependence. The gauge theory side of the duality often operates at strong 't Hooft coupling, so its non-protected observables cannot be computed perturbatively. To compute such quantities, one often resorts to other harder techniques such as integrability \cite{Beisert:2010jr}.

In recent years, research on \emph{twisted holography} has generated many toy models of holography that capture precisely the protected BPS subsectors of their more physical counterparts \cite{Bonetti:2016nma,Ishtiaque:2018str,Costello:2018zrm,Costello:2020jbh,Fernandez:2024tue}. These dualities arise by twisting both the gauge and gravitational sides of standard AdS/CFT dualities. Importantly, they can also be intrinsically defined as open-closed dualities of topological string theories in the spirit of \cite{Dijkgraaf:2002fc,Gopakumar:1998ki} without resorting to twisting. So they provide a self-contained subsector in which holography can be tested and proved in a systematic manner. 

However, despite their success, twisted holographic dualities rarely give us a handle on non-BPS observables. This motivates the search for simpler versions of holography in which non-BPS quantities do exist, but they can be computed at weak or zero 't Hooft coupling while retaining a holographic description. Such constructions would naturally fill the gap between twisted holography and physical holography. In this work, we propose such a duality.

\medskip

A common approach to holography at weak coupling has been the exploration of the tensionless limit of the AdS/CFT correspondence in string theory. This has blossomed into a highly successful paradigm for AdS$_3$/CFT$_2$. For example, it gives rise to a precise worldsheet dual of the symmetric product CFT $\mathrm{Sym}^N(T^4)$ in its free limit \cite{Eberhardt:2019ywk,Eberhardt:2018ouy,Gaberdiel:2018rqv}. This raises the natural question: can we also construct holographic duals of higher-dimensional conformal field theories at weak coupling?

In the duality relating type IIB strings on AdS$_5\times S^5$ to $\cN=4$ super Yang-Mills (sYM) theory, as one sends the string length $\al'\to\infty$, the 't Hooft coupling $\lambda\to0$. There are multiple ways of taking the zero coupling limit of gauge theory. The standard $\lambda\to0$ limit gives rise to free super Yang-Mills theory, but its bulk dual inhabits a highly stringy regime involving infinitely many massless higher-spin particles \cite{Haggi-Mani:2000dxu,Sundborg:2000wp,Sezgin:2001zs,Mikhailov:2002bp,Sezgin:2002rt}. This proves to be extremely hard to bring under control.

Fortunately, there are alternatives. In \cite{Witten:2003nn}, Witten proposed a \emph{chiral} free limit taken after rescaling the fields of the $\cN=4$ vector multiplet by suitable factors of the gauge coupling. This corresponds to the \emph{self-dual} limit of $\cN=4$ sYM. As our main result, we will find that the gravitational dual of this theory is a topological string theory with a conventional target space description defined on a semiclassical bulk. It will be a twistor string theory.

\paragraph{Self-duality and holography.} The self-dual sector of four-dimensional gauge theory has always been a fascinating arena to study quantum field theory in a controlled setting. Self-dual Yang-Mills (sdYM) arises from standard Yang-Mills through a zero-coupling limit if, as we review in section \ref{ssec:chiral}, the limit is implemented in a chiral manner so that the theory retains a chiral subset of its interaction vertices. This allows it to feature rich nonlinear dynamics even in the absence of a Yang-Mills coupling. At the same time, it boasts an integrable structure that allows one to explore many perturbative and non-perturbative phenomena in fine detail. Owing to these features, the self-dual limit of $\cN=4$ sYM provides a tantalizing alternative to free sYM for studying and proving a simplified model of holography. 

The tensionless string dual of free $\cN=4$ sYM was recently investigated by Gaberdiel and Gopakumar in \cite{Gaberdiel:2021qbb,Gaberdiel:2021jrv}. It had the feature of exhibiting a stringy, non-geometric regime that could not be captured by a target space effective supergravity theory. Instead, it was described in terms of a worldsheet sigma model. The target space of this sigma model was \emph{not} $\AdS_5\times S^5$, but was instead closely related to the twistor space of $\AdS_5$ \cite{Adamo:2016rtr}. A target space string field theory description of this sigma model remains to be found.

This string theory in twistor space is reminiscent of Witten and Berkovits' twistor string theories \cite{Witten:2003nn,Berkovits:2004hg} and their more recent ambitwistor cousins \cite{Mason:2013sva,Geyer:2014fka}. However, in contrast to the proposal of Gaberdiel and Gopakumar, twistor strings are usually known to describe self-dual, rather than free, $\cN=4$ sYM. This motivates us to look for a closed string dual of the self-dual theory instead of the free theory.

There are a multitude of reasons to be interested in this. For one, it pushes us to develop a precise understanding of the string field theory of twistor strings that has been lacking in the literature. We also foresee a variety of interesting applications of our duality to holographic computations of the amplitudes and correlators of $\cN=4$ sYM perturbatively around its self-dual sector. For example, a concrete application that is the focus of section \ref{sec:giant} is the duality between half-BPS determinant correlators and giant gravitons in the bulk. 

In the rest of this section, we discuss each of these ideas in turn, while taking the opportunity to briefly describe the results of this paper.

\paragraph{A fresh look at twistor strings.} The highlight of twistor string theory was the discovery of worldsheet formulae for scattering amplitudes \cite{Witten:2003nn,Berkovits:2004hg,Roiban:2004yf}. It reproduced all tree-level amplitudes of $\cN=4$ sYM in a perturbative expansion around its self-dual sector. However, its closed string sector was generally believed to describe conformal gravity instead of Einstein gravity \cite{Berkovits:2004jj}.\footnote{A closed twistor string that reproduces the amplitudes of Einstein supergravity was subsequently discovered in \cite{Skinner:2013xp}, but finding its target space effective action on twistor space has remained an open problem.} In this work, we revisit certain subtleties of the open string sector of the B-model in twistor space that drastically change this conclusion (see also \cite{Bittleston:2024efo}).

The topological B-model on a Calabi-Yau manifold $X$ is an $\cN=(2,2)$ supersymmetric sigma model twisted by a nilpotent supercharge $Q$ \cite{Witten:1991zz}. The fields of its worldsheet theory consist of bosonic target space complex coordinates $x^i,\bar x^\bi$, as well as a set of fermionic $0$-forms $\theta_i,\psi^\bi$ and $1$-forms $\rho^i$. The supercharge $Q$ acts on these as
\be
    Qx^i = 0\,,\qquad Q\bar x^\bi = \psi^\bi\,,\qquad Q\psi^\bi = Q\theta_i = 0\,,\qquad Q\rho^i = \d x^i
\ee
and is taken to be the BRST operator of the worldsheet sigma model. The fermions $\theta_i$ and $\psi^\bi$ transform as the worldsheet pullbacks of target space $(1,0)$-vectors $\p_i$ and $(0,1)$-covectors $\d\bar x^\bi$ respectively, while $Q$ generates the action of the $\dbar$-operator on $X$. So operators in $Q$-cohomology of the worldsheet sigma model are in 1:1 correspondence with the Dolbeault cohomology groups $H^{0,p}(X,\wedge^qT_X)$ of the target space, summed over $(p,q)$. In passing to the topological string, as explained in \cite{Costello:2019jsy}, coupling to worldsheet topological gravity requires that the operators must additionally preserve the Calabi-Yau volume form $\Omega$ and so are divergence-free. These constitute the fields of the target space theory.

One of the novelties of twistor string theory was that its target space was taken to be a Calabi-Yau \emph{super}manifold. This was (a certain subspace of) complex projective superspace $\P^{3|4}$ known as supertwistor space, described by four bosonic homogeneous coordinates $Z^A$ and four fermionic homogeneous coordinates $\eta^I$. Branes filling this supertwistor space engineer a twistor uplift of the self-dual sector of $\cN=4$ sYM on $\R^4$. Witten further argued that D-instanton perturbation theory in this setting generates an expansion of full sYM around its self-dual sector \cite{Witten:2003nn}. In Berkovits' formulation and in the heterotic formulation, these D-instantons were instead viewed as worldsheet instantons \cite{Berkovits:2004hg,Mason:2007zv}.

As for a bosonic Calabi-Yau, worldsheet B-twisted supersymmetry tells us that each bosonic field $Z^A$ is accompanied by a spin-0 superpartner $\veps_A$, whose zero-mode represents the tangent vector $\p/\p Z^A$. Likewise, on a Calabi-Yau supermanifold, each fermionic coordinate $\eta^I$ of the target space comes with a bosonic superpartner we shall call $W_I$ and whose zero-mode represents $\p/\p\eta^I$. In particular, under scalings $(Z^A,\eta^I)\mapsto r(Z^A,\eta^I)$ of the homogeneous coordinates of supertwistor space, $(\veps_A,W_I)\mapsto r^{-1}(\veps_A,W_I)$ so each have weight $-1$. The $(\veps_A,W_I)$ play no role in studying open strings on branes filling supertwistor space, because the standard Dirichlet boundary conditions $\veps_A|_{\p\Sigma}=0=W_I|_{\p\Sigma}$ of the B-model ensure that these fields vanish on-shell (or in BRST cohomology) throughout the worldsheet $\Sigma$. However, they do appear in the closed string sector. 

In \cite{Berkovits:2004jj}, it was argued that closed twistor string theory also lives on supertwistor space and describes the twistor uplift of  $\cN=4$ conformal supergravity on $\R^4$. From the B-model perspective, this conclusion was reached in part because the Beltrami differential arising as a state in the B-model on supertwistor space could be identified with an $\cN=4$ conformal supergraviton multiplet on $\R^4$ via the Penrose transform. This posed a big hurdle for the application of twistor string techniques to loop amplitudes, which were expected to receive unwelcome contributions from conformal graviton loops. Some of these setbacks have been partially overcome (see \emph{e.g.} \cite{Skinner:2013xp,Mason:2013sva,Adamo:2013tsa,Geyer:2015bja,Geyer:2016wjx,Seet:2025mes}) by modifying the worldsheet theory in various ways, but unlike the B-model, the target space effective actions and brane dynamics of these theories remain underdeveloped.

The true nature of closed twistor string theory turns out to be much more nuanced. Zero modes of the extra spin-0 worldsheet bosons $W_I$ give rise to four extra complex, bosonic dimensions of the target space. So closed strings do not see the target as a complex 3-fold like twistor space $\PT$, but instead as a complex 7-fold. Since the $W_I$ each have scaling weight $-1$, this 7-fold is the total space of the rank 4 vector bundle $\CO(-1)^{\oplus4}\to\PT$, with $Z^A$ and $W_I$ acting as coordinates on its base and fibers respectively. 

One way to understand the emergence of these additional fiber directions from the supertwistor perspective is to note that higher polyvectors on a supermanifold space should be \emph{graded} anti-symmetric, so in particular they are symmetric functions of the $W_I$. Unlike for polyvectors on a bosonic manifold, where $\wedge^qT_X$ vanishes if $q> \dim_\C(X)$, on a supermanifold the range of $q$ does not need to truncate. We can combine the polyvectors $H^{0,*}(\PT,\wedge^qT_{\P^{3|4}})$ which point along the fermionic tangent directions of $\P^{3|4}$ to obtain a vertex operator
\[
    \cV(Z,W) = V(Z) + V^I(Z)W_I + \frac{1}{2}\,V^{IJ}(Z)W_IW_J + \cdots
\]
 which behaves just as a vertex operator on the bosonic 7-fold $X=\CO(-1)^{\oplus4}$. This behaviour of the higher polyvectors on supertwistor space was overlooked in the original constructions, which focussed mostly on the supertwistor Beltrami differential. Allowing vertex operators that also depend on the zero modes of the fermionic fields $\eta^I$ and $\veps_A$ can then be interpreted as giving the standard polyvectors $H^{0,*}(X,\wedge^q T_X)$ on this bosonic 7-fold $X$. (As we discuss in sections \ref{ssec:open} and \ref{ssec:bcov}, following~\cite{Alexandrov:1995kv} these in turn can be interpreted as $(0,*)$-forms whose coefficients are functions on a $7|7$-dimensional supermanifold $\Pi T^*_X$.)

In section \ref{ssec:bcov}, we discuss closed twistor string theory as a B-model on this 7-complex-dimensional target. Topological strings do not possess a critical dimension (in the usual sense of a worldsheet conformal anomaly), so the high dimensionality poses no risk. The closed string field theory of the B-model is a higher-dimensional generalization of Kodaira-Spencer gravity \cite{Bershadsky:1993cx} that we refer to as BCOV theory following \cite{Costello:2012cy}. Closed string excitations are not restricted to $W_I=0$, but can instead propagate into the bulk of $\CO(-1)^{\oplus4}$. The conformal gravitons of Berkovits \& Witten \cite{Berkovits:2004jj} are recovered when they are restricted to the zero section $W_I=0$, much as how Einstein gravity in the bulk of AdS$_5$ restricts to conformal gravity on the boundary \cite{Ferrara:1998ej,Liu:1998bu}. But in general, closed twistor strings are not confined to twistor space at all!

In this work, we wrap the zero section of $\CO(-1)^{\oplus4}\to\PT$ with $N$ D5 branes. The theory living on these $N$ D5s is precisely the holomorphic super Chern-Simons theory that describes $\cN=4$ sdYM on $\R^4$. In section \ref{backreaction} we study the backreaction of the D5s on the closed string sector in the large $N$ limit. Unlike twisted holography on a 3-fold \cite{Costello:2018zrm}, the backreaction on a 7-fold is not a Beltrami differential, but rather a higher-form flux given by a $(0,3)$-form valued in $(3,0)$-vectors, see equation \eqref{xiN}. Thus the complex structure of $\CO(-1)^{\oplus4}\to\PT$ is not deformed, though there is a sense in which the associated supergeometry does become deformed as we briefly explore in section \ref{ssec:deformed}. This motivates our holographic conjecture:
\[\colorbox{gray!15}{
$
\hspace{1cm}
\begin{aligned}
\rule[-0.3ex]{0pt}{2.5ex}\\[-1em]
\parbox[c]{3cm}{\centering $\U(N)$ $\cN=4$ sdYM on $\R^4$}
&\overset{N\to\infty}{=}\quad\!
\parbox[c]{3.5cm}{\centering B-model on $\til X$ with $N$ units of flux}\\[-1em]\rule[-0.3ex]{0pt}{2.5ex}
\end{aligned}
\hspace{1cm}$}\]
where $\til X$ denotes the total space of $\CO(-1)^{\oplus4}\to\PT$ with its zero section removed.

\paragraph{A non-standard holographic dictionary.} The key novelty of this approach is that it doesn't directly engineer 4d $\cN=4$ sdYM as a brane worldvolume theory, but only its 6d twistor uplift. Consequently, our duality obeys the rules of a standard holographic duality relating a 6d theory to a gravitational theory in 14 real dimensions. But the dictionary relating the 4d theory to the 14d theory proves to be much more involved.

Single-trace local operators of the 6d theory are in 1:1 correspondence with single-particle states of the 14d BCOV theory. The 6d theory is related to the 4d theory by compactification along the fibers of the twistor fibration $\P^1\hookrightarrow\PT\to\R^4$. But 6d local operators rarely descend to 4d local operators under the compactification. This is because the standard 6d local operators of interest are $0$-forms on $\PT$, but physical fields of the 4d theory arise as Penrose transforms of the $(0,1)$-form fields of the theory on $\PT$. Instead, local operators on $\R^4$ are often uplifted to $\PT$ as non-local operators, which has proven to be a helpful technique for computing their form factors in the past \cite{Costello:2022wso,Bogna:2023bbd,Koster:2016fna,Chicherin:2014uca,Adamo:2011dq,Adamo:2011cd,Mason:2010yk,Koster:2016ebi}.

We discuss this subtlety in detail in section \ref{sec:currents}. In section \ref{ssec:locality}, we review the construction of the holographic dictionary in twisted holography, and explain its shortcomings for our purposes. One of the key cases that can nevertheless be handled by this standard holographic dictionary is that of conserved currents. In section \ref{ssec:global}, we derive a dictionary relating $(0,1)$-form-valued BCOV states and superconformal symmetry currents of the 6d theory. If a symmetry of the 6d theory is induced by a symmetry of the 4d theory, then the corresponding Noether current of the theory on $\PT$ necessarily descends to the Noether current of the same symmetry on $\R^4$. 

In section \ref{ssec:super}, we describe the computation of the pushforward of 6d symmetry currents to 4d currents. This furnishes a dictionary between superconformal currents of the 4d theory and the corresponding BCOV states. For example, the 4d stress tensor is dual to complex structure deformations of the 7-fold that point along twistor space, while the R-symmetry currents are dual to complex structure deformations transverse to twistor space. We do not attempt to compute the correlators of such currents, but anticipate that they will be dual to Witten diagrams of BCOV states in the large $N$ expansion.

\paragraph{Determinants and giant gravitons.} This begs the question: what is the bulk dual of a general local operator on $\R^4$? We content ourselves with studying the cases of Lagrangian insertions and half-BPS operators. It turns out that answering this question thrusts us headfirst into the realm of determinant operators and giant graviton branes.

In section \ref{sec:dict}, we develop a dictionary relating certain brane configurations in the bulk to local operators of $\cN=4$ sdYM. We begin with the classical case of Witten's D-instantons, which were originally thought to be D1 branes in twistor space. In section \ref{ssec:D9}, we show that the D-instantons actually arise as D9 branes wrapping a twistor line in $\PT$ as well as all four of the $\CO(-1)^{\oplus4}$ fibers above it. The 5-9 and 9-5 strings engineer certain current algebra systems that were used by Nair \cite{Nair:1988bq} and Witten \cite{Witten:2003nn} to compute MHV gluon amplitudes. 

When the bulk fields are decoupled, integrating out the 5-9 and 9-5 strings inserts a determinant operator in the path integral of $\cN=4$ sdYM. This is not gauge invariant by itself, but its gauge anomaly can be cancelled by dressing it with the gravitational field sourced by the D9 brane. Integrating this determinant over chiral superspace produces an insertion of the non-self-dual interaction Lagrangian $L_\text{int}$ that deforms the self-dual theory into full $\cN=4$ sYM. This determinant has previously been used to motivate the twistor action of $\cN=4$ sYM \cite{Boels:2006ir}. One can also sew together such determinants to form a null-polygonal Wilson loop whose expectation value computes planar $\cN=4$ sYM amplitudes via the Amplitude/Wilson loop duality \cite{Mason:2010yk}. In this sense, our holographic conjecture also allows us to find the gravitational duals of observables at weak but \emph{nonzero} 't Hooft coupling. We briefly comment on this application in section \ref{sec:outlook}.

In the rest of the paper, we focus on half-BPS operators in $\cN=4$ sdYM. Following the work of Caron Huot, et al.\ \cite{Caron-Huot:2023wdh}, we organize half-BPS operators into generating functionals that take the form of determinant operators. For instance, bosonic half-BPS operators can be arranged into a bosonic determinant $\det(1+y\cdot\Phi)$, where $\Phi_{IJ}=-\Phi_{JI}$ are the six scalars of the $\cN=4$ vector multiplet. Every half-BPS sector is labeled by a null vector $y^{IJ}\in\wedge^2\C^4=\C^6$, with $y\cdot\Phi$ acting as its sole letter. When such a determinant is expanded out, it produces a generating functional for the chiral primaries $\Tr(y\cdot\Phi)^n$. So the correlators of such chiral primaries can be extracted as coefficients in the $y$-expansion of determinant correlators. %This idea is motivated by previous investigations in the context of the physical AdS$_5$/CFT$_4$ duality \cite{Jiang:2019xdz}.

In section \ref{ssec:D5}, we show that such a half-BPS determinant operator is dual to a giant graviton D$5'$ brane that intersects the D5s wrapping twistor space in a twistor line and wraps two out of the four fibers of $\CO(-1)^{\oplus4}$ above it. In this case, the theory living on the brane intersection is a current algebra system on a superline $\P^{1|2}$. This coincides precisely with the $\P^{1|2}$ model of \cite{Caron-Huot:2023wdh}. Decoupling the bulk modes and integrating out the 5-$5'$ and $5'$-5 strings inserts determinants such as $\det(1+y\cdot\Phi)$ in the path integral of $\cN=4$ sdYM, as desired. This establishes a highly efficient holographic dictionary between half-BPS local operators and giant graviton branes in the 14d bulk.

In section \ref{sec:giant}, we study the correlator of bosonic determinant operators $\det(1+y_i\cdot\Phi)(x_i)$ for $i=1,\dots,k$, with an eye toward finding its holographic description. The correlators of half-BPS operators in a shared half-BPS sector are protected. But beyond two and three points, correlators of half-BPS operators in different half-BPS sectors are in general not protected. These are fairly hard to compute in the standard AdS$_5$/CFT$_4$ duality. But owing to the absence of a 't Hooft coupling, in the self-dual theory, they can be reduced at arbitrary multiplicity to the saddle point expansion of a matrix integral \cite{Caron-Huot:2023wdh}. In the non-self-dual theory, this matrix model is expected to get further loop corrections perturbative in the Yang-Mills coupling \cite{Jiang:2019xdz}.

In section \ref{ssec:rho}, we briefly review this matrix model. It is a one-matrix model governing a $k\times k$ matrix $\rho_{ij}$, but with a $\GL_k(\C)$ ``gauge'' symmetry that has been completely broken down to its maximal torus. In section \ref{ssec:giant}, we work out the open string field theory living on a stack of $k$ giants. To begin with, the giants are taken to be coincident. Inspired by the broken $\GL_k(\C)$ symmetry, the $i^\text{th}$ giant is then sent to its desired fiber over the $i^\text{th}$ superline $\P^{1|2}$ by Higgsing this theory. Having separated the giants, we backreact by the D5s wrapping twistor space. This adds a potential energy to the theory on the giants, which allows the open strings stretching across non-intersecting giants to carry nontrivial massless states. 

In section \ref{ssec:map}, we construct a map from saddles of the equations of motion of $\rho_{ij}$ to open string configurations on the giants in the presence of said backreaction. Remarkably, we find a very explicit bulk interpretation of the $\rho_{ij}$: they comprise the states of the strings stretching from the $i^\text{th}$ giant to the $j^\text{th}$ giant. Such open string configurations act as nontrivial vacua of the giants, which is the sense in which the backreaction deforms their ``geometry'' in this higher-dimensional context. Our computation of the correlators of D5$'$ giants is to be contrasted with the case of D$1'$ giants that occur in twisted holography on $\SL_2(\C)$ \cite{Budzik:2021fyh}, whose bosonic geometry genuinely deforms in the presence of backreaction. We review the $\SL_2(\C)$ case in appendix \ref{app:giants}, while also developing a novel, systematic technique for handling backreaction in that setup.

%%%%%%%%%%%%%%%%%%%%%%%%%%%%
%%%%%%%%%%%%%%%%%%%%%%%%%%%%

\section{$\cN=4$ self-dual Yang-Mills}
\label{sec:sdym}

In this section, we provide a a brief review of the self-dual (sd) sector of $\U(N)$ $\cN=4$ super Yang-Mills (sYM) theory. We will refer to this as $\cN=4$ sdYM. We discuss how it arises as a particular free limit of the non-self-dual theory \cite{Siegel:1992xp,Siegel:1992wd,Witten:2003nn}. Following this, we describe its twistor string uplift, which will be used in later sections to engineer its large $N$ holographic dual.

\subsection{A chiral free limit}
\label{ssec:chiral}

Let us begin by considering $\U(N)$ gauge theory on $\R^4$ in Euclidean signature. Self-dual Yang-Mills (sdYM) theory describes a gauge field $A\in\Omega^1(\R^4,\fu(N))$ whose field strength $F = \d A+A^2$ satisfies the self-duality equation,
\be
F = *F\,,
\ee
where $*$ denotes the Hodge dual with respect to the Euclidean metric. Let $F_\pm = (F\pm *F)/2$ denote the self-dual (sd) and anti-self-dual (asd) parts of $F$. The self-duality equation arises as the equation of motion of the action
\be\label{sdac}
S_\sdYM[A,B] = -\int_{\R^4} \Tr\,B\wedge F_-\,,
\ee
where $B\in\Omega^2_-(\R^4,\fu(N))$ is an asd 2-form Lagrange multiplier.

This theory may be interpreted as a free limit of regular gauge theory taken in a chiral manner. To see this, one starts with the Yang-Mills action (including the $\theta$-term)
\be
S_\YM[A] = -\frac{1}{2g_\YM^2}\int_{\R^4}\Tr\,F\wedge *F + \frac{i\theta}{16\pi^2}\int_{\R^4}\Tr\,F\wedge F\,.
\ee
It is useful to express this in terms of $F_\pm$,
\be
S_\YM[A] = \frac{i\tau_+}{8\pi}\int_{\R^4}\Tr\,F_+^2+\frac{i\tau_-}{8\pi}\int_{\R^4}\Tr\,F_-^2\,,
\ee
where wedge products are suppressed for clarity, and we have introduced the complex gauge couplings
\be
\tau_\pm = \frac{\theta}{2\pi} \pm \frac{4\pi i}{g_\YM^2}\,.
\ee
If $\theta$ and $g_\YM$ are real, then $\tau_-$ becomes the complex conjugate of $\tau_+$.

To construct a chiral free limit, we generalize this to the case when $\tau_\pm$ are independent complex couplings. Next, we integrate in our asd 2-form $B$ to express the Yang-Mills action as
\be
S_\YM[A,B] = \frac{i\tau_+}{8\pi}\int_{\R^4}\Tr\,F_+^2 - \int_{\R^4}\Tr\left(B F_- - \frac{2\pi i}{\tau_-}\,B^2\right)\,.
\ee
We recover the action \eqref{sdac} of sdYM from this in the limit
\be
\tau_+\to0\,,\quad\tau_-\to\infty\,.
\ee
Thus, the self-dual theory arises as a chiral limit of the non-self-dual theory. Since sdYM does not possess a coupling constant, it acts as a chiral \emph{free} limit. This makes it a desirable target for studying holography in a simplified setting. 

At the same time, we can study the non-self-dual theory perturbatively around its self-dual limit by sending $\tau_+\to0$ while keeping $\tau_-$ large but finite. This produces the Chalmers-Siegel action for Yang-Mills \cite{Chalmers:1996rq},
\be\label{CSac}
S_\YM[A,B]\bigr|_{\tau_+=0} = -\int_{\R^4}\Tr\left(B F_- + \frac{g_\YM^2}{4}\,B^2\right)\,,
\ee
having eliminated $\theta$ by solving $\tau_+=0$. Since the $\theta$-term is topological, the choice of $\theta$ that sets $\tau_+=0$ yields the same perturbative physics as the choice $\theta=0$. This idea lies at the core of the highly successful MHV formalism that governs scattering amplitudes \cite{Cachazo:2004kj,Cachazo:2004zb,Cachazo:2004by}.

At this stage, it is convenient to pass to spinor notation. The isometry group of $\R^4$ with the flat Euclidean metric is $\SO(4)\cong(\SU(2)\times\SU(2))/\Z_2$. Let $\al=0,1$ and $\dal=\dot0,\dot1$ index the fundamental representation of these two $\SU(2)$ factors, \emph{i.e.} be indices for left- and right-Weyl spinors of $\SO(4)$. We will raise and lower them using 2d Levi-Civita symbols following the conventions of \cite{Adamo:2017qyl}: $\lambda^\al = \eps^{\al\beta}\lambda_\beta$, $\kappa_\beta = \kappa^\al\eps_{\al\beta}$, and similarly for dotted indices. Spinor inner products are indicated by angle or square brackets: $\la\lambda\kappa\ra = \lambda^\al\kappa_\al$, $[\mu\rho]=\mu^\dal\rho_\dal$, \emph{etc}.

The spinor equivalents of $F_{\pm}^{\mu\nu}$, $B^{\mu\nu}$ are of the form $F_+^{\al\dal\beta\dot\beta}=\eps^{\al\beta}F^{\dal\dot\beta}$, $F_-^{\al\dal\beta\dot\beta}=\eps^{\dal\dot\beta}F^{\al\beta}$ and $B^{\al\dal\beta\dot\beta} = \eps^{\dal\dot\beta}B^{\al\beta}$ respectively. In terms of these, the Chalmers-Siegel formulation \eqref{CSac} of Yang-Mills theory can be written as
\be
S_\YM[A,B]\bigr|_{\tau_+=0} = \int_{\R^4}\d^4x\;\Tr\left(B^{\al\beta}F_{\al\beta} + \frac{g_\YM^2}{4}\,B^{\al\beta}B_{\al\beta}\right)\,.
\ee
Self-dual Yang-Mills is obtained by sending the Yang-Mills coupling to zero.

Such a chiral free limit readily generalizes to $\cN=4$ sYM. The $\cN=4$ vector multiplet consists of a gauge field $A_{\al\dal}$, four gauginos $\til\Psi_{I\dal},\Psi^I_{\al}$ of either helicity $\pm1/2$, and six scalars $\Phi_{IJ}$. Continuing to work with the $\theta$ angle that sets $\tau_+$ to zero, the $\cN=4$ sYM action is given by
\begin{multline}
S_{\cN=4} = -\frac{1}{g_\YM^2}\int_{\R^4}\d^4x\;\Tr\,\bigg(F^{\al\beta}F_{\al\beta} + 4\,\til\Psi_{I\dal}D^{\al\dal}\Psi^I_\al + \frac12\,D^{\al\dal}\Phi_{IJ}\,D_{\al\dal}\Phi^{IJ}\\
+ 4\,\Phi^{IJ}\til\Psi_{I\dal}\til\Psi^\dal_J + 4\,\Phi_{IJ}\Psi^I_\al\Psi^{J\al}-2\,\Phi_{IJ}\Phi^{JK}\Phi_{KL}\Phi^{LI}\bigg)\,,
\end{multline}
where we have raised $\SU(4)$ R-symmetry indices using the Levi-Civita symbol $\eps^{IJKL}$,
\be
\Phi^{IJ} \vcentcolon= \frac12\,\eps^{IJKL}\Phi_{KL}\,.
\ee
As in the non-supersymmetric case, we can again integrate in the asd field $B_{\al\beta}$.

To obtain the self-dual limit, as explained in \cite{Witten:2003nn} one rescales some of the fields by appropriate factors of $g_\YM$ and sends $g_\YM\to0$. In detail, assign the fields $A_{\al\dal}$, $\til\Psi_{I\dal}$, $\Phi_{IJ}$, $\Psi^I_\al$, $B_{\al\beta}$ the R-charges $r=0,-1,-2,-3,-4$ respectively. If we rescale $\til\Psi_{I\dal}$, $\Phi_{IJ}$, $\Psi^I_\al$ by the corresponding factors of $(g_\YM/2)^{-r/2}$, the $\cN=4$ sYM action splits as a sum of two terms,
\be
S_{\cN=4} = S_{-4}+\frac{g_\YM^2}{4}\,S_{-8}\,.
\ee
The first term has total R-charge $r=-4$,
\be\label{S-4}
S_{-4} = \int_{\R^4}\d^4x\;\Tr\left(B^{\al\beta}F_{\al\beta} - \til\Psi_{I\dal}D^{\al\dal}\Psi^I_\al- \frac18\,D^{\al\dal}\Phi_{IJ}\,D_{\al\dal}\Phi^{IJ} - \Phi^{IJ}\til\Psi_{I\dal}\til\Psi^\dal_J\right)\,,
\ee
while the second term has R-charge $r=-8$,
\be\label{S-8}
S_{-8} = \int_{\R^4}\d^4x\;L_\text{int} = \int_{\R^4}\d^4x\;\Tr\left(B^{\al\beta}B_{\al\beta} - \Phi_{IJ}\Psi^I_\al\Psi^{J\al} + \frac12\,\Phi_{IJ}\Phi^{JK}\Phi_{KL}\Phi^{LI}\right)\,.
\ee
$S_{-4}$ describes the self-dual sector of $\cN=4$ sYM, which we will abbreviate as $\cN=4$ sdYM. It is obtained in the chiral free limit $g_\YM\to0$. The second term containing $S_{-8}$ describes perturbations away from self-duality. 

%%%%%%%%%%%%%%%%%%%%%%%%%%%%
%%%%%%%%%%%%%%%%%%%%%%%%%%%%

\subsection{Open twistor string theory}
\label{ssec:open}

The crucial property of $\cN=4$ sdYM -- as opposed to free $\cN=4$ sYM -- that allows us to find an intrinsic definition of its holographic dual is its twistor string embedding. Twistor string theory was constructed by Berkovits and Witten in \cite{Witten:2003nn,Berkovits:2004hg,Berkovits:2004jj}. Over the years, it has been adapted to generate a variety of worldsheet formulae for scattering amplitudes. The reader may consult the review \cite{Geyer:2022cey} for a comprehensive list of references. In this section, we outline some of its salient aspects, while adapting it to a more modern formulation.

\paragraph{Twistor geometry.} Let $\P^3$ denote complex projective 3-space with homogeneous coordinates $Z^A$. Split these coordinates as $Z^A=(\mu^\dal,\lambda_\al)$. The twistor space of Euclidean $\R^4$ is the subspace of $\P^3$ obtained by excising the ``line at infinity'' $\P^1=\{\lambda_\al=0\}$,
\be
\PT = \P^3-\P^1\,.
\ee
Let $x^{\al\dal}$ be the spinor equivalent of Cartesian coordinates $x^\mu\in\R^4$. Every spacetime point $x^{\al\dal}$ corresponds to a projective line $\P^1_x\subset\PT$ cut out by Penrose's incidence relations \cite{Penrose:1967wn},
\be
\P^1_x\;:\quad \mu^\dal=x^{\dal\al}\lambda_\al\,.
\ee
For $x^{\dal\al}$ obeying Euclidean reality conditions, these lines are found to be invariant under the fixed-point-free, antiholomorphic involution
\be
\label{involution}
\mu^\dal\mapsto\hat\mu^\dal \vcentcolon= (-\overline{\mu^{\dot1}},\overline{\mu^{\dot0}})\,,\qquad
\lambda_\al\mapsto\hat\lambda_\al \vcentcolon= (-\overline{\lambda_1},\overline{\lambda_0})\,. 
\ee
For every twistor $Z^A$, this determines a point $x\in\R^4$ that corresponds to the line passing through $Z^A$ and $\hat Z^A=(\hat\mu^\dal,\hat\lambda_\al)$. Consequently, $\PT$ fibers over $\R^4$, with the fiber over $x$ being $\P^1_x$. A helpful introduction to this geometry may be found in \cite{Adamo:2017qyl}.

We can describe supertwistor space in a similar manner. One starts with projective superspace $\P^{3|4}$ with four bosonic coordinates $Z^A$ and four fermionic coordinates $\eta^I$ defined up to $\C^*$ rescalings of the supertwistor $\cZ^\bA = (Z^A|\eta^I)$. Supertwistor space is obtained by excising the subset $\P^{1|4}=\{\lambda_\al=0\}$,
\be
\pt = \P^{3|4}-\P^{1|4}\,.
\ee
The on-shell superspace of 4d $\cN=4$ supersymmetry consists of left- and right-moving Grassmann coordinates $\theta^{\al I}$, $\til\theta^\dal_I$. In Euclidean signature, these are independent complex variables, and $\til\theta^\dal_I$ should not be confused with the hermitian conjugates of $\theta^{\al I}$. To accommodate the Grassmann dependence, the incidence relations are generalized to \cite{Ferber:1977qx}
\be
\P^1_{x,\theta}\;:\quad\mu^\dal = (x^{\dal\al}+\til\theta^\dal_I\theta^{\al I})\lambda_\al\,,\quad\eta^I=\theta^{\al I}\lambda_\al\,.
\ee
These projective lines (as well as their bosonic counterparts) are commonly referred to as \emph{twistor lines}. As written, they are adapted to chiral superspace, and we will simplify them by working on the slice $\til\theta^\dal_I=0$ throughout this work.

The excision of the line $\{\lambda_\al=0\}$ equips $\PT$ with the structure of a vector bundle over a Riemann sphere $\P^1$ carrying homogeneous coordinates $\lambda_\al$. It can be identified with the rank 2 holomorphic vector bundle $\CO(1)\oplus\CO(1)\to\P^1$, with fiber coordinates $\mu^\dal$. Similarly, supertwistor space $\pt$ can be viewed as the total space of $\CO(1)^{\oplus2}\oplus\Pi\CO(1)^{\oplus4}\to\P^1$.

\paragraph{Twistor strings.} The utility of twistor theory arises from the fact that we can uplift 4d self-dual theories to 6d holomorphic theories on twistor space. Famously, twistor string theory engineers the twistor uplift of $\cN=4$ sdYM as an open topological string theory \cite{Witten:2003nn}. To describe this construction, we will follow the modern treatment of \cite{Bittleston:2024efo}.

We will study the B-model on the Calabi-Yau 7-fold $\CO(-1)^{\oplus 4}\to\PT$. As before, let $Z^A$ denote homogeneous coordinates on $\PT$, and let $W_I$ denote complex coordinates on the $\C^4$ fibers of $\CO(-1)^{\oplus 4}$. These are defined up to the scaling equivalence
\be
(Z^A,W_I)\sim (tZ^A,t^{-1}W_I)\quad\forall\;t\in\C^*\,.
\ee
By virtue of being Calabi-Yau, this 7-fold comes equipped with the weightless volume form
\be
\Omega = \D^3Z\wedge\d^4W\,,
\ee
where $\D^3Z = \frac{1}{3!}\,\eps_{ABCD}Z^A\d Z^B\wedge\d Z^C\wedge\d Z^D$ is a weight $+4$ volume form on $\PT$ and $\d^4W$ is a weight $-4$ volume form on the fibers.

As reviewed in the introduction, in the B-model the bosonic fields $(Z^A,W_I)$ are accompanied by two sets of fermionic worldsheet scalars. Zero-modes of the first set $(\veps_A,\eta^I)$ represent a basis of the holomorphic tangent bundle $T_X$, while the second set (which we shall mostly suppress) similarly give a basis of the antiholomorphic cotangent bundle $\overline{T}^*_X$. Expanding in powers of these fermions, the BRST-cohomology of the worldsheet CFT then corresponds to the Dolbeault cohomology groups $H^{0,p}(X,\wedge^qT_X)$.

In the open string sector, worldsheet boundary conditions compatible with B-twisted supersymmetry require that fermions representing holomorphic tangent vectors along the brane and (0,1)-forms transverse to the brane are both set to zero. Fermions representing the holomorphic normal directions and (0,1)-forms along the brane are not required to vanish. In particular, if we wrap the 6-real-dimensional zero section of $\CO(-1)^{\oplus 4}\to\PT$ with $N$ D5-branes, then we must set $\veps_A$, but not $\eta^I$, to zero. The open string field on the branes is then
\be
\label{open-superfield}
\cA(Z,\bar{Z},\eta) = \bs{a}(Z,\bar{Z}) + \til{\bs{\psi}}_I(Z,\bar{Z})\eta^I + \frac12\,\bs{\vphi}_{IJ}(Z,\bar{Z})\eta^I\eta^J + \bs{\psi}^I(Z,\bar{Z})\eta^3_I + \bs{b}(Z,\bar{Z})\,\eta^4\,,
\ee
where $\eta^3_L = \eps_{IJKL}\,\eta^I\eta^J\eta^K / 3!$ and $\eta^4 = \eps_{IJKL}\,\eta^I\eta^J\eta^K\eta^L/4!$. That is, in the BV formalism the off-shell string field can be viewed as a superfield
\[
\cA\in\Pi\Omega^{0,*}(\pt,\gl_N(\C))
\]
on supertwistor space $\pt$. This is a $(0,*)$-polyform on $\PT$ with coefficients that are functions of both $Z^A$ and $\eta^I$.\footnote{More generally, suppose $M$ is a supermanifold equipped with the structure of a vector bundle $M\to M_0$ on a bosonic base $M_0$ with fermionic fibers. Then by a $(0,p)$-form on $M$, we will mean a $(0,p)$-form on $M_0$ with coefficients that are functions of both the bosonic and (holomorphic) fermionic coordinates of $M$.}. In particular, the physical gauge field is the (0,1)-form component of $\cA$. The generators $\d\hat Z^A$ of $(0,*)$-forms are taken to be Grassmann odd (\emph{e.g.} they arise from worldsheet fermions), so to ensure the physical gauge field has the usual statistics we must take the overall Grassmann parity of $\cA$ to be odd. $\cA$ should be invariant and basic under scalings $(Z^A,\eta^I)\mapsto (rZ^A,r\eta^I)$.

The open string field theory living on the D5 branes is a holomorphic Chern-Simons theory with action
\be\label{hCS}
S_\text{hCS}[\cA] = \int_{\pt}\D^{3|4}\cZ\;\,\Tr\left(\frac12\,\cA\,\dbar\cA + \frac13\,\cA^3\right)\,,
\ee
where $\D^{3|4}\cZ = \D^3Z\,\d^4\eta$ is a weightless volume form (Berezinian) on supertwistor space. This is exactly the (BV form of the) string field theory for perturbative open strings used in the original twistor string \cite{Witten:2003nn}, but here obtained from D5 branes wrapping bosonic twistor space inside the bosonic Calabi-Yau 7-fold $\CO(-1)^{\oplus4}$. From the bosonic perspective, we view $\cA$ as a polyform valued in antisymmetric powers of the normal bundle $N_{\PT/X}$ to $\PT$ inside $X$, so that the Chern-Simons form is likewise $\wedge^*N_{\PT/X}$-valued. The Berezinian $\D^{3|4}\cZ$ then amounts to saying that we contract this Chern-Simons form into the bosonic Calabi-Yau form $\Omega$ before integrating over the D-brane locus.

As usual, this action has a $\GL_N(\C)$ gauge symmetry, which is a complexified\footnote{The reduction of the gauge group from $\GL_N(\C)$ to its real form $\U(N)$ follows from an extension of Euclidean reality conditions \eqref{involution} to Lie algebra-valued fields. We will mostly be content to continue working with the complex gauge group $\GL_N(\C)$.} uplift of the $\U(N)$ gauge symmetry of $\cN=4$ sdYM. In the BV formalism, the associated target space BRST operator acts on the polyform field $\cA$ as
\be
Q\cA = \dbar\cA + \cA^2\,.
\ee
Nilpotence of $Q$ then arises as a Bianchi-type identity. We can expand $\cA$ in $(0,p)$-forms of definite degree as
\be
\cA = \cA_0 + \cA_1 + \cA_2 + \cA_3\,,
\ee
where $\cA_p$ is a $(0,p)$-form on $\PT$. The components of $\cA_p$ are functions of the supertwistor coordinates $(Z^A,\eta^I)$ and have Grassmann parity $(-1)^{p+1}$. The BRST variations of $\cA_0$ and $\cA_1$ read
\be
Q\cA_0 = \cA_0^2 = \frac{1}{2}\,[\cA_0,\cA_0]\,,\qquad Q\cA_1 = \dbar\cA_0 + [\cA_1,\cA_0]\,,
\ee
allowing us to interpret the $0$-form field $\cA_0$ as containing the ghosts of $\GL_N(\C)$ gauge symmetries. In turn, the physical $(0,1)$-form field $\cA_1$ has its standard BRST variation.  For completeness, let us write out the superfield expansion of the physical field as
\be
\label{A1}
\cA_1 = a + \til\psi_I\eta^I +\frac12\,\vphi_{IJ}\eta^I\eta^J + \psi^I\eta^3_I + b\,\eta^4\,,
\ee
which will be used later. In particular, the field $a$ plays the role of a $\GL_N(\C)$ gauge field, and the other component fields in the $\cA_1$ supermultiplet are valued in the adjoint. The remaining fields $\cA_2,\cA_3$ comprise the BV antifields of the physical fields and the ghosts respectively.

\medskip

That open strings on D5 branes wrapping $\PT\subset \CO(-1)^{\oplus4}$ are the same thing as open strings on D-branes filling the supermanifold $\pt$ has been pointed out previously in \cite{Bittleston:2024efo}. As in the AKSZ formalism \cite{Alexandrov:1995kv}, instead of identifying zero-modes of the fermionic fields $(\veps_A,\eta^I)$ with a basis of $T_X$ and using them to form polyvectors, we can just view them as coordinates on the fibers of the parity-reversed holomorphic cotangent bundle $\Pi T^*_X$, where the parity-reversal operation $\Pi$ recognises that these coordinates are fermionic. Likewise the fermions representing $(0,p)$-forms on $X$ can be thought of a coordinates on $\Pi \br{T}_X$. Thus the space $\Omega^{0,*}(X,\wedge^*T_X)$ can equally be thought of as a space of functions on the total space of the bundle $\Pi T^*_X\oplus\Pi\br{T}_X\to X$, where we are taking the fiberwise direct sum. 

Concentrating as usual on holomorphic directions, let $\pi : X\to\PT$ be the projection that forgets the fiber coordinates $W_I$. Then in our case the holomorphic tangent bundle $T_X$ is defined by the exact sequence
\be
    0 \to T_{X/\PT} \to T_X \to \pi^* T_\PT \to 0\,,
\ee
where $T_{X/\PT}\cong\CO_\PT(-1)^{\oplus4}$ is the vertical tangent bundle. Dualising to the cotangent bundle and applying the parity reversal gives
\be
    0 \to \Pi(\pi^*T^*_\PT) \to \Pi T^*_X \to  \Pi (\CO(1)^{\oplus4}_{\PT}) \to 0\,.
\ee
Explicitly, if we use homogeneous coordinates $(Z^A,W_I)\sim (rZ^A,r^{-1}W_I)$ on $X$ and fermionic coordinates $(\veps_A,\eta^I)\sim(r^{-1}\veps_A,r\eta^I)$ on the fibers, then $\Pi T^*_X$ is defined by the relation 
\be
\label{euler}
 Z^A\veps_A - W_I \eta^I =0\,.
\ee
($\Pi(\pi^*T^*_\PT)$ is the subbundle given by $\eta^I=0$ while the projection to $\Pi(\CO^{\oplus4}_\PT(1))$ forgets the $\eta^I$s.) The relation \eqref{euler} amounts to the requirement that forms are basic, \emph{i.e.} have no component along the Euler vector field $Z^A\p_A - W_I\p^I$ associated to scalings of the homogeneous coordinates. From the worldsheet perspective, this constraint arises as a superpartner to the constraint that, after twisting, the worldsheet fields must lie in the minimum of the worldsheet superpotential defining $X$ as a symplectic quotient. 

The coordinates $(Z^A,W_I\,|\,\veps_A,\eta^I)$ may be split into a supertwistor and a parity-reversed dual supertwistor,
\be\label{ZW}
\cZ^\bA = (Z^A|\eta^I)\,,\qquad\cW_\bA = (\veps_A|W_I)\,.
\ee
$\Pi T^*_X$ is cut out by the parity-reversed quadric
\be
\cZ\cdot\cW \equiv Z^A\veps_A-W_I\eta^I = 0\,.
\ee
In this notation, we can also introduce a convenient weightless volume form on $\Pi T^*_X$,
\be\label{bomega}
\bs\Omega = \underset{\cZ\cdot\cW=0}{\text{Res}}\frac{\D^{3|4}\cZ\,\d^{4|4}\cW}{\cZ\cdot\cW}\,.
\ee
Finally, supertwistor space is recovered as the submanifold 
\be
\pt=\{\cW_\bA = 0\}\subset \Pi T^*_X\,.
\ee
We discuss this supergeometry further in section \ref{ssec:deformed}.

\paragraph{Reduction to spacetime.} Details of the reduction to $\R^4$ of the $\cN=4$ holomorphic Chern-Simons action \eqref{hCS} can be found in \cite{Boels:2006ir,Adamo:2011pv}. Taking the perspective emphasised in \cite{Costello:2021bah,Costello:2022wso}, in essence, the idea is that since $\PT\cong S^2\times \R^4$ as a smooth 6-manifold, we can envisage performing a Kaluza-Klein expansion of the twistor fields. Since the Chern-Simons action does not involve any metric on $\PT$, we are free to gauge fix by choosing a metric that makes vol$(S^2)$ arbitrarily small. With this gauge fixing, all but the lightest KK modes (representing cohomology along the $S^2$ fibers) decouple, showing that the holomorphic theory on $\PT$ is gauge equivalent to a four-dimensional theory. Understanding some of the details of this reduction will enable us to relate twistor space operators to local operators on $\R^4$, so let us briefly collect the facts that will use later. For our purposes, it will suffice to keep only the physical $(0,1)$-form fields $\cA_1$ and set the ghosts and antifields to zero.

Working on the slice $\til\theta^\dal_I=0$, we first introduce a basis of $(0,1)$-forms on $\PT$ that are adapted to the fibration $\PT \to \R^4$. We let
\be
\bar e^0 = \frac{\la\hat\lambda\,\d\hat\lambda\ra}{\|\lambda\|^4}\,,\qquad\bar e^\dal = -\frac{\hat\lambda_\al\d x^{\dal\al}}{\|\lambda\|^2}\,,
\ee
be our basis, while 
\be
\dbar_0 = \|\lambda\|^2\lambda_\al\frac{\p}{\p\hat\lambda_\al}\,,\qquad\dbar_\dal = \lambda^\al\p_{\dal\al}
\ee
form the corresponding dual basis of $(0,1)$-vectors. The factors of $\|\lambda\|^2 = \la\hat\lambda\lambda\ra = |\lambda_0|^2+|\lambda_1|^2$ in these expressions ensure that these differential forms have zero homogeneity in $\hat\lambda_\al$. 

Twistor space has a larger gauge redundancy than $\R^4$ as gauge transformations involve a choice of a smooth $\mathfrak{u}(N)$-valued function of 6 real variables. It is this larger redundancy that allows a six-dimensional theory to be equivalent to a four-dimensional one. We can fix the additional redundancy by demanding the $(0,1)$-parts of $\cA$ are harmonic along the $S^2$ fibers of $\PT\to\R^4$ with respect to the Fubini-Study metric on $S^2\cong\P^1$, as in \cite{Woodhouse:1985id}. This leaves the standard gauge redundancy on $\R^4$ unfixed. In this gauge, the component physical fields in \eqref{A1} can be expressed as
\begingroup
\allowdisplaybreaks
\begin{align}
    a &= a_{\dal}\bar e^\dal\,,\label{aw}\\
    \til\psi_I &= \til\psi_{I\dal}\bar e^\dal\,,\label{psitw}\\
    \vphi_{IJ} &= \Phi_{IJ}\bar e^0 + \vphi_{IJ\dal}\bar e^\dal\,,\label{phiw}\\
    \psi^I &= \frac{2\,\hat\lambda^\al\Psi^I_\al\bar e^0}{\|\lambda\|^2} + \psi^I_\dal\bar e^\dal\,,\label{psiw}\\
    b &= \frac{3\,\hat\lambda^\al\hat\lambda^\beta B_{\al\beta}\bar e^0}{\|\lambda\|^4} + b_\dal\bar e^\dal\,,\label{bw}
\end{align}
\endgroup
where $\Phi_{IJ},\Psi^I_\al,B_{\al\beta}$ are identified with some of the familiar spacetime fields of the $\cN=4$ vector multiplet. The extra data $a_\dal(x,\lambda),\til\psi_{I\dal}(x,\lambda),\vphi_{IJ\dal}(x,\lambda)$, $\psi_\dal^I(x,\lambda)$ and $b_\dal(x,\lambda)$ are not fixed by this gauge choice alone. Instead, we can determine their on-shell values by solving the vertical equations of motion, which amounts to finding their lowest KK modes. This leads to~\cite{Mason:2005zm,Boels:2006ir,Adamo:2017qyl}
\begingroup
\allowdisplaybreaks
\begin{align}
    a_\dal &= \lambda^\al A_{\dal\al}\,,\label{adal}\\
    \til\psi_{I\dal} &= \til\Psi_{I\dal}\,,\label{psitdal}\\
    \vphi_{IJ\dal} &= \frac{\hat\lambda^\al D_{\dal\al}\Phi_{IJ}}{\|\lambda\|^2}\,,\label{phidal}\\
    \psi^I_\dal &= \frac{\hat\lambda^\al\hat\lambda^\beta D_{\dal\al}\Psi^I_\beta}{\|\lambda\|^4}\,,\label{psidal}\\
    b_\dal &= \frac{\hat\lambda^\al\hat\lambda^\beta \hat\lambda^\gamma D_{\dal\al}B_{\beta\gamma}}{\|\lambda\|^6}\,,\label{bdal}
\end{align}
\endgroup
where $D_{\dal\al} = \p_{\dal\al}+A_{\dal\al}$ is the covariant derivative on $\R^4$. Inserting \eqref{aw}-\eqref{bdal} into the Chern-Simons action, all the dependence on the $S^2$ fibers becomes explicit and one can integrate them out. Doing so, one obtains the action~\eqref{S-4} for $\cN=4$ sdYM (see \cite{Mason:2005zm,Boels:2006ir,Adamo:2017qyl} for further details).

The spectrum of local operators is to be constructed modulo the equations of motion. In the holomorphic Chern-Simons theory (neglecting the antifields) this is
\be
    \dbar\cA_1 + \cA_1^2 =0
\ee
stating that we have a holomorphic bundle on $\pt$. Expanding out in terms of component fields \eqref{A1} this becomes
\begingroup
\allowdisplaybreaks
\begin{align}
    &\dbar a + a^2 = 0\,,\label{aeq}\\
    &\dbar_a\til\psi_I = 0\,,\\
    &\dbar_a\vphi_{IJ} + [\til\psi_I,\til\psi_J] = 0\,,\\
    &\dbar_a\psi^I + [\til\psi_J,\vphi^{IJ}] = 0\,,\\
    &\dbar_ab + [\psi^I,\til\psi_I] + \frac14\,[\vphi^{IJ},\vphi_{IJ}] = 0\,,\label{beq}
\end{align}
\endgroup
where $\dbar_a\til\psi_I=\dbar\til\psi_I + [a,\til\psi_I]$, \emph{etc.}\ denote covariant derivatives acting in the adjoint. In particular, \eqref{aeq} gives a standard holomorphic bundle on $\PT$ which is equivalent to a self-dual YM bundle on $\R^4$ by the Penrose-Ward construction \cite{Ward:1977ta}. We used the vertical parts of these equations of motion -- in other words the equations $\bar{e}^\dal\wedge(\dbar\cA_1+\cA_1^2)=0$ -- to arrive at \eqref{adal}-\eqref{bdal}. The remaining equation is $\bar{e}^0\wedge(\dbar \cA_1+\cA_1^2)=0$ which yields the equations
\begingroup
\allowdisplaybreaks
\begin{align}
    &F_{\al\beta} = 0\,,\label{Aeq}\\
    &D_{\al\dal}\til\Psi_I^\dal = 0\,,\label{Psiteq}\\
    &D^2\Phi_{IJ} - 2\,[\til\Psi_{I\dal},\til\Psi_J^\dal] = 0\,,\label{Phieq}\\
    &D_{\al\dal}\Psi^{I\al} + [\Phi^{IJ},\til\Psi_{J\dal}] = 0\,,\label{Psieq}\\
    &D_{\beta\dal}B^\beta_\al + [\Psi^I_\al,\til\Psi_{I\dal}] - \frac14\,[\Phi^{IJ},D_{\al\dal}\Phi_{IJ}] = 0\,.\label{Beq}
\end{align}
\endgroup
on the fields on $\R^4$. These agree with the equations of motion of the action \eqref{S-4} describing $\cN=4$ sdYM.

%%%%%%%%%%%%%%%%%%%%%%%%%%%%
%%%%%%%%%%%%%%%%%%%%%%%%%%%%

\section{The gravitational dual}
\label{sec:dual}

In the previous section, we introduced the B-model topological string on the Calabi-Yau 7-fold $X=\CO(-1)^{\oplus4}\to\PT$. We reviewed how the open string field theory on a stack of $N$ D5 branes wrapping the zero section $\PT$ engineers the twistor action of $\U(N)$ $\cN=4$ sdYM. In this section, we describe the closed string theory of this B-model. At large $N$, the D5 branes get replaced by their backreaction. This allows us to conjecture a holographic duality between $\cN=4$ sdYM and closed twistor string theory in the presence of the backreaction. It provides a chiral analog of the more standard AdS/CFT dualities introduced by Maldacena \cite{Maldacena:1997re}.

\subsection{Closed twistor string theory}
\label{ssec:bcov}

In the past, perturbative closed twistor strings have been argued to describe models of self-dual conformal supergravity on $\R^4$ \cite{Berkovits:2004jj}. In our context, in principle this could be done by restricting the closed string excitations to the zero section of $X$, for instance, by an appropriate compactification of the $\CO(-1)^{\oplus4}$ fibers. However, we instead want to study the backreaction of branes wrapping the zero section $\PT$, so we want the directions normal to $\PT$ to remain non-compact. This entails that the closed string theory be a theory living in the full 7-complex-dimensional geometry $X$. It is a (topological) string theory in 14 real dimensions!

The closed string field theory of our B-model is the BCOV theory of \cite{Costello:2012cy}. This is a higher-dimensional generalization of the theory of Kodaira-Spencer gravity introduced by Bershadsky, Cecotti, Ooguri and Vafa in \cite{Bershadsky:1993cx}. As mentioned above, the spectrum of BCOV theory consists of $(0,q)$-forms valued in $(p,0)$-polyvectors. Let 
\be
\PV^{q,p}(X) \vcentcolon= \Omega^{0,p}(X,\wedge^q T_X)\,,
\ee
where $T_X$ denotes the holomorphic tangent bundle of $X$. The complete closed string spectrum of BCOV theory on $X$ can be packaged as a field
\be
\Bsi\in\bigoplus_{p,q}\PV^{q,p}(X)\,.
\ee
Contracting with the Calabi-Yau volume form on $X$ allows us dualize polyvectors into differential forms as
\be
\label{polyvectors2diffforms}
\Bsi\ip\Omega\in\bigoplus_{p,q}\Omega^{7-q,p}(X)\,.
\ee
This provides an isomorphic presentation of the spectrum. We can use this isomorphism to define a divergence operator $\p_\Omega:\PV^{q,p}\to\PV^{q-1,p}$ by $(\p_\Omega\Bsi)\ip\Omega = \p(\Bsi\ip\Omega)$, where $\p$ is the standard holomorphic exterior derivative. The level-matching constraint of the worldsheet theory of the closed string B-model translates into the constraint that $\Bsi$ be divergence-free \cite{Bershadsky:1993cx,Costello:2019jsy}, so
\be
\p(\Bsi\ip\Omega) = 0\,.
\ee
In particular, for $p=q=1$ this is the requirement that the target space Beltrami differential $\beta\in\Omega^{0,1}(X,T_X)$ Lie derives the Calabi-Yau volume form.

An equivalent but occasionally more convenient description of the same spectrum can be formulated in terms of the target superspace $\Pi T^*_X$. The Grassmann coordinates $(\veps_A,\eta^I)$ along the fibres of $\Pi T^*_X\to X$ can be reinterpreted as frames of the holomorphic tangent bundle by making the identifications
\be\label{fermitovec}
\frac{\p}{\p Z^A}\leftrightarrow\veps_A\,,\qquad \frac{\p}{\p W_I}\leftrightarrow\eta^I
\ee
in all polyvectors. With this identification, polyvectors on $X$ can be treated as functions on $\Pi T^*_X$, as in \cite{Alexandrov:1995kv}. For instance, a polyvector $\beta^A\p_A+\beta_I\p^I$ can be represented as $\beta^A\veps_A + \beta_I\eta^I$. Wedge products of vectors become ordinary products of Grassmann variables. We will use the two formulations interchangeably. For example, in the second formulation the divergence operator  is written as 
\be
\p_\Omega = \frac{\p}{\p\veps_A}\frac{\p}{\p Z^A} + \frac{\p}{\p\eta^I}\frac{\p}{\p W_I}\,.
\ee
in terms of coordinates on $\Pi T_X^*$. In particular, this makes it clear that $\p_\Omega^2=0$ so that the divergence operator $\p_\Omega$ is a nilpotent.

The divergence-free constraint is hard to impose in a local manner, so one often uses nilpotence to replace the condition that $\Bsi\in\ker\p_\Omega$ by the assumption that $\Bsi\in\text{im}\,\p_\Omega$. That is, at least locally on $X$ there exists a potential $\bs{\Gamma}$ satisfying
\be
\Bsi = \p_\Omega\bs{\Gamma}\,.
\ee
Intuitively, $\Bsi$ is analogous to RR form field strengths in type IIB string theory, and $\bs\Gamma$ is analogous to their RR form gauge potentials. One consequence of this assumption is that $\Bsi$ does not have any components valued in $\PV^{7,*}(X)$, as any such components would have belonged to the image of $\PV^{8,*}(X)$ under $\p_\Omega$, which is trivial since $\PV^{q,*}(X)=0$ for $q\geq8$. %Henceforth, we abuse notation slightly by denoting $\p_\Omega$ simply as $\p$. The identification of $\PV^{q,p}$ with $\Omega^{7-q,p}(X)$ in \eqref{polyvectors2diffforms} ensures there is no real ambiguity here.

\medskip

The classical action of BCOV theory is a mildly non-local functional of the closed string field $\Bsi$,
\be\label{bcov}
S_\text{BCOV} = \int_X\Omega\wedge\left(\frac12\,\p_\Omega^{-1}\Bsi\,\dbar\Bsi + \frac{1}{3!}\,\Bsi^3\right)\ip\Omega\,.
\ee
It may be written as a local action for the potential $\bs\Gamma$ if needed. Its fields are taken up to the (infinitesimal) gauge transformations
\be
\delta\Bsi = \dbar\bs{f} + [\Bsi,\bs{f}]\,,
\ee
where $\bs{f}\in\mathrm{im}\,\p_\Omega$ is another polyvector-valued polyform. Occasionally, it is helpful to write this action as an integral over the superspace $\Pi T^*_X$,
\be\label{bcovsuper}
S_\text{BCOV} = \int_{\Pi T^*_X}\bs{\Omega}\wedge\left(\frac12\,\p_\Omega^{-1}\Bsi\,\dbar\Bsi + \frac{1}{3!}\,\Bsi^3\right)\,,
\ee
where $\bs\Omega$ is the volume form on $\Pi T^*_X$ given in \eqref{bomega}. In this expression, $\Bsi$ is treated as a superfield in $\veps_A,\eta^I$ instead of as a polyvector in the span of $\p_A,\p^I$. This rewriting puts closed twistor string theory on the same footing as the supertwistor action \eqref{hCS} describing the open twistor string.

The equation of motion following from the BCOV action is a Maurer-Cartan equation,
\be\label{bcoveq}
\dbar\Bsi + \frac12\,[\Bsi,\Bsi] = 0\,.
\ee
In this equation, $[-,-]$ denotes the Schouten bracket of polyvector fields.\footnote{The distinction between the Schouten bracket in the closed string case and the Lie algebra bracket in the open string case will hopefully be clear from context.} On a pair of polyvectors $\Bsi,\Bsi'$ of definite Grassmann degree, this is given by
\be
[\Bsi,\Bsi'] = \p_\Omega(\Bsi\wedge\Bsi') - \p_\Omega\Bsi\wedge\Bsi' - (-1)^{\deg\Bsi}\Bsi\wedge\p_\Omega\Bsi'\,.
\ee
If both $\Bsi,\Bsi'$ are divergence-free, then this reduces to the simple expression:
\be
\p_\Omega\Bsi = \p_\Omega\Bsi' = 0\quad \implies\quad [\Bsi,\Bsi']=\p_\Omega(\Bsi\wedge\Bsi')\,.
\ee
This allows one to recover \eqref{bcoveq} from a variation of \eqref{bcov}. Since \eqref{bcoveq} is a local field equation, perturbative BCOV theory remains perfectly well-defined.

It is helpful to make this equation more explicit. To do this, let us write $\Bsi$ as a sum of polyvectors of increasing degree,
\be\label{Psiexp}
\Bsi = \al + \beta + \gamma + \xi + \til\gamma + \til\beta + \til\al\,, %\zeta + \chi + \omega\,,
\ee
where $\al\in\PV^{0,*}$, $\beta\in\PV^{1,*}$, $\gamma\in\PV^{2,*}$, $\xi\in\PV^{3,*}$, $\til\gamma\in\PV^{4,*}$, $\til\beta\in\PV^{5,*}$ and $\til\al\in\PV^{6,*}$. The total field $\Bsi$ is taken to be graded even, and the frames $\{d\hat Z^A,\ldots\}$ of $(0,p)$-forms and frames $\{\veps_A,\ldots\}$ of $(q,0)$-vectors are understood as being Grassmann odd. This allows one to read off the Grassmann parity of the coefficients of each component field in this expansion. Similarly, the operator $\dbar$ is viewed as being Grassmann odd.

The Schouten bracket $[v,w]$ of two fields $v\in\PV^{k,r}(X),w\in\PV^{l,s}(X)$ belongs to $\PV^{k+l-1,r+s}(X)$. Therefore, the classical equations \eqref{bcoveq} expand out to give
% \begingroup
% \allowdisplaybreaks
\begin{align}
    &\dbar\al + [\beta,\al] = 0\,,\\
    &\dbar\beta + [\al,\gamma] + \frac12\,[\beta,\beta] = 0\,,\\
    &\dbar\gamma + [\al,\xi] + [\beta,\gamma] = 0\,,\\
    &\dbar\xi + [\al,\til\gamma] + [\beta,\xi] + \frac12\,[\gamma,\gamma] = 0\,,\label{xieq}\\
    &\dbar\til\gamma + [\al,\til\beta] + [\beta,\til\gamma] + [\gamma,\xi] = 0\,,\\
    &\dbar\til\beta + [\al,\til\al] + [\beta,\til\beta] + [\gamma,\til\gamma] + \frac12\,[\xi,\xi] = 0\,,\\
    &\dbar\til\al + [\beta,\til\al] + [\gamma,\til\beta] + [\xi,\til\gamma] = 0\,.
\end{align}
%\endgroup
As a simple situation, consider the case when only a field $\beta\in\PV^{1,1}(X)$ is turned on. Such a field is a Beltrami differential and represents a complex structure deformation of $X$. In the absence of $\al,\gamma$, we see that its equation of motion reduces to $\dbar\beta+\frac12\,[\beta,\beta]=0$. When this is obeyed, the deformed dbar operator $\dbar_\beta\equiv\dbar+[\beta,-]$ describes an integrable almost complex structure. This is the sense in which BCOV theory can be thought of as a ``gravitational'' theory. The divergence-free constraint  ensures that $\beta$ Lie derives the CY volume form so that $X$ is deformed as a CY manifold, rather than just as a generic complex manifold. 

%%%%%%%%%%%%%%%%%%%%%%%%%%%%
%%%%%%%%%%%%%%%%%%%%%%%%%%%%

\subsection{Brane backreaction}
\label{backreaction}

We can replace our stack of $N$ D5 branes wrapping $\PT\subset X$ by their backreaction. Then the closed string B-model in the presence of this backreaction will be holographically dual to the twistor uplift of $\cN=4$ sdYM.

This is an instance of \emph{twisted holography}, \emph{i.e.}, a holographic duality for a topological string \cite{Costello:2018zrm}. A key simplification that happens in twisted holography -- as opposed to standard holography -- is that the decoupling (near-horizon) limit is trivialized. This is a consequence of the B-model localising on constant maps, so that open strings are fully confined to the brane and cannot probe the bulk geometry. As a result, holography in the B-model can be usefully thought of as an open-closed duality.

To compute the backreaction, we recall that just as D-branes in the RNS string source RR forms, topological D-branes in the B-model source various components of the closed string field $\Bsi$. Our $N$ D5 branes wrapping $\PT$ add a tadpole to the closed string action,
\be\label{tadpole}
N\int_{\PT}\p^{-1}(\Bsi\ip\Omega) = N\int_{\PT}\p^{-1}(\xi\,\ip\Omega)\,,
\ee
where $\xi$ is the $\PV^{3,*}(X)$ component of $\Bsi$, and the integral over $\PT$ picks the $(3,3)$-form part of $\p^{-1}(\xi\,\ip\Omega)\in\Omega^{3,*}(X)$.

This tadpole adds a delta function to the right-hand side of the equation of motion \eqref{xieq} of $\xi$. We look for solutions in which only $\xi$ is turned on and all other closed string fields are set to zero. With this ansatz, the equation obeyed by $\xi$ reduces to 
\be
\dbar\xi\,\ip\Omega = N\,\bar\delta^4(W)\,\d^4W\,.
\ee
Note that this is a consistent equation because $\xi\,\ip\Omega\in\Omega^{4,*}(X)$. In a standard harmonic gauge, the solution for $\xi\,\ip\Omega$ can be written in terms of the Bochner-Martinelli kernel on the $\C^4$ fibers of $\CO(-1)^{\oplus 4}$,
\be\label{xiON}
\xi\,\ip\Omega = \frac{N\,\D^3\br W\,\d^4W}{\|W\|^8}\,.
\ee
Here, $\|W\|^2=\br W^I W_I$ denotes the Euclidean norm on $\C^4$, with $\br W^I$ being the hermitian conjugate of $W_I$, and we have abbreviated $\D^3\br W = \frac{1}{3!}\,\eps_{IJKL}\br W^I\d\br W^J\d\br W^K\d\br W^L$. We are suppressing the wedge products by thinking of $\d W_I,\d\br W^I$ as Grassmann variables. We have also dropped a normalization factor of $3!/(2\pi i)^4$ that occurs in this Bochner-Martinelli kernel, as it may always be absorbed in a rescaling of the volume form. 

While \eqref{xiON} gives the form of $\xi\ip\Omega$, explicit representatives of the polyvector $\xi\in\PV^{3,3}(X)$ itself can be found patch-by-patch on $X$ by remembering that $\Omega=\D^3Z\,\d^4W$. For example, pick an affine patch $\{A\cdot Z\equiv A_AZ^A\neq 0\}$ of twistor space, where $A_A$ is a `reference' dual twistor, and pick the associated patch of $\CO(-1)^{\oplus4}$. On this patch, we can write
\be\label{xiN}
\xi = \xi_N \equiv \frac{N}{A\cdot Z}\frac{\D^3\br W}{\|W\|^8}\,\eps^{ABCD}A_A\p_B\p_C\p_D\,.
\ee
It is only the $(4,3)$-form $\xi_N\ip\Omega$ in which the data of the reference dual twistor manifestly drops out. This is compatible with the isomorphism $\PV^{3,3}(X)\cong\Omega^{0,3}(X,\wedge^3T_X)$ in the sense that two different choices of reference $A_A$ lead to polyvectors $\xi_N$ that differ non-projectively on $\C^8$, but descend to the same polyvector on $X$ (on overlaps where both are defined) up to terms that are pure gauge or are proportional to the Euler vector \eqref{euler}.

The $(4,3)$-form backreaction found in \eqref{xiON} has a singularity at the zero section $\PT=\{W_I=0\}$ where the D5 branes had been wrapped. To describe the gravitational side of our holographic duality, we excise this locus from our target space $X=\CO(-1)^{\oplus 4}$ to define
\be
\til X \vcentcolon= X - \PT\,.
\ee
At this stage, we can finally present our holographic proposal in the large $N$ limit.
\[\colorbox{gray!15}{
$
\hspace{1cm}
\begin{aligned}
\rule[-0.3ex]{0pt}{2.5ex}\\[-1em]
\parbox[c]{3cm}{\centering $\U(N)$ $\cN=4$ sdYM on $\R^4$}
&\overset{N\to\infty}{=}\quad\!
\parbox[c]{4cm}{\centering B-model on $\til X$ with $N$ units of $(4,3)$-form flux}\\[-1em]\rule[-0.3ex]{0pt}{2.5ex}
\end{aligned}
\hspace{1cm}$}\]
We reiterate that the gauge theory side of this duality has no 't Hooft coupling. This is consistent with the fact that the gravitational side is a topological string with no $\al'$ dependence. The only nontrivial parameter is $N$, the rank of the gauge group.

\paragraph{Boundary structure.} The geometry of the 14-real-dimensional space $\til X$ has a natural boundary structure. $\til X$ is obtained by removing the zero section of $\CO(-1)^{\oplus 4}\to\PT$. This can be made manifest by splitting $W_I$ into a coordinate $n$ that captures its scale and a coordinate $Y_I$ that is defined projectively:
\be
W_I = \frac{1}{n}\,Y_I\,,
\ee
where $n\in\C^*$ and $Y_I\in\P^3$. The excised locus is $n=\infty$, so we can take $n,Z^A,Y_I$ as new coordinates on $\til X$. These are defined up to the rescalings
\be
(n,Z^A,Y_I)\sim (st n,sZ^A,tY_I)\quad\forall\;s,t\in\C^*\,.
\ee
This allows us to identify $\til X$ with the total space of the line bundle $\CO(1,1)\to\PT\times\P^3$
minus its zero section $n=0$ where $W_I=\infty$. The region of $\til X$ where $W_I\to\infty$ forms its asymptotic boundary.

It is useful to partially compactify $\til X$ by explicitly attaching back the zero section $n=0$. Hence we define
\be
\br X = \CO(1,1)\to\PT\times\P^3\,.
\ee
The boundary divisor at $n=0$ is a copy of $\PT\times\P^3$. This is similar to how the boundary of (the Poincar{\'e} patch of) AdS$_5\times S^5$ is given by $\R^{4}\times S^5$. The $\P^3$ factor in $\PT\times\P^3$ is analogous to the R-symmetry sphere $S^5$, whereas the $\PT$ factor supports the boundary dual comprised of the $\cN=4$ sdYM theory. Lastly, $n$ acts as the emergent radial coordinate pointing inward into the bulk.

\paragraph{Matching global symmetries.} With this rewriting of the geometry, it is easy to match the symmetries of the bulk and boundary theories. As we reviewed before, the boundary $\cN=4$ sdYM theory on $\R^4$ is obtained by compactifying the open string theory on $\PT$ along the twistor fibration $\P^1\hookrightarrow\PT\to\R^4$. This theory has an $\SO(1,5)$ Euclidean conformal symmetry with double cover $\Spin(1,5)$, which uplifts to a complexified $\SL_4(\C)$ action on $\PT$.\footnote{Strictly speaking, the conformal symmetry acts on the compactification $\P^3$, which is the twistor space of $S^4$ instead of $\R^4$. Since $\cN=4$ sYM is a CFT, this distinction is innocuous.} The twistor coordinates $Z^A$ transform in the fundamental of this $\SL_4(\C)$. The fibration of $\PT\to\R^4$ involves choosing a real structure on $\PT$ which breaks $\SL_4(\C)$ down to $\Spin(1,5)$.

Similarly, $\cN=4$ sdYM has an $\SU(4)$ R-symmetry. This uplifts to an $\SL_4(\C)$ symmetry that acts on the internal $\P^3$ with coordinates $Y_I$. This gets broken to the real form $\SU(4)$ by the presence of the Euclidean norm $\|W\|^2$ in the expression for the backreaction \eqref{xiON}. The supersymmetries on spacetime similarly uplift to act on supertwistor coordinates $\eta^I$ in a well-known fashion.

%%%%%%%%%%%%%%%%%%%%%%%%%%%%
%%%%%%%%%%%%%%%%%%%%%%%%%%%%

\subsection{The deformed supergeometry}
\label{ssec:deformed}

To better understand the closed string side of this duality, it is useful to explore the geometric effect of turning on the backreaction \eqref{xiON}. Expanding the BCOV action \eqref{bcov} around the brane backreaction cancels the brane tadpole and yields a new action on $\til X$,
\be\label{bcovN}
S_\text{BCOV} = \int_{\til X}\Omega\wedge\left(\frac12\,\p^{-1}_\Omega\Bsi\,\dbar_N\Bsi + \frac{1}{3!}\,\Bsi^3\right)\ip\Omega\,.
\ee
In this action, the kinetic operator has been deformed from $\dbar$ to
\be
\dbar_N = \dbar + [\xi_N,-]\,.
\ee
Explicitly, this has been obtained by expanding $\xi\mapsto\xi_N+\xi$, where $\xi_N$ denotes the backreaction \eqref{xiN}. Varying it produces the background-coupled Maurer-Cartan equation,
\be
\dbar_N\Bsi + \frac12\,[\Bsi,\Bsi] = 0\,.
\ee
States of the gravitational theory can be found by solving its linearization $\dbar_N\Bsi = 0$. Since $\dbar_N^2=0$ on $\til X$, such states are only defined modulo linearized gauge transformations $\Bsi\sim\Bsi + \dbar_N\bs{f}$, so this reduces to a cohomology problem. 

We can systematically obtain solutions of the linearized equation by noting that it is very similar to a Cauchy-Riemann equation. In the quintessential example of twisted holography developed in \cite{Costello:2018zrm}, the authors find that D1 branes in the B-model on $\C^3$ backreact to generate a Beltrami differential background. In that case, the linearized field equation becomes an actual Cauchy-Riemann equation. In our case, the backreaction is valued in $\PV^{3,3}(X)$, so it does not correspond to a deformation of purely the bosonic geometry of $X$. Instead, we must think of $\dbar_N$ as giving rise to a deformation of the supergeometry.

Previously, we mentioned that the target space of the B-model on $X$ is the supermanifold $\Pi T^*_X\oplus\Pi\br{T_X}\to X$. In this framework, $(Z^A,W_I)$ and $(\veps_A,\eta^I)$ acted as coordinates on the base and fibers of $\Pi T^*_X\to X$. In the same way, we can formally identify the generators $\d\hat Z^A,\d\br W^I$ of $(0,*)$-forms as Grassmann coordinates on the fibers of $\Pi\br{T_X}$. Both $\veps_A,\eta^I$ and $\d\hat Z^A,\d\br W^I$ arise from zero modes of spin-0 worldsheet fermions, which places them on an equal footing.

Before turning on the backreaction, all the coordinates $(Z^A,W_I\,|\,\veps_A,\eta^I,\d\hat Z^A,\d\br W^I)$ on our target superspace were $\dbar$-closed, \emph{i.e.}, were holomorphic. The backreaction deforms $\dbar\mapsto\dbar_N$. Because the Schouten bracket does not act on $\d\hat Z^A,\d\br W^I$, these Grassmann coordinates continue to be $\dbar_N$-closed. Similarly, because $\xi_N$ only points along $\PT$ as a polyvector, the fiber coordinates $W_I$ remain $\dbar_N$-closed. However, the base coordinates $Z^A$ and the Grassmann coordinates $\veps_A,\eta^I$ are no longer $\dbar_N$-closed. Instead, they get deformed into a new set of $\dbar_N$-closed quantities which we may call ``holomorphic''. In this sense, the backreaction deforms the complex supergeometry of the target space. This phenomenon has also been encountered in the study of twisted holography on $\AdS_3\times S^3\times M_4$ for $M_4 = T^4$ or $K3$, where the backreaction gives rise to a deformed superconifold \cite{Costello:2020jbh,Fernandez:2024tue}.

To obtain deformed holomorphic coordinates, it is useful to work with weightless quantities like $W_IZ^A$, etc. For instance, a straightforward computation results in
\be
\dbar_N(W_IZ^A) = [\xi_N,W_IZ^A] = \frac{3NW_I\D^3\br W}{\|W\|^8}\frac{\eps^{ABCD}A_B\veps_C\veps_D}{A\cdot Z}\,.
\ee
The right-hand side of this equation may be expressed as a $\dbar$-closed form using the identity
\be
\dbar\bigg(\frac{\eps_{IJKL}\br W^J\bar e^K\bar e^L}{\|W\|^2}\bigg) = \frac{6\,W_I\D^3\br W}{\|W\|^8}\,,
\ee
where we have introduced a basis of projective $(0,1)$-forms on the fibers of $\CO(-1)^{\oplus4}$,
\be
\bar e^I \vcentcolon= \p^I\dbar\log\|W\|^2 = \dbar\bigg(\frac{\br W^I}{\|W\|^2}\bigg)\,.
\ee
In fact, we can employ this identity to conclude that
\be
\dbar_N(W_IZ^A) = \dbar_N\bigg(\frac{N}{2}\frac{\eps_{IJKL}\br W^J\bar e^K\bar e^L}{\|W\|^2}\frac{\eps^{ABCD}A_B\veps_C\veps_D}{A\cdot Z}\bigg) \,.
\ee
This is because $[\xi_N,-]$ vanishes when acting on the argument of $\dbar_N$ on the right, as the result would have had degree 5 in $\d\br W^I$.

Thus, we  find 16 new holomorphic bosonic quantities
\be\label{gAI}
\msf{G}^A{}_I = Z^AW_I - \frac{N}{2}\frac{\eps_{IJKL}\br W^J\bar e^K\bar e^L}{\|W\|^2}\frac{\eps^{ABCD}A_B\veps_C\veps_D}{A\cdot Z}\,.
\ee
These satisfy $\dbar_N\msf{G}^A{}_I=0$ away from the excised locus $W_I=0$ and in the patch $A\cdot Z\neq0$. The $\msf{G}^A{}_I$ already include the four $W_I$ through the relation $W_I = A_B\msf{G}^B{}_I/A\cdot Z$, so we do not need to think of $W_I$ as independent coordinates. 

Similarly, the fermionic coordinates $\veps_A$ and $\eta^I$ are no longer holomorphic. The new holomorphic fermionic coordinates can be found by starting with the $\sl_4(\C)$ R-symmetry and $\sl_4(\C)$ conformal generators of supertwistor space and suitably deforming them to incorporate the backreaction,
\begin{align}
    &\msf{G}^I{}_J = W_J\eta^I - \frac14\,\delta^I_J\,W_K\eta^K + \frac{N}{2}\frac{\eps_{JKLM}\br W^I\br W^K\bar e^L\bar e^M}{\|W\|^4}\frac{\eps^{ABCD}A_A\veps_B\veps_C\veps_D}{A\cdot Z}\,,\label{gIJ}\\
    &\msf{G}^A{}_B = Z^A\veps_B - \frac14\,\delta^A_B\,Z^C\veps_C + \frac{N}{2}\frac{\eps_{IJKL}\br W^I\bar e^J\bar e^K\eta^L}{\|W\|^2}\frac{\eps^{ACDE}A_BA_C\veps_D\veps_E}{(A\cdot Z)^2}\,.\label{gAB}
\end{align}
In particular, the proof of $\dbar_N\msf{G}^A{}_B=0$ crucially uses the constraint $Z^A\veps_A=W_I\eta^I$. These expressions are easily checked to be trace-free: $\msf{G}^A{}_A = \msf{G}^I{}_I = 0$. The fact that such deformations exist is consistent with the expectation that these conformal and R-symmetries persist as symmetries of the deformed supergeometry (up to reality conditions).

Continuing the pattern, we can also construct a further set of bosonic holomorphic quantities by deforming the fermion bilinears $\eta^I\veps_A$. These are found to be
\be\label{gIA}
\msf{G}^I{}_A = \eta^I\veps_A + \frac{N}{2}\frac{\eps_{JKLM}\br W^I\br W^J\bar e^K\bar e^L\eta^M}{\|W\|^4}\frac{\eps^{BCDE}A_AA_B\veps_C\veps_D\veps_E}{(A\cdot Z)^2}\,.
\ee
Altogether, these quantities may be arranged into a parity-reversed element of $\psl(4|4,\C)$,
\be\label{GG}
\msf{G}^\bA{}_{\bB} = \left(
\begin{array}{c|c}
\msf{G}^A{}_B\hspace{0.1em} & \hspace{0.1em}\msf{G}^I{}_B \\ [2pt] \hline \\[-11.25pt]
\msf{G}^A{}_J\hspace{0.1em} & \hspace{0.1em}\msf{G}^I{}_J
\end{array}
\right)\,,
% \begin{pmatrix}
%     \msf{G}^A{}_B&&\msf{G}^I{}_B\\
%     \msf{G}^A{}_J&&\msf{G}^I{}_J
% \end{pmatrix}
\ee
where $\bA,\bB$ are supertwistor indices as before. Taken alongside $\d\hat Z^A,\d\br W^I$, these provide holomorphic coordinates on the backreacted target superspace.

Before backreaction, our supergeometry had $7$ bosonic coordinates $(Z^A,W_I)$, $7$ fermionic coordinates $(\veps_A,\eta^I)$, and $7$ fermionic coordinates coming from $(0,1)$-forms $(\d\hat Z^A,\d\br W^I)$, all taken projectively. After backreaction, we seem to have acquired an excess of riches as far as holomorphic quantities are concerned! Hence, there must exist relations among the deformed coordinates $\msf{G}^\bA{}_\bB$ that reduce the number of their independent components.

A helpful way to motivate such relations is to start with the $N=0$ case of vanishing backreaction. In this case, our definition of $\msf{G}^A{}_I$ reduces to $\msf{G}^A{}_I =  Z^AW_I$. When projectivized, this describes a patch of the Segre embedding $\P^3_Z\times\P^3_W\hookrightarrow\P^{15}$. It is well-known that this embedding is a determinantal variety inside $\P^{15}$ and can equally well be described through the vanishing of all the $2\times 2$ minors of $\msf{G}^A{}_I$,
\be
\msf{G}^A{}_I\msf{G}^B{}_J - \msf{G}^B{}_I\msf{G}^A{}_J=0\,.
\ee
Due to Cayley's syzygies, not all of these minors are independent. Non-projectively on $\C^{16}$, one finds that on the support of these constraints, the $16$ quantities $\msf{G}^A{}_I$ contain precisely $7$ independent degrees of freedom. 

This idea can be straightforwardly generalized to the super-coordinates $\msf{G}^\bA{}_\bB$ by employing Segre embeddings of projective superspaces \cite{lebrun1990projective,Fioresi:2017cax}. When $N=0$, we are working with a trace-free analog of such a super Segre embedding, $\msf{G}^\bA{}_\bB = \cZ^\bA\cW_\bB-\frac14\,\delta^\bA_\bB\cZ^\bC\cW_\bC$ (employing the notation \eqref{ZW}). Equivalently, this can be described as a determinantal variety in parity-reversed $\psl(4|4,\C)$ cut out by quadratic constraints.

The role of our backreaction is simply to deform this determinantal variety: a \emph{geometric transition}. For instance, with some work, it may be checked that the deformed coordinates \eqref{gAI} satisfy the relations
\be\label{ggN}
%\begin{split}
    \msf{G}^A{}_I\msf{G}^B{}_J - \msf{G}^B{}_I\msf{G}^A{}_J=-N\eps^{ABCD}\eps_{IJKL}\bar e^K\bar e^Le_Ce_D\,,
    %\\
    %&\msf{G}^A{}_B\msf{G}^B{}_I - \msf{G}^A{}_J\msf{G}^J{}_I = 0\,.
%\end{split}
\ee
where we have introduced a basis of $(1,0)$-vectors on $\PT$ adapted to the patch $A\cdot Z\neq0$,
\be
e_A = \veps_A - \frac{A_AZ^B}{A\cdot Z}\,\veps_B\,.
\ee
These vectors satisfy $Z^A e_A=0$, so this definition picks out three independent vectors pointing along $\PT$ out of the four non-projective vectors $\p/\p Z^A$. The $e_A$ are not $\dbar_N$-holomorphic by themselves, but products like $\bar e^I e_A$ are so due to $\bar e^I\wedge\xi_N \propto\bar e^I\wedge\D^3\br W=0$.

In deriving relations like \eqref{ggN}, it is useful to keep in mind that due to the algebraic identity $W_I\bar e^I=0$, only three out of the four $\bar e^I$'s are independent, whence the wedge products of four or more $\bar e^I$'s vanishes. This is why we never encounter terms quadratic in $N$. A similar calculation can be used to obtain relations satisfied by the other graded minors of $\msf{G}^\bA{}_\bB$, but we will not need them for what follows.

Having the quantities $\msf{G}^\bA{}_\bB$ helps with solving the Cauchy-Riemann equation $\dbar_Nf=0$ for holomorphic functions $f$ on our supergeometry. This is because such functions are often naturally expressed as functions of these deformed quantities instead of the original coordinates $Z^A,W_I,\veps_A,\eta^I$.\footnote{This is similar to how chiral primary superfields on 4d $\cN=1$ superspace with coordinates $(x^{\dal\al}\,|\,\til\theta^\dal,\theta^\al)$ appear as functions of $y^{\dal\al}\equiv x^{\dal\al}+\til\theta^\dal\theta^\al$ instead of the original bosonic coordinates $x^{\dal\al}$.} At the same time, their dependence on the $(0,1)$-forms $\d\hat Z^A,\d\br W^I$ can be arbitrary. Such functions are in one-to-one correspondence with $\dbar_N$-closed elements of $\PV^{*,*}(X)$. After accounting for the linearized gauge symmetry $f\sim f+\dbar_Ng$, they generate the Hilbert space of single-particle states in the bulk.

%%%%%%%%%%%%%%%%%%%%%%%%%%%%
%%%%%%%%%%%%%%%%%%%%%%%%%%%%

\section{Dictionary for local twistorial operators}
\label{sec:currents}

Having detailed the gauge and gravitational sides of our chiral holographic duality, we come to the construction of its holographic dictionary. Here we encounter a deep and curious puzzle that sits at the core of twistorial applications to spacetime physics.

The bulk-boundary dictionary for both standard and twisted holographic dualities relates states of the bulk theory to local operators of the gauge theory on the branes. In our case, it relates single-particle states of the 14d BCOV theory on $\CO(1,1)\to\PT\times\P^3$ to single-trace local operators of the 6d gauge theory on $\PT$. This 6d gauge theory is a holomorphic Chern-Simons theory that is only related to the 4d $\cN=4$ sdYM theory by a $\P^1$ compactification. So there is no easy procedure to directly relate 4d local operators to single-particle gravitational states.

We have nonetheless found at least two simple instances in which well-known 4d local operators are dual to states of our gravitational theory. The first case is that of symmetries of the spacetime theory that uplift to symmetries of the twistor action. This ensures that the Noether currents of spacetime symmetries like the $\cN=4$ superconformal algebra directly uplift to local currents on twistor space. In turn, these can be related to specific gravitational modes in the bulk by studying the couplings of open strings on D5 branes to fields of the BCOV theory before backreaction. 

In this section, we focus on this piece of the dictionary. Our proposals for the bulk BCOV fields dual to $\cN=4$ superconformal currents are collected in table \ref{tab:dict}, and the rest of the section is dedicated to systematically justifying this dictionary through open-closed couplings.

The second case is much broader and relates half-BPS operators and Lagrangian insertions of the 4d theory to giant graviton states in the bulk by means of determinant operators. Witten's D-instantons arise as an example of this construction. We visit this aspect of the dictionary in sections \ref{sec:dict} and \ref{sec:giant}.

%%%%%%%%%%%%%%%%%%%%%%%%%%%%
%%%%%%%%%%%%%%%%%%%%%%%%%%%%

\subsection{The problem of locality}
\label{ssec:locality}

It is worth developing a dictionary between 6d operators and 14d light states before we embark on relating 4d operators to 14d heavy states. In twisted holography, there exists a systematic procedure to set up such a dictionary by means of open-closed couplings.

The general idea is the following. As before, let $X=\CO(-1)^{\oplus4}\to\PT$ denote our bulk geometry before backreaction, with $N$ D5 branes wrapped along its zero section $\PT$. Using worldsheet disk amplitudes \cite{Hofman:2002cw}, one can construct string field theory interaction vertices coupling closed strings propagating on $X$ to open strings on $\PT$. These can be used to relate closed string states to open string operators.

If one computes a disk amplitude with one closed string and no open strings, one obtains the brane tadpole \eqref{tadpole}. More generally, we can look at the disk amplitude of one closed string vertex and multiple open string vertices. These give rise to open-closed couplings of the form\footnote{These are the B-model analogs of the Chern-Simons couplings between RR form gauge fields and open string fields on D-branes encountered in the RNS string.}
\be\label{oc}
\int_{\pt}\D^{3|4}\cZ\;\p^n\Bsi\;\cO_n\,,
\ee
where $\p^n\Bsi$ denotes some combination of (bosonic or fermionic) derivatives normal to $\pt$ acting on the closed string field $\Bsi$, and $\cO_n$ denotes a single-trace operator of the holomorphic Chern-Simons theory on supertwistor space. In this integral, $\p^n\Bsi$ has been pulled back to supertwistor space by setting $W_I=\veps_A=0$.

States of the bulk theory before backreaction are classical profiles  $\Bsi=\Bsi_0$ given by solutions of $\dbar\Bsi_0=0$. Plugging such a classical solution into \eqref{oc} picks out a specific operator in the gauge theory. On the other hand, the chosen state $\Bsi_0$ can be perturbatively corrected into a state $\Bsi_N$ satisfying $\dbar_N\Bsi_N=0$.\footnote{The fact that we can solve for $\Bsi_N$ as a perturbative expansion in $N$ around $\Bsi_0$ is a simplifying feature of twisted holography \cite{Costello:2020jbh}. Such expansions almost always truncate, so issues of convergence do not arise.} This furnishes a state-operator dictionary
\be
\Bsi_N\quad\longleftrightarrow\quad\sum_n\int_\pt\D^{3|4}\cZ\;\p^n\Bsi_0\;\cO_n\,.
\ee
Generalizing the observations of \cite{Budzik:2023xbr}, in the BV formalism, the operators $\cO_n$ can typically be arranged into polyforms obeying the descent equations
\be
Q\cO_n=\dbar\cO_n\,.
\ee
This ensures that operators built from states satisfying $\dbar\Bsi_0=0$ are $Q$-closed, and those built from pure gauge states $\Bsi_0=\dbar\bs f$ are $Q$-exact. Finally, Witten diagrams involving bulk states $\Bsi_N$ compute gauge theory correlators of their dual operators. This procedure builds an analog of Witten's ``differentiate dictionary'' \cite{Witten:1998qj} for our duality and also works for other similar twisted holographic dualities.

In our context, a shortcoming of this approach is that it relates states of BCOV theory to local operators on \emph{twistor space} instead of on spacetime. For example, we typically work with twistorial local operators $\cO_n$ that are $0$-forms. Recall that the open string field on $\pt$ was given by a $\gl_N(\C)$-valued $(0,*)$ polyform
\be
\cA = \bs{a} + \til{\bs{\psi}}_I\eta^I + \frac12\,\bs{\vphi}_{IJ}\eta^I\eta^J + \bs{\psi}^I\eta^3_I + \bs{b}\,\eta^4\,.
\ee
The fields of the $\cN=4$ vector multiplet on $\R^4$ are obtained by Penrose transforming the $(0,1)$-form parts of this string field. But a $0$-form operator on twistor space would be constructed by gauge invariant combinations of the $0$-form components of the polyforms $\bs{a},\til{\bs{\psi}}_I$, \emph{etc}. These $0$-form parts are the \emph{ghosts} of gauge symmetries of the twistor action and, except for the basic ghost associated to standard gauge symmtery, have no spacetime counterpart. So, in general, local scalar operators on twistor space cannot correspond to the familiar spacetime local operators like $\Tr\,\Phi^n$ that carry zero ghost number. 

Ideally, we would have wanted to build a holographic dictionary between states of BCOV theory and local operators on \emph{spacetime}. However, most local operators on $\R^4$ do not correspond to local operators on twistor space. This is a well-known issue, arising fundamentally because a point $x\in\R^4$ corresponds to a Riemann sphere $\P^1_x\subset\PT$. There have been numerous studies of the uplift of local spacetime operators to twistor space, as this aids in the computation of their form factors. See for instance the works \cite{Costello:2022wso,Bogna:2023bbd,Koster:2016fna,Chicherin:2014uca,Adamo:2011dq,Adamo:2011cd,Mason:2010yk} and references therein. Operators like $\Tr\,\Phi^n(x)$ can be uplifted to twistor space by plugging in the Penrose integral for the scalars $\Phi_{IJ}$ \cite{Koster:2016ebi}. These give rise to operators on $\PT$ that are polylocal along the twistor line $\P^1_x$. In sections \ref{sec:dict} and \ref{sec:giant}, we will show that a generating functional for such polylocal operators can be identified with heavy states in the bulk, such as giant gravitons. 

Very few gauge invariant operators that are local on twistor space can be built purely from the physical fields in the $(0,1)$-form part of $\cA$. However, those which can are particularly important. Symmetries of the self-dual theory on $\R^4$ are generated by local currents on spacetime. If these symmetries uplift to symmetries of the twistorial theory, then they can also be represented by ghost number zero local currents on twistor space. Let us begin by looking for such currents and their gravitational duals.

%%%%%%%%%%%%%%%%%%%%%%%%%%%%
%%%%%%%%%%%%%%%%%%%%%%%%%%%%

\subsection{Global symmetries}
\label{ssec:global}

Global 0-form symmetries of the twistorial theory give rise to conserved charges of the form
\be\label{Jc}
\int_\Sigma\D^3Z\wedge J\,,
\ee
where $J$ is a weight $-4$ $(0,*)$-polyform on $\PT$, and $\Sigma$ is a real-codimension-$1$ submanifold of $\PT$. The integral over $\Sigma$ picks out the $(0,2)$-form part of the symmetry current $J$. The BV-BRST variation of $J$ generally takes the form
\be
QJ = \dbar J\,,
\ee
which ensures that the charge \eqref{Jc} is conserved and gauge invariant if $\p\Sigma=0$. In particular, the $0$-form part of $J$ (if nonzero) gives a BRST invariant local operator.

If $J$ describes the twistor uplift of a spacetime symmetry, then before the addition of BV ghosts and antifields, its physical $(0,2)$-form part must be expressible purely in terms of the physical fields in $\cA$. This is a necessary condition for it to give rise to a conserved spacetime current that is expressible purely in terms of the ghost number $0$ fields $A,\til\Psi_I,\Phi_{IJ},\Psi^I,B$ on $\R^4$. The physical fields on $\PT$ are precisely the $(0,1)$-form parts $\cA_1\subset\cA$. So the $(0,2)$-form part of $J$ must be quadratic in these. After adding back ghosts and antifields, $J$ deforms to a polyform that is quadratic in the full open string field $\cA$. 

To find the bulk state dual to such a current $J$, we should look for the open-closed couplings of bulk fields to single-trace operators quadratic in $\cA$. There is a standard trick to obtain such couplings. Turning on a closed string background $\Bsi$ deforms the kinetic operator in the holomorphic Chern-Simons action from $\dbar$ to $\dbar + [\Bsi,-]$, where $[-,-]$ denotes the Schouten bracket. To accomplish this, the first-order coupling of $\Bsi$ to holomorphic Chern-Simons theory that is quadratic in $\cA$ must take the form
\be\label{2coup}
\frac12\int_\pt\D^{3|4}\cZ\;\Tr\,\cA\,[\Bsi,\cA]\,.
\ee
This implements the shift $\cA\,\dbar\cA\mapsto\cA\,(\dbar\cA + [\Bsi,\cA])$ in the kinetic term.

Quite beautifully, these couplings allow us to find both the expressions for a very interesting class of conserved currents as well as the fields of BCOV theory that they are dual to in one go. About half of the quadratic single-trace operators that we obtain will correspond to the currents generating the $\cN=4$ superconformal symmetry of the gauge theory. We expect that the remaining operators also fit into the stress tensor supermultiplet, but we leave their analysis to the future.

We can expand $\Bsi$ in components to make the single-trace operators appearing in \eqref{2coup} more explicit. The Schouten bracket $[\Bsi,\cA]$ is given by
\be
[\Bsi,\cA] = \p^A\Bsi\,\p_A\cA + \p_A\Bsi\,\p^A\cA + \p_I\Bsi\,\p^I\cA + \p^I\Bsi\,\p_I\cA\,,
\ee
where $\p_A\equiv\p/\p Z^A$, $\p^I\equiv\p/\p W_I$ as before, and we have also abbreviated the Grassmann derivatives as $\p^A\equiv\p/\p\veps_A$, $\p_I\equiv\p/\p\eta^I$. We need to evaluate this along $W_I=\veps_A=0$. As this expression contains at most one $\veps_A$ derivative acting on $\Bsi$, it suffices to expand $\Bsi$ as a polyvector in $\p_A\equiv\veps_A$ to first order, while keeping its $\p^I\equiv\eta^I$ dependence implicit within the coefficients,
\be
\Bsi = \Bsi_0 + \Bsi_1^A\veps_A + \rO(\veps^2)\,.
\ee
We also note that $\p^A\cA=\p^I\cA=0$, because $\cA$ only depends on the supertwistor coordinates $\cZ^\bA = (Z^A|\eta^I)$. Using these properties, we expand \eqref{2coup} to obtain the open-closed coupling
\be\label{gencur}
\frac12\int_\pt\D^{3|4}\cZ\;\big\{\Bsi_1^A\,\Tr(\cA\,\p_A\cA) + \p^I\Bsi_0\,\Tr(\cA\,\p_I\cA)\big\}\,.
\ee
In simplifying this, we used the facts that $\Bsi_0$ was Grassmann even and $\Bsi_1^A,\cA$ were Grassmann odd.

Unfortunately, the individual boundary operators $\Tr\,\cA\,\p_A\cA$ and $\Tr\,\cA\,\p_I\cA$ are not yet in a form in which they obey $Q\cO=\dbar\cO$. Instead, one can use the definition $Q\cA = \dbar\cA+\cA^2$ to show that
\be\label{QAdA}
\begin{split}
    Q\,\Tr(\cA\,\p_\bA\cA) &= \dbar\,\Tr(\cA\,\p_\bA\cA) - \frac13\,\p_\bA\Tr(\cA^3)\,,
\end{split}
\ee
where $\p_\bA = (\p_A|\p_I)$ collectively denotes a supertwistor derivative. This subtlety arises from the fact that we have not diagonalized the modes of the bulk fields $\Bsi_0,\Bsi_1^A$ to which these operators couple. The bulk fields are interlinked by the divergence-free constraint $\p^A\p_A\Bsi + \p_I\p^I\Bsi=0$. This imposes the constraint
\be\label{cons}
\p_A\Bsi_1^A = \p_I\p^I\Bsi_0
\ee
on the modes of interest.

To make progress, it is helpful to break up $\Bsi$ into components of definite polyvector degree. Let us recall the expansion of $\Bsi$ given in \eqref{Psiexp},
\be
\Bsi = \al + \beta + \gamma + \xi + \til\gamma + \til\beta + \til\al\,,
\ee
where the displayed fields are $(0,*)$-polyforms listed in order of increasing polyvector degree ranging from $0$ to $6$. Making the polyvector dependence explicit, we can write out the superfield expansions of $\Bsi_0,\Bsi_1^A$,
\begin{align}
    \Bsi_0 &= \al + \beta_I\eta^I + \frac12\,\gamma_{IJ}\eta^I\eta^J + \xi^I\eta^3_I+ \til\gamma\,\eta^4\,,\label{Bsi0exp}\\
    \Bsi_1^A &= \beta^A + \gamma^A{}_I\eta^I + \frac12\,\xi^A{}_{IJ}\eta^I\eta^J + \til\gamma^{AI}\eta^3_I + \til\beta^A\eta^4\,.\label{Bsi1exp}
\end{align}
We are mildly abusing notation by denoting the $\eta^4$ component of $\til\gamma$ by the same symbol. Its meaning will hopefully be clear from context. Furthermore, since we are building the couplings to $(0,2)$-form gauge theory currents, eventually we will only need to keep the $(0,1)$-form parts of $\Bsi_0,\Bsi_1^A$.

By plugging the expansions \eqref{Bsi0exp} and \eqref{Bsi1exp} into \eqref{gencur}, one obtains component expressions for the open-closed couplings. For example, the simplest is the coupling to $\al$, as $\al$ has no vector degree,
\be
\int_\PT\D^3Z\;\p^I\al\;\bs{H}_I\,.
\ee
The associated gauge theory current $\bs{H}_I$ is given by
\be
\label{H_Idef}
\bs{H}_I = \frac12\int\d^4\eta\;\Tr(\cA\,\p_I\cA)\,.
\ee
The BRST variation of $\Tr(\cA\,\p_I\cA)$ was provided in \eqref{QAdA}. Using this expression, the BRST variation of $\bs{H}_I$ is readily derived:
\be
Q\bs{H}_I = \dbar\bs{H}_I\,.
\ee
Crucially, the boundary term $-\frac13\,\p_I\Tr(\cA^3)$ found in \eqref{QAdA} drops out on performing the superspace integrals in \eqref{H_Idef}. This ensures that $\bs{H}_I$ gives rise to a conserved current in the twistorial theory. 

To find the physical part of the $(0,2)$-form currents, we turn off the ghost and antifield parts of $\cA$. This amounts to simply replacing every occurrence of $\cA$ in operators like $\bs H_I$ by its $(0,1)$-part,
\be
\cA_1 = a + \til\psi_I\eta^I + \frac12\,\vphi_{IJ}\eta^I\eta^J + \psi^I\eta^3_I + b\,\eta^4\,.
\ee
Continuing with our example, this replacement shows that the physical part of $\bs{H}_I$ is
\be\label{HI}
H_I = \Tr(\til\psi_I b+\vphi_{IJ}\psi^J)\,.
\ee
This can be identified with a supersymmetry current. We will see this by Penrose transforming such currents in the next section.

The subtlety with the divergence-free constraint is first encountered when computing the couplings to the Beltrami differential $\beta$. At first sight, these take the form
\be\label{betacoup}
\frac12\int_\pt\D^{3|4}\cZ\;\Big\{\beta^A\,\Tr(\cA\,\p_A\cA) + \p^I\beta_J\,\eta^J\,\Tr(\cA\,\p_I\cA)\Big\}\,.
\ee
Substituting the expansions \eqref{Bsi0exp}, \eqref{Bsi1exp} into the constraint \eqref{cons} gives rise to the divergence-free constraint on the Beltrami,
\be
\p_A\beta^A + \p^I\beta_I = 0\,.
\ee
This relates the trace of $\p^I\beta_J$ to $\beta^A$. Hence, we can diagonalize the coupling \eqref{betacoup} by replacing $\p^I\beta_J$ with its decomposition into a trace-free part plus the trace,
\be
\p^I\beta_J = \p^I\beta_J - \frac14\,\delta^I_J\,\p^K\beta_K - \frac14\,\delta^I_J\,\p_A\beta^A\,.
\ee
We then integrate by parts in \eqref{betacoup} to combine the couplings to $\beta^A$.

This allows us to recast the Beltrami coupling as a sum of two independent couplings,
\be
\int_\PT\D^3Z\left(\beta^A\bs{T}_A - \p^I\beta_J\bs{R}^J{}_I\right)\,,
\ee
wherein we have obtained twistor uplifts of the stress tensor and R-symmetry currents,
\begin{align}
    \bs T_A &= \frac12\int\d^4\eta\;\Tr\left\{\cA\,\p_A\cA - \frac14\,\p_A(\cA\,\eta^I\p_I\cA)\right\}\,,\\
    \bs R^I{}_J &= \frac12\int\d^4\eta\;\Tr\left\{\cA\left(\eta^I\p_J-\frac14\,\delta^I_J\,\eta^K\p_K\right)\cA\right\}\,.
\end{align}
As desired, since $\bs{R}^I{}_J$ is trace-free, the coupling $\p^I\beta_J\bs{R}^J{}_I$ no longer contains the trace of $\p^I\beta_J$. After an application of \eqref{QAdA} and judicious integrations by parts, the BRST variations of $\bs T_A$ and $\bs R^I{}_J$ reduce to
\be
Q\bs T_A = \dbar\bs T_A\,,\qquad Q\bs R^I{}_J = \dbar\bs R^I{}_J\,.
\ee
This is a general pattern. Before restricting them to physical fields, all our gauge theory operators $\cO$ will obey BRST transformation laws of the form $Q\cO = \dbar\cO$.

The physical parts of $\bs T_A$ and $\bs R^I{}_J$ are easily extracted after performing the superspace integrals and keeping only the $(0,1)$-form fields in $\cA$,
\begin{align}
T_A &= \Tr\left(b\,\p_A a + \frac14\,\vphi^{IJ}\p_A\vphi_{IJ} + \frac34\,\psi^I\p_A\til\psi_I - \frac14\,\til\psi_I\p_A\psi^I\right)\,,\label{TA}\\
R^I{}_J &= \Tr\left(\frac{1}{2}\,\vphi^{IK}\vphi_{JK} + \psi^I\til\psi_J - \frac{1}{4}\,\delta^I_J\,\psi^K\til\psi_K\right)\,.\label{RIJ}
\end{align}
The stress tensor $T_A$ looks like the 6d version of the stress tensor of a 2d $\beta\gamma$ or $bc$ system. It obeys $Z^AT_A=0$, which ensures that the $(1,2)$-form $T_A\d Z^A$ is projectively well-defined. The R-symmetry currents $R^I{}_J$ are traceless, signifying that the R-symmetry is $\SU(4)$ and not $\U(4)$. 

We can perform a similar analysis for the open-closed couplings involving the remaining fields in $\Bsi_0,\Bsi_1^A$. When the dust settles, we find the following set of diagonalized couplings,
\begingroup
\allowdisplaybreaks
\begin{align}
    &\int_\PT\D^3Z\left(\gamma^A{}_I\bs G^I{}_A + \frac12\,\p^K\gamma_{IJ}\bs S^{IJ}{}_K\right)\,,\\
    &\int_\PT\D^3Z\left(\frac12\,\xi^{AIJ}\til{\bs S}_{IJA} + \p^I\xi^J\til{\bs R}_{IJ}\right)\,,\\
    &\int_\PT\D^3Z\;\til\gamma^{AI}\til{\bs G}_{IA}\,,\\
    &\int_\PT\D^3Z\;\til\beta^{BA}\til{\bs H}_{AB}\,.
\end{align}
\endgroup
The coupling to $\til\gamma$ has been eliminated completely using the divergence-free constraint $\p^I\til\gamma + \p_A\til\gamma^{AI} = 0$ so as to get a coupling purely to $\til\gamma^{AI}$. Similarly, the divergence-free constraint on $\til\beta^A$ reduces to $\p_A\til\beta^A=0$. We have solved this by introducing a potential $\til\beta^{AB}=-\til\beta^{BA}$:
\be
\til\beta^A = \p_B\til\beta^{AB}\,.
\ee
The coupling to $\til\beta^A$ has been written in terms of this potential, which is necessary to make it manifestly BRST invariant.

The twistorial operators that we obtain through this procedure are listed below:
\begingroup
\allowdisplaybreaks
\begin{align}
    \bs G^I{}_A &= -\int\d^4\eta\;\eta^I\,\Tr\left(\frac12\,\cA\,\p_A\cA - \frac{1}{6}\,\p_A(\cA\,\eta^J\p_J\cA)\right)\,,\\
    \bs S^{IJ}{}_K &= \int\d^4\eta\;\Tr\left(\frac12\,\eta^I\eta^J\cA\,\p_{K}\cA - \frac13\,\delta^{[I}_K\eta^{J]}\cA\,\eta^L\p_L\cA\right)\,,\\
    \til{\bs S}_{IJA} &= \frac{1}{4!}\,\eps_{IJKL}\int\d^4\eta\;\eta^K\eta^L\;\Tr\left(\frac12\,\cA\,\p_A\cA - \frac{1}{4}\,\p_A(\cA\,\eta^M\p_M\cA)\right)\,,\\
    \til{\bs R}_{IJ} &= -\int\d^4\eta\;\Tr\left(\frac12\,\eta^3_I\cA\,\p_{J}\cA + \frac18\,\eps_{IJKL}\,\eta^K\eta^{L}\cA\,\eta^M\p_M\cA\right)\,,\\
    \til{\bs G}_{IA} &= -\frac12\int\d^4\eta\;\eta^3_I\;\Tr\,\big(\cA\,\p_A\cA - \p_A(\cA\,\eta^J\p_J\cA)\big)\,,\\
    \til{\bs H}_{AB} &= \frac12\int\d^4\eta\;\eta^4\;\tr(\p_A\cA\,\p_B\cA)\,.
\end{align}
\endgroup
All the operators in this list obey BRST covariance conditions such as $Q\bs{S}^{IJ}{}_K=\dbar\bs{S}^{IJ}{}_K$, $Q\bs{G}^I{}_A = \dbar\bs G^I{}_A$, \emph{etc}. Their $(0,2)$-form physical parts are obtained after performing the superspace integrals and dropping ghosts and antifields,
\begingroup
\allowdisplaybreaks
\begin{align}
    G^I{}_A &= \Tr\left(\psi^I\p_Aa + \frac23\,\vphi^{IJ}\p_A\til\psi_J - \frac13\,\til\psi_J\p_A\vphi^{IJ}\right)\,,\label{GIA}\\
    S^{IJ}{}_K &= \Tr\left(\vphi^{IJ}\til\psi_K - \frac23\,\delta^{[I}_K\vphi^{J]L}\til\psi_L\right)\,,\\
    \til S_{IJA} &= \frac12\,\Tr\left(\vphi_{IJ}\p_Aa + \til\psi_{[I|}\p_A\til\psi_{|J]}\right)\,,\\
    \til R_{IJ} &= \frac12\,\Tr(\til\psi_I\til\psi_J)\,,\\
    \til G_{IA} &= \Tr(\til\psi_I\p_Aa)\,,\\
    \til H_{AB} &= \frac12\,\Tr(\p_A a\,\p_B a)\,.
\end{align}
\endgroup
The operators carrying twistor indices satisfy $Z^A G^I{}_A = Z^A\til S_{IJA} = Z^A\til G_{IA} = Z^A\til H_{AB} = 0$, ensuring that they are projectively well-defined. Similarly, $S^{IJ}{}_K$ is trace-free: $S^{IJ}{}_J=0$. 

Off-shell, we have seen that in the BV formalism, the operators $\bs T_A$, $\bs R^I{}_J,\,\ldots$ satisfy descent equations of the form $Q\bs T_A=\dbar\bs T_A$, $Q\bs R^I{}_J = \dbar\bs R^I{}_J$ and so on. Since $Q\cA=\dbar\cA+\cA^2$, the action of the BRST charge identically vanishes on-shell. As a result, we find that all such operators become $\dbar$-closed on the support of the equations of motion. This is the sense in which they are conserved currents in the twistorial theory. When the BV ghosts and antifields are dropped, one obtains the conservation laws
\be
\dbar H_I = \dbar T_A = \dbar R^I{}_J = \dbar G^I{}_A = \cdots = \dbar\til H_{AB} = 0\,.
\ee
It is also possible to directly verify these conservation laws on the support of the twistor uplifts \eqref{aeq} -- \eqref{beq} of the field equations of $\cN=4$ sdYM.

Out of these conserved operators, we can identify the generators of superconformal symmetries. $T_A$ plays the role of the stress tensor and $R^I{}_J$ the role of R-symmetry currents. The charges generating complexified $\SL_4(\C)$ conformal symmetry and $\SL_4(\C)$ R-symmetry are respectively given by
\be
    Q^A{}_B = \int_\Sigma\D^3Z\;Z^AT_B\,,\qquad Q^I{}_J = \int_\Sigma\D^3Z\;R^I{}_J\,.
\ee
The integrands of these integrals are weightless because $T_A$ carries weight $-5$ whereas $R^I{}_J$ has weight $-4$. Similarly, the supersymmetries and conjugate supersymmetries are generated by the graded-odd currents $H_I,G^I{}_A$ through the charges
\be
    Q^A{}_I = \int_\Sigma\D^3Z\;Z^AH_I\,,\qquad Q^I{}_A = \int_\Sigma\D^3Z\;G^I{}_A\,,
\ee
which are meaningful integrals because $H_I$ has weight $-5$ and $G^I{}_A$ has weight $-4$. Together, these charges generate the familiar $\mathrm{PSL}(4|4,\C)$ superconformal symmetry. Along these lines, it would be desirable to understand if the dual superconformal symmetry of $\cN=4$ sYM and its Yangian extension \cite{Drummond:2008vq,Drummond:2009fd} also play a role in our holographic duality.

Among the remaining operators, $S^{IJ}{}_K$ and $\til S_{IJA}$ have weight $-3$, whereas $\til R_{IJ}$, $\til G_{IA}$ and $\til H_{AB}$ have weight $-2$. To form conserved charges out of them, we would have to integrate them against holomorphic functions of $Z^A$ carrying weight $-1$ or $-2$ respectively (so as to cancel the $+4$ weight of $\D^3Z$). There are no such holomorphic functions. So these operators do not appear to give rise to any further global symmetries of $\cN=4$ sdYM.\footnote{In principle, one may be able to construct conserved charges out of these currents by integrating their holomorphic derivatives normal to $\Sigma$ against positive weight functions. We do not explore this here.}

In the next section, we will convert these twistorial local operators into local operators on $\R^4$. This will give rise to conserved currents of $\cN=4$ sdYM. 

%%%%%%%%%%%%%%%%%%%%%%%%%%%%
%%%%%%%%%%%%%%%%%%%%%%%%%%%%

\subsection{Superconformal currents}
\label{ssec:super}

In the previous section, we identified a set of conserved currents $T_A$, $H_I$, $G^I{}_A$ and $R^I{}_J$ on $\PT$ that generate the superconformal algebra. In this section, we study their pushforward to $\R^4$, identifying these twistorial currents with their standard counterparts in $\cN=4$ sdYM.

For the reader's convenience, we repeat here the combined effect of the (Woodhouse) gauge choice \eqref{aw}-\eqref{bw} that the restriction of $\cA_1$ to any $\P^1_x$ be harmonic together with the solutions \eqref{adal}-\eqref{bdal} of the vertical equations of motion:
\begingroup
\allowdisplaybreaks
\begin{align}
    a &= \lambda^\al A_{\al\dal}\bar e^\dal\,,\label{wa}\\
    \til\psi_I &= \til\Psi_{I\dal}\bar e^\dal\,,\label{wpsit}\\
    \vphi_{IJ} &= \Phi_{IJ}\bar e^0 + \frac{\hat\lambda^\al D_{\al\dal}\Phi_{IJ}\bar e^\dal}{\|\lambda\|^2}\,,\label{wphi}\\
    \psi^I &= \frac{2\,\hat\lambda^\al\Psi^I_\al\bar e^0}{\|\lambda\|^2} + \frac{\hat\lambda^\al\hat\lambda^\beta D_{\al\dal}\Psi^I_\beta\bar e^\dal}{\|\lambda\|^4}\,,\label{wpsi}\\
    b &= \frac{3\,\hat\lambda^\al\hat\lambda^\beta B_{\al\beta}\bar e^0}{\|\lambda\|^4} + \frac{\hat\lambda^\al\hat\lambda^\beta \hat\lambda^\gamma D_{\al\dal}B_{\beta\gamma}\bar e^\dal}{\|\lambda\|^6}\,.\label{wb}
\end{align}
\endgroup
To perform the pushforward, we plug these representatives into the expressions for our currents.

First consider the currents $H_I,R^I{}_J$ that do not carry any twistor indices. On the support of the field equations, they are $\dbar$-closed. It is helpful to study the general structure of such $\dbar$-closed currents before working out explicit examples. 

Let $J\in\Omega^{0,2}(\PT,\CO(-n))$ be such a current. We assume that $n\geq4$ so that this current can give rise to conserved charges. Being a $(0,2)$-form on $\PT$, it may be expanded as
\be
J = J_\dal\,\bar e^0\wedge\bar e^\dal + \frac12\,J_0\,\bar e_\dal\wedge\bar e^\dal\,.
\ee
The conservation law $\dbar J=0$ translates into a single equation on its coefficients,
\be\label{dbarJ}
\dbar_0J_0 = \dbar_\dal J^\dal\,.
\ee
In Woodhouse gauge, both $H_I$ and $R^I{}_J$ will give rise to coefficients of the form
\be
J_{\dot\gamma} = \frac{(n-2)\,\hat\lambda^\al\cdots\hat\lambda^\beta\hat\lambda^\gamma J_{\al\cdots\beta\gamma\dot\gamma}(x)}{\|\lambda\|^{2(n-3)}}\,,
\ee
where $J_{(\al\cdots\beta\gamma)\dot\gamma}(x)$ is a spacetime current with $n-3$ symmetrized undotted spinor indices. In such cases, the coefficient $J_0$ is completely fixed by \eqref{dbarJ} to be
\be
J_0 = \frac{\hat\lambda^\al\cdots\hat\lambda^\beta\hat\lambda^\gamma\hat\lambda^\delta\p_{\delta\dot\delta}J_{\al\cdots\beta\gamma}{}^{\dot\delta}(x)}{\|\lambda\|^{2(n-2)}}\,.
\ee
Additionally, we obtain the spacetime conservation law
\be
\p^{\gamma\dot\gamma}J_{\al\cdots\beta\gamma\dot\gamma} = 0\,,
\ee
which identifies $J_{\al\cdots\beta\gamma\dot\gamma}$ with a symmetry current on $\R^4$.

This tells us that we only need to compute the $\bar e^0\wedge\bar e^\dal$ components of $H_I,R^I{}_J$, which tend to be simpler. On plugging the Woodhouse representatives into the expressions \eqref{HI}, \eqref{RIJ} for these currents, we obtain the structures
\begin{align}
    &H_I = -\frac{3\,\hat\lambda^\al\hat\lambda^\beta H_{I\al\beta\dot\beta}}{\|\lambda\|^4}\,\bar e^0\wedge\bar e^{\dot\beta} - \frac{\hat\lambda^\al\hat\lambda^\beta\hat\lambda^\gamma\p_{\al\dal}H_{I\beta\gamma}{}^\dal}{\|\lambda\|^6}\,\frac{\bar e_{\dot\beta}\wedge\bar e^{\dot\beta}}{2}\,,\\
    &R^I{}_J = -\frac{2\,\hat\lambda^\al R^I{}_{J\al\dal}}{\|\lambda\|^2}\,\bar e^0\wedge\bar e^{\dal} - \frac{\hat\lambda^\al\hat\lambda^\beta\p_{\beta\dal} R^I{}_{J\al}{}^{\dal}}{\|\lambda\|^4}\,\frac{\bar e_{\dot\beta}\wedge\bar e^{\dot\beta}}{2}\,.
\end{align}
The spacetime currents extracted from these expressions read
\begin{align}
H_{I\al\beta\dot\beta} &= \Tr\left(\til\Psi_{I\dot\beta}B_{\al\beta} + \frac23\,\Psi^J_{(\al}D_{\beta)\dot\beta}\Phi_{IJ} -\frac13\,\Phi_{IJ}D_{(\beta|\dot\beta}\Psi^J_{|\al)}\right)\,,\\
R^I{}_{J\al\dal} &= \Tr\left\{\Psi^I_\al\til\Psi_{J\dal} - \frac14\,\delta^I_J\,\Psi^K_\al\til\Psi_{K\dal} - \frac14\left(\Phi^{IK}D_{\al\dal}\Phi_{JK}-\Phi_{JK}D_{\al\dal}\Phi^{IK}\right)\right\}\,.
\end{align}
The tensor equivalent of the Grassmann-odd current $H_{I\al\beta\dot\beta}$ has index structure $H_{I\al\mu}$. This makes it manifest that it is a supersymmetry current, as it carries one spinor index $\al$ and one vector index $\mu$. It is a chiral analog of the current that would have been dual to a gravitino mode in Maldacena's AdS$_5$/CFT$_4$ duality. Similarly, the current $R^I{}_{J\al\dal}$ can be expressed as a 1-form $R^I{}_{J\mu}$. This is the chiral counterpart of the R-symmetry current that would have been dual to the $\SO(6)$ gauge field in the AdS$_5$ $\cN=8$ supergravity multiplet.

The pushforwards of the stress tensor $T_A$ and the right-moving supercurrent $G^I{}_A$ are slightly more involved. The first step is to express these currents as $(1,2)$-forms:
\be\label{Tform}
T = T_A\d Z^A\,,\qquad G^I = G^I{}_A\d Z^A\,.
\ee
% \begin{align}
%     &T = T_A\d Z^A = \Tr\left(b\,\p a + \frac14\,\vphi^{IJ}\p\vphi_{IJ} + \frac34\,\psi^I\p\til\psi_I - \frac14\,\til\psi_I\p\psi^I\right)\,,\label{Tform}\\
%     &G^I = G^I{}_A\d Z^A = \Tr\left(\psi^I\p a + \frac23\,\vphi^{IJ}\p\til\psi_J - \frac13\,\til\psi_J\p\vphi^{IJ}\right)\,,\label{Gform}
% \end{align}
%where $\p = \d Z^A\,\p_A$ is the holomorphic exterior derivative on $\PT$. The idea is to evaluate these derivatives 
The idea is to evaluate these forms in a frame adapted to the fibration $\P^1\hookrightarrow\PT\to\R^4$. So, instead of the holomorphic frame $\d Z^A$ of $T^*_\PT$, introduce the spacetime-adapted frame
\be
e^0 = \la\lambda\,\d\lambda\ra \equiv \D\lambda\,,\qquad e^\dal = \lambda_\al\d x^{\al\dal}\,.
\ee
These span the same $(1,0)$-forms on $\PT$ as $\d Z^A$. Because $Z^AT_A=Z^AG^I{}_A=0$, the $(1,0)$-forms $T$ and $G^I$ are projectively well-defined on $\PT$. Therefore they will indeed belong to the span of $e^0,e^\dal$.

Let us first evaluate $T$ using the formula for $T_A$ obtained in \eqref{TA}. When the various Woodhouse representatives are substituted into $T$, its pushforward exhibits the following structure:
\be
%\begin{split}
    T %&= T_0\wedge e^0 + T_\dal\wedge e^\dal\\
    = \left(T_{0\dal}\,\bar e^0\wedge\bar e^\dal + \frac12\,T_{00}\,\bar e_{\dot\beta}\wedge\bar e^{\dot\beta}\right)\wedge e^0 + \left(T_{\dal\dot\beta}\,\bar e^0\wedge\bar e^{\dot\beta} + \frac12\,T_{\dal0}\,\bar e_{\dot\beta}\wedge\bar e^{\dot\beta}\right)\wedge e^\dal\,.
%\end{split}
\ee
Holomorphicity $\dbar T = 0$ again fixes $T_{00}$, $T_{\dal0}$ in terms of $T_{0\dal}$, $T_{\dal\dot\beta}$ by imposing the constraints $\dbar_0T_{00}=\dbar_\dal T_0{}^\dal-T_\dal{}^\dal$ and $\dbar_0T_{\dal0}=\dbar_{\dot\beta}T_\dal{}^{\dot\beta}$. The unfixed components $T_{0\dal}$, $T_{\dal\dot\beta}$ take the form
\be
\begin{split}
    T_{0\dal} &= -\frac{4\,\hat\lambda^\al\hat\lambda^\beta\hat\lambda^\gamma}{\|\lambda\|^6}\, K_{\al\beta\gamma\dal}(x)\,,\\
    T_{\dal\dot\beta} &= - \frac{3\,\hat\lambda^\al\hat\lambda^\beta}{\|\lambda\|^4} \left(T_{\al\dal\beta\dot\beta}(x) + \p^\gamma_{\dot\beta}K_{\al\beta\gamma\dal}(x)\right) - \frac{4\,\lambda^{(\al}\hat\lambda^\beta\hat\lambda^\gamma\hat\lambda^{\delta)}}{\|\lambda\|^6}\,L_{\al\beta\gamma\delta\dal\dot\beta}(x)\,.
\end{split}
\ee
The spacetime tensor $T_{\al\dal\beta\dot\beta}\leftrightarrow T_{\mu\nu}$ extracted from the second of these is found to be
\begin{multline}
T_{\al\dal\beta\dot\beta} = \Tr\left(B_{\al\beta}F_{\dal\dot\beta} + \frac12\,\Psi^I_{(\al}D_{\beta)\dal}\til\Psi_{I\dot\beta} + \frac12\,\til\Psi_{I(\dal}D_{\dot\beta)(\al}\Psi^I_{\beta)}\right.\\
\left. + \frac16\,D_{(\al|\dal}\Phi^{IJ}D_{|\beta)\dot\beta}\Phi_{IJ} - \frac{1}{12}\,\Phi^{IJ}D_{(\al|\dal}D_{|\beta)\dot\beta}\Phi_{IJ}\right)\,.
\end{multline}
On the support of the field equations \eqref{Aeq} -- \eqref{Beq}, this becomes symmetric and obeys the conservation law $\p^\mu T_{\mu\nu}=0$. So it can be identified with the stress tensor of $\cN=4$ sdYM. In fact, it is independently symmetric under exchanging $\al\leftrightarrow\beta$ and $\dal\leftrightarrow\dot\beta$. This implies that $T_{\mu\nu}$ is trace-free, which is consistent with 4d conformal symmetry.

Applying \eqref{GIA}, the pushforward of the supercurrent $G^I$ admits a similar expansion,
\be
G^I = \left(G^I_{0\dal}\,\bar e^0\wedge\bar e^\dal + \frac12\,G^I_{00}\,\bar e_{\dot\beta}\wedge\bar e^{\dot\beta}\right)\wedge e^0 + \left(G^I_{\dal\dot\beta}\,\bar e^0\wedge\bar e^{\dot\beta} + \frac12\,G^I_{\dal0}\,\bar e_{\dot\beta}\wedge\bar e^{\dot\beta}\right)\wedge e^\dal\,.
\ee
Due to $\dbar G^I=0$, the coefficients of $\bar e_{\dot\beta}\wedge\bar e^{\dot\beta}$ are again fully determined by constraints like $\dbar_0G^I_{00}=\dbar_\dal G_0{}^\dal - G_\dal{}^\dal$ and $\dbar_0G_{\dal0}=\dbar_{\dot\beta}G_\dal{}^{\dot\beta}$. The data of the spacetime supercurrent can be extracted from remaining coefficients $G^I_{0\dal}$, $G^I_{\dal\dot\beta}$, which take the form
\be
\begin{split}
    G^I_{0\dal} &= -\frac{3\,\hat\lambda^\al\hat\lambda^\beta}{\|\lambda\|^4}\, K^I_{\al\beta\dal}(x)\,,\\
    G^I_{\dal\dot\beta} &= -\frac{2\,\hat\lambda^\al}{\|\lambda\|^2}\left(G^I_{\al\dal\dot\beta}(x)+\p_{\dot\beta}^\beta K^I_{\al\beta\dal}(x)\right) - \frac{3\,\lambda^{(\al}\hat\lambda^\beta\hat\lambda^{\gamma)}}{\|\lambda\|^4}\,L^I_{\al\beta\gamma\dal\dot\beta}(x)\,.
\end{split}
\ee
With our Woodhouse representatives, we obtain the following right-handed supersymmetry current on $\R^4$,
\be
G^I_{\al\dal\dot\beta} = \Tr\left(\Psi^I_\al F_{\dal\dot\beta} - \frac23\,\til\Psi_{J(\dot\beta|}D_{\al|\dal)}\Phi^{IJ} + \frac13\,\Phi^{IJ}D_{\al\dal}\til\Psi_{J\dot\beta}\right)\,.
\ee
On-shell, this is symmetric in its dotted indices. Its tensor equivalent has the index structure $G^I_{\mu\dal}$ and obeys $\p^\mu G^I_{\mu\dal}=0$. In the standard duality, its non-chiral counterpart would have arisen as the operator dual to another gravitino mode in AdS$_5$ supergravity.

\begin{table}[t]
\centering
\renewcommand{\arraystretch}{1.25}
\begin{tabular}{c|c|c|c|c}
{\bf Symmetry current}& {\bf $\R^4$ operator} & {\bf $\PT$ operator} & {\bf Bulk field} & {\bf Bulk degree}\\
\hline
Supercurrent& $H_{I\al\mu}$ & $H_I$ & $\p^I\al$ & $\PV^{0,1}$\\
Stress tensor & $T_{\mu\nu}$ & $T_A$ & $\beta^A$ & $\PV^{1,1}$\\
R-symmetry current & $R^I{}_{J\mu}$ & $R^I{}_J$ & $\p^J\beta_I$ & $\PV^{1,1}$\\
Supercurrent& $G^I_{\mu\dal}$ & $G^I{}_A$ & $\gamma^A{}_I$ & $\PV^{2,1}$\\
\end{tabular}
\caption{The bulk-boundary dictionary relating superconformal currents to BCOV fields.}
\label{tab:dict}
\end{table}

In total, we find the symmetry currents $T_{\mu\nu},G^I_{\mu\dal},H_{I\al\mu},R^I{}_{J\mu}$ governing the $\cN=4$ superconformal algebra on spacetime. This entry of our bulk-boundary dictionary is more or less analogous to the standard AdS/CFT dictionary. We find that the stress tensor $T_{\mu\nu}$ is dual to the Beltrami mode $\beta^A\p_A\in\PV^{1,1}(X)$ that describes complex structure deformations pointing along twistor space. The R-symmetry currents $R^I{}_{J\mu}$ are dual to the trace-free part of the Beltrami mode $\p^I\beta_J\p^J\in\PV^{1,1}(X)$. This describes complex structure deformations in the normal $W_J$ directions. Similarly, the supercurrents $H_{I\al\mu}$ and $G^I_{\mu\dal}$ are respectively dual to the modes $\p^I\al\in\PV^{0,1}(X)$ and $\gamma^A{}_I\p_A\p^I\in\PV^{2,1}(X)$ of the bulk fermions $\al,\gamma$. This dictionary is tabulated in table \ref{tab:dict}. 

In the original interpretation of twistor strings \cite{Berkovits:2004jj}, the couplings of these currents to the Beltrami differential on supertwistor space $\pt$ was interpreted as the coupling of $\cN=4$ sdYM to \emph{dynamical} $\cN=4$ conformal supergravity on $\R^4$. In our construction, closed strings are not confined to the branes which are not space-filling. We prefer to interpret these couplings in terms of a holographic dictionary. The situation is exactly analogous to the standard holographic dictionary, where the leading terms in the Fefferman-Graham expansion of $\cN=8$ Einstein supergravity in $AdS_5$ correspond to the $\cN=4$ conformal supergravity multiplet on $\R^4$ \cite{Ferrara:1998ej,Liu:1998bu}.

%%%%%%%%%%%%%%%%%%%%%%%%%%%%
%%%%%%%%%%%%%%%%%%%%%%%%%%%%

\section{Dictionary for local spacetime operators}
\label{sec:dict}

In the previous section, we discussed a holographic dictionary between local operators on $\PT$ and closed string states in the bulk. In special cases, we were able to use it to find the bulk fields dual to symmetry currents, such as the stress tensor of $\cN=4$ sdYM. But in general, this remains unsatisfactory since we are looking for a duality between local operators on $\R^4$ and closed string states in the bulk. In this section, we argue that local operators on spacetime are not dual to single-particle gravitational states. Instead, they are naturally encoded in \emph{heavy} objects like D-branes in the bulk. 

As an illustration, we show how Witten's D-instantons arise as D9 branes wrapping the fibres of our 7-fold $X = \CO(-1)^{\oplus4}\to\PT$. These encode Lagrangian insertions $L_\text{int}$. We follow this with a discussion of half-BPS operators. Recently, in \cite{Caron-Huot:2023wdh}, Caron-Huot \emph{et al.}\  arranged the half-BPS operators of $\cN=4$ sdYM into a generating functional that took the form of a determinant operator. We show that such determinants are dual to giant graviton D5 branes that intersect the original D5 branes wrapping supertwistor space in twistor superlines $\P^{1|2}$.

%%%%%%%%%%%%%%%%%%%%%%%%%%%%
%%%%%%%%%%%%%%%%%%%%%%%%%%%%

\subsection{D-instantons are D9 branes}
\label{ssec:D9}

Twistor string theory originated in \cite{Witten:2003nn}, where Witten argued that D-instanton perturbation theory can be used to compute $\cN=4$ sYM amplitudes in an expansion around its self-dual sector. In that work, the target space was thought of as being supertwistor space $\pt$ instead of our 7-fold $X$, while a D-instanton was thought to be a D1 brane wrapping a rational curve in $\pt$. The simplest D-instanton wrapped a twistor line at a fixed point $(x^{\dal\al},\theta^{I\al})$ in chiral\footnote{To simplify expressions in chiral superspace, we will work on the slice $\til\theta^\dal_I=0$, which is possible in Euclidean signature without setting $\theta^{I\al}=0$.} $\cN=4$ superspace $\R^{4|8}$. Indeed, considering the effect of strings stretched between the $N$ branes filling $\pt$ and such D-instantons gave rise to remarkable worldsheet formulae for $\cN=4$ sYM scattering amplitudes \cite{Roiban:2004yf,Roiban:2004vt,Berkovits:2004hg}. However, the effect of D1-D1 strings was largely overlooked. This motivates us to reinterpret Witten's D-instantons as branes in our twistor string on $X$. 

For simplicity, we focus on degree 1 D-instantons wrapping twistor lines, which contribute to MHV amplitudes of $\cN=4$ sYM. In Witten's picture, the theory governing the strings connecting the branes filling $\pt$ to the D-instanton was taken to be a 2d chiral gauge theory with action
\be\label{chiac}
S = \int_{\P^1_{x,\theta}}\D\lambda\;\til\chi\,\dbar_\cA\chi\,,
\ee
where $\D\lambda = \lambda^\al\d\lambda_\al$, the kinetic operator $\dbar_\cA = \dbar + \cA$ is the gauge-covariant derivative involving the supertwistor gauge field $\cA$, and $\chi,\til\chi$ are weight $-1$ fermions that transform in the fundamental and antifundamental of $\GL_N(\C)$ respectively. Since $\chi,\til\chi$ are $0$-forms on $\P^1$, the components of $\cA$ that survive in this action are necessarily the physical $(0,1)$-parts of $\cA$.

\paragraph{D5/D9 gauge theory.} We would like to identify a brane in $X$ that can reproduce this action. In our formulation, the correct brane that accomplishes this is a not a D1 brane but rather a D9 brane that wraps the holomorphic locus $\mu^{\dal}=x^{\dal\al}\lambda_\al$ in $X$. This brane has 5 complex bosonic dimensions, as it wraps a $\P^1\subset\PT$ as well as the fibers of $\CO(-1)^{\oplus4}$ above this line.

We must also specify boundary conditions on the worldsheet fermions. To obtain an allowed D-brane in the B-model on $X$, one must ensure that the shifted holomorphic symplectic potential on $\Pi T^*_X$, 
\be\label{Theta}
\Theta = \veps_A\d Z^A + \eta^I\d W_I\,,
\ee
becomes exact when pulled back to the brane. This is a necessary condition for the boundary state defined by such a brane to preserve the scalar supersymmetry of the B-twisted worldsheet theory: the supersymmetry variation of the worldsheet action gives rise to an integral of $\Theta$ around the boundary of the worldsheet \cite{Hori:2003ic,Ikeda:2013wh}, which vanishes precisely when $\Theta$ is exact on the boundary.\footnote{The boundary term is readily obtained by an application of Noether's theorem to the worldsheet action of a 2d $\cN=(2,2)$ supersymmetric sigma model \cite{Hori:2003ic}. The D5 branes filling twistor space have $\veps_A=0$, $W_I=0$, so $\Theta$ is trivially exact.} 

Motivated by Witten's expression, altogether we choose the boundary conditions
\be\label{D9emb}
\begin{split}
    \mu^\dal &= x^{\dal\al}\lambda_\al\,,\\
    \eta^I &= \theta^{I\al}\lambda_\al\,,\\
    \veps^\al &= W_I\theta^{I\al}-\veps_\dal x^{\dal\al}
\end{split}
\ee
on the fermionic fields, where we have split $\veps_A = (\veps_\dal,\veps^\al)$. It is easily checked that
\be
\begin{split}
    \Theta\bigr|_{\D9} &= (W_I\theta^{I\al} - \veps_\dal x^{\dal\al})\,\d\lambda_\al + \veps_\dal x^{\dal\al}\d\lambda_\al + \d W_I\,\theta^{I\al }\lambda_\al \\
    &= \d(W_I\theta^{I\al}\lambda_\al )\,.
\end{split}
\ee
As a further consistency check, we observe that the embedding \eqref{D9emb} satisfies the projectivization constraint \eqref{euler}. Thus, from the point of view of the target superspace $\Pi T^*_X$, the D9 brane wraps the super-directions $\veps_\dal$ complimentary to $\mu^\dal$, while staying localized in the $\eta^I,\veps^\al$ directions.

Open strings confined to this D9 brane are described by a superfield
\be
\cB \in \Pi\Omega^{0,*}(\D9,\gl_1(\C))\,,
\ee
which is Abelian since we only have a single D9. Here, `D9' is short for the locus \eqref{D9emb} that the D9 brane wraps. These strings are governed by another holomorphic Chern-Simons action
\be
S_{99} = \frac12\int_{\D9}\D\lambda\,\d^4W\,\d^2\veps\;\cB\,\dbar\cB
\ee
living on the D9 worldvolume. In particular, the integral over $\d^2\veps$ is along the $\veps_\dal$ directions. 

Finally, we have 5-9 and 9-5 strings living on the intersection of the D5 and D9 branes. This is precisely the line $\P^1_{x,\theta}$ in the zero section $W_I=\veps_A=0$. They correspond to a pair of fields 
\be
\begin{split}
    &\bs{\chi}\in\Pi\Omega^{0,*}(\P^1_{x,\theta},\CO(-1)\otimes F_N)\,,\\
    &\til{\bs{\chi}}\in\Pi\Omega^{0,*}(\P^1_{x,\theta},\CO(-1)\otimes F^\vee_N)\,,
\end{split}
\ee
where $F_N$ and $F_N^\vee$ denote the fundamental and antifundamental of $\GL_N(\C)$, and where the 5-9 and 9-5 strings have opposite charges under the gauge field living on the D9 branes. These strings are governed by the action principle
\be
S_{59} = \int_{\P^1_{x,\theta}}\D\lambda\;\til{\bs{\chi}}\,(\dbar + \cA - \cB)\bs{\chi}\,,
\ee
which is a BV completion of the action \eqref{chiac}. The $0$-form components of $\bs{\chi},\bs{\til\chi}$ get identified with $\chi,\til\chi$. In holographic applications, we will decouple the 9-9 strings $\cB$ by turning off the $\bs{\til\chi}\cB\bs{\chi}$ coupling. Thus, we can identify Witten's D-instanton with D9 branes wrapping a twistor line in twistor space as well as the four fibre directions of $\CO(-1)^{\oplus4}$.

\paragraph{Determinants and their anomalies.} Next, let us explore what operators these branes give rise to in the boundary theory. We will do this  by decoupling the 9-9 strings, and considering purely the theory of the 5-9 and 9-5 strings as an insertion in the path integral of $\cN=4$ sdYM. This is a common technique in twisted holography \cite{Budzik:2021fyh}.

However, before we decouple the 9-9 strings, we must address an important subtlety. If we use the gauge freedom in the $\P^1$ directions to set the $(0,1)$-form parts of $\bs{\chi},\til{\bs\chi}$ to zero, the 5-9 and 9-5 theory reduces to the path integral associated with the action \eqref{chiac}. Integrating out $\chi$ and $\til\chi$ then produces a functional determinant insertion in the path integral of the 5-5 strings,
\be\label{detA}
\det\!\big(\dbar_\cA|_{\P^1_{x,\theta}}\big)\,.
\ee
This determinant suffers from a well-understood 1-loop gauge anomaly,
\be\label{detanomaly}
Q\log\det\!\big(\dbar_\cA|_{\P^1_{x,\theta}}\big) = \frac12\int_{\P^1_{x,\theta}}\Tr(\cA\,\p\cA)\,,
\ee
which may be cancelled by a standard anomaly inflow mechanism. This mechanism was first discussed in \cite{Berkovits:2004jj}, and it can be easily adapted to twistor strings viewed as a B-model on $\CO(-1)^{\oplus4}\to\PT$. (We also remark in passing that gauge-invariant operators closely related to~\eqref{detA} include the integral of its logarithm over an appropriate combination of $\theta^{I\al}$s \cite{Boels:2006ir}, and a holomorphic analogue of a Wilson line. The Wilson loop construction is closely related to the computation of scattering amplitudes \cite{Mason:2010yk}, as we briefly touch upon in section \ref{sec:outlook}.) 

Briefly, the combined theory of closed strings and the D5/D9 open strings must be anomaly-free. In the presence of the D5 branes, the BRST transformation of closed strings receives a correction supported on the D5 worldvolume $\pt$. At the same time, the D9 brane produces a tadpole for the closed strings which is an integral over the D9 worldvolume. The BRST variation of the D9 tadpole receives a contribution from the D5 correction which localizes on the D5/D9 intersection. This precisely cancels the 1-loop gauge anomaly of the theory of the 5-9 and 9-5 strings.

This anomaly inflow mechanism works in the context of brane intersections in the B-model. But in the holographic context, we would like to study determinants like \eqref{detA} purely as operator insertions in the path integral of $\cN=4$ sdYM, after having decoupled the 9-9 strings and closed strings. Fortunately, in this case, we can still cancel the gauge anomaly \eqref{detanomaly} by turning on a local counterterm that takes the form of a \emph{gravitational dressing}, in the spirit of \cite{Dirac:1955uv,Donnelly:2015hta,Donnelly:2016rvo}.

Such a counterterm can be motivated by formally looking at the backreaction of a single D9 brane.\footnote{We emphasize that we will treat this backreaction purely as a formal gadget to implement a version of anomaly inflow within the D5 brane gauge theory. We are \emph{not} studying holography at finite $N$.} This is a closed string background $\Bsi = \Bsi_{x,\theta}\in\PV^{*,1}(X)$ that solves
\be\label{Psixt}
\dbar\Bsi_{x,\theta} = \delta_{\D9}\,,
\ee
where $\delta_{\D9}$ is a super delta function on $\Pi T^*_X$ that is supported along the D9 locus \eqref{D9emb}, see \eqref{delta9} below.\footnote{Strictly speaking, we should demand
\be
\dbar\Bsi_{x,\theta} + \frac12\,[\Bsi_{x,\theta},\Bsi_{x,\theta}] = \delta_{\D9}\,,
\ee
but as usual we can solve this equation through a Bochner-Martinelli kernel $\Bsi_{x,\theta}\propto \D\hat\mu$ which ensures that $[\Bsi_{x,\theta},\Bsi_{x,\theta}]\propto\D\hat\mu\wedge\D\hat\mu=0$.} Employing the open-closed couplings discussed in section \ref{ssec:global}, we can couple this as a fixed background to the theory on the D5 branes, and thereafter turn off the closed string fluctuations as well as the 9-9 strings. This amounts to inserting the combination
\be
\exp\left(-\frac12\int_\pt\D^{3|4}\cZ\;\Tr\,\cA\,[\Bsi_{x,\theta},\cA]\right)\;\det\!\big(\dbar_\cA|_{\P^1_{x,\theta}}\big)\,,
\ee
into the path integral of the D5 brane gauge theory. This \emph{gravitational dressing} of the naive determinant operator furnishes an anomaly-free operator insertion.

It is possible to find an explicit expression for $\Bsi_{x,\theta}$ in harmonic or axial gauge, but we will only need \eqref{Psixt} to verify the anomaly cancellation. Using $Q\cA=\dbar\cA+\cA^2$ and integrating by parts, the BRST variation of our gravitational dressing is found to be
\be\label{Qdress}
\begin{split}
    Q\int_\pt\D^{3|4}\cZ\;\Tr\,\cA\,[\Bsi_{x,\theta},\cA] &= \int_\pt\D^{3|4}\cZ\;\Tr\,\cA\,[\dbar\Bsi_{x,\theta},\cA]\\
    &= \int_\pt\D^{3|4}\cZ\;\Tr\,\cA\,[\delta_{\D9},\cA]\,.
\end{split}
\ee
To compute this integral, let us work in the patch $\iota^\al\lambda_\al\neq0$ for some reference spinor $\iota_\al$ and choose the representative
\be\label{delta9}
\delta_{\D9} = \bar\delta^{2|4}(\mu-x\lambda\,|\,\eta-\theta\lambda)\,\frac{\iota_\al}{\la\lambda\iota\ra}\left(\veps^\al+x^{\dal\al}\veps_\dal-\theta^{I\al}W_I\right)
\ee
for the D9 delta function.\footnote{Keep in mind that $\delta(\theta)=\theta$ for a Grassmann variable $\theta$. So $\delta_{\D9}$ imposes $\iota_\al(\veps^\al+x^{\dal\al}\veps_\dal-\theta^{I\al}W_I)=0$. By the quadric constraint $Z^A\veps_A-W_I\eta^I=0$, we automatically also have $\lambda_\al(\veps^\al+x^{\dal\al}\veps_\dal-\theta^{I\al}W_I)=0$ on the support of $\delta_{\D9}$. Since $\iota_\al,\lambda_\al$ form a spinor basis, $\delta_{\D9}$ indeed imposes $\veps^\al=\theta^{I\al}W_I-x^{\dal\al}\veps_\dal$ on $\Pi T^*_X$.} This is proportional to a delta function that localizes supertwistor integrals to integrals over the line $\P^1_{x,\theta}$. With this representative, we obtain
\be\label{d9a}
\begin{split}
    [\delta_{\D9},\cA] &= \p^A\delta_{\D9}\,\p_A\cA - \p^I\delta_{\D9}\,\p_I\cA\\
    &= \bar\delta^{2|4}(\mu-x\lambda\,|\,\eta-\theta\lambda)\,\frac{\iota_\al}{\la\lambda\iota\ra}\left(\frac{\p}{\p\lambda_\al}+x^{\dal\al}\frac{\p}{\p\mu^\dal}+\theta^{I\al}\frac{\p}{\p\eta^I}\right)\cA\\
    &= \bar\delta^{2|4}(\mu-x\lambda\,|\,\eta-\theta\lambda)\,\frac{\iota_\al}{\la\lambda\iota\ra}\frac{\p}{\p\lambda_\al}\big(\cA\big|_{\P^1_{x,\theta}}\big)\,.
\end{split}
\ee
The independence of this expresion from $\iota_\al$ may be verified using
\be
\lambda_\al\frac{\p}{\p\lambda_\al}\big(\cA\bigr|_{\P^1_{x,\theta}}\big) = \cZ^\bA\,\frac{\p\cA}{\p\cZ^\bA}\biggr|_{\P^1_{x,\theta}} = 0\,,
\ee
which is the statement that $\cA$ has homogeneity zero in $\cZ^\bA$. Finally, plugging \eqref{d9a} into \eqref{Qdress}, we get
\be
\begin{split}
    Q\int_\pt\D^{3|4}\cZ\;\Tr\,\cA\,[\Bsi_{x,\theta},\cA] &= \int_{\P^1_{x,\theta}}\d\lambda_\beta\,\frac{\lambda^\beta\iota_\al}{\la\lambda\iota\ra}\;\Tr\left(\cA\,\frac{\p}{\p\lambda_\al}\big(\cA\big|_{\P^1_{x,\theta}}\big)\right)\\
    &=\int_{\P^1_{x,\theta}}\d\lambda_\al\;\Tr\left(\cA\,\frac{\p}{\p\lambda_\al}\big(\cA\big|_{\P^1_{x,\theta}}\big)\right)\\
    &= \int_{\P^1_{x,\theta}}\;\Tr(\cA\,\p\cA)\,.
\end{split}
\ee
This precisely cancels the anomalous BRST variation \eqref{detanomaly}.

\paragraph{Lagrangian insertions.}  Having cancelled its anomaly, we can study the spacetime content of the determinant insertion \eqref{detA}. We can expand it in $\cA$ by taking a logarithm and applying $\log\det = \Tr\log$,
\begin{multline}\label{detexp}
\log\det\!\big(\dbar_\cA|_{\P^1_{x,\theta}}\big) - \log\det\!\big(\dbar|_{\P^1_{x,\theta}}\big) \\
= -\sum_{n=1}^\infty\frac{1}{n}\left(\frac{i}{2\pi}\right)^n\int_{(\P^1_{x,\theta})^n}\frac{\D\lambda_1\,\D\lambda_2\cdots\D\lambda_n}{\la12\ra\la23\ra\cdots\la n1\ra}\;\Tr(\cA_1\cA_2\cdots\cA_n)\,.
\end{multline}
Here, we have abbreviated $\la\lambda_i\lambda_j\ra\equiv\la ij\ra$, and denoted by $\cA_i$ the value of $\cA$ evaluated at the point $\lambda_{i\al}$ on the line $\P^1_{x,\theta}$. Terms in the $\theta$-expansion of this determinant have previously been studied in \cite{Boels:2006ir}. They contribute to the Lagrangian of non-self-dual $\cN=4$ sYM. The $n=1$ term in the sum in \eqref{detexp} is a self-energy divergence that may be renormalized by a suitable counterterm.

As explained in \cite{Boels:2006ir}, we can reduce this determinant to spacetime by working in Woodhouse gauge. In this gauge, we find
\be
\cA|_{\P^1_{x,\theta}} = \bar e^0\biggr(\frac12\,\Phi_{IJ}\eta^I\eta^J - \frac{2\,\hat\lambda^\al\Psi_{\al}^I\eta^3_I}{\|\lambda\|^2} + \frac{3\,\hat\lambda^\al\hat\lambda^\beta B_{\al\beta}\eta^4}{\|\lambda\|^4}\biggr)\biggr|_{\eta=\theta\lambda}\,,
\ee
so the $\theta$-expansion of $\cA$ starts at order $\theta^2$. Consequently, the sum over $n$ in \eqref{detexp} only runs over $n\leq4$. The result can be written as a superfield expansion in $\theta^{I\al}$, running up to order $\theta^8$. In particular, performing a $\d^8\theta$ integral extracts its top component, which precisely reproduces the non-self-dual interaction $L_\text{int}$ that deforms $\cN=4$ sdYM into $\cN=4$ sYM,
\be
\begin{split}
\int\d^8\theta\,\log\det\!\big(\dbar_\cA|_{\P^1_{x,\theta}}\big)&=-\frac12\,\Tr\left(B^{\al\beta}B_{\al\beta} + \Phi_{IJ}\Psi^{I\al}\Psi^J_{\al} + \frac{1}{2}\,\Phi_{IJ}\Phi^{JK}\Phi_{KL}\Phi^{LI}\right)\\
&\equiv L_\text{int}\,.
\end{split}
\ee
Exponentiating \eqref{detexp}, we can express our determinant as
\be
    \det\!\big(\dbar_\cA|_{\P^1_{x,\theta}}\big) \propto \exp\left(A\,\theta^2 + \cdots+ L_\text{int}\theta^8\right)
\ee
up to a proportionality constant independent of $(x,\theta)$. Each term in the exponential on the right is a single-trace term. The expansion of the exponential also truncates at order $\theta^8$. The top component of this expansion takes the form
\be
\left(L_\text{int} + \text{multi-trace}\right)\theta^8\,.
\ee
As a result, Witten's D-instantons encode the non-self-dual Lagrangian insertions, up to multi-trace terms. This allowed Witten to compute tree-level MHV amplitudes of $\cN=4$ sYM perturbatively around its self-dual sector using such determinant insertions \cite{Witten:2003nn}.

%%%%%%%%%%%%%%%%%%%%%%%%%%%%
%%%%%%%%%%%%%%%%%%%%%%%%%%%%

\subsection{Half-BPS operators and D5 giants}
\label{ssec:D5}

As an application of our duality, we would like to develop a similar dictionary for single-trace half-BPS operators. The recent work \cite{Caron-Huot:2023wdh} teaches us that when restricted to chiral superspace, half-BPS multiplets in the self-dual theory can be cleverly repackaged into heavy determinant operators. In this section, we show that such determinants arise from D5 brane giant gravitons wrapping a twistor line in $\PT$ and two out of the four fiber directions of $\CO(-1)^{\oplus4}$.

To set up such a geometry, let's split the coordinates on $X$ into pairs:
\be
Z^A = (\mu^\dal,\lambda_\al)\,,\qquad W_I = (\nu^{\dot a},\tau_a)\,,
\ee
where $a=0,1$ and $\dot a=\dot0,\dot1$ are another set of 2-component indices that are R-symmetry analogs of the spinor indices $\al,\dal$. The twistor $Z^A$ is split along the lines of the familiar decomposition $\SO(4)=(\SU(2)\times\SU(2))/\Z_2$ of the Euclidean rotation group. Similarly, the splitting of $W_I$ is meant to break the $\SU(4)$ R-symmetry down to $\SU(2)\times\SU(2)$. We can also split the dual odd coordinates $\veps_A$ and $\eta^I$ into pairs:
\be
\veps_A = (\veps_\dal,\veps^\al)\,,\qquad\eta^I = (\eta_{\dot a},\eta^a)\,.
\ee
As the reader might already suspect, our brane will wrap essentially half of these bosonic directions and a complementary half of the fermionic directions.

A half-BPS operator on spacetime gives rise to a superfield that depends on only half of the eight superspace coordinates $\theta^{I\al}=(\theta^\al_{\dot a},\theta^{\al a})$ and their right-handed counterparts $\til\theta^\dal_I$. In the spirit of \cite{Caron-Huot:2023wdh}, we will restrict attention to the pullbacks of such operators to chiral superspace $\til\theta^\dal_I=0$. The resulting superfields will depend on a choice of four Grassmann coordinates, say
\be
\zeta^{\al a} = \theta^{\al a} + y^{\da a}\theta^\al_{\dot a}
\ee
for some matrix $y^{\da a}\in\C^{2\times 2}$.

We propose that this data can be encoded into a D5 brane wrapped along the following locus in the target supermanifold $\Pi T^*_X$:
\be\label{D5'}
\begin{split}
    \mu^\dal &= x^{\dal\al}\lambda_\al\,,\\
    \nu^{\dot a} &= y^{\da a}\tau_a\,,\\
    \veps^\al &= \zeta^{\al a}\tau_a - x^{\dal\al}\veps_\dal\,,\\
    \eta^a &= \zeta^{\al a}\lambda_\al - y^{\da a}\eta_{\dot a}\,.
\end{split}
\ee
One may verify that this embedding satisfies the projectivization constraint \eqref{euler}. It is an allowed brane in the B-model because the symplectic potential \eqref{Theta} again evaluates to an exact quantity,
\be\label{ThetaD5'}
\begin{split}
    \Theta\bigr|_{\D5'} &= (\zeta^{\al a}\tau_a - x^{\dal\al}\veps_\dal)\,\d\lambda_\al + \veps_\dal x^{\dal\al}\d\lambda_\al + \eta_{\dot a}y^{\da a}\d\tau_a + (\zeta^{\al a}\lambda_\al - y^{\da a}\eta_{\dot a})\,\d\tau_a\\
    &= \d(\zeta^{\al a}\lambda_\al\tau_a)\,.
\end{split}
\ee
Here and in what follows, to distinguish such \emph{giant graviton branes} from our original stack of $N$ D5 branes wrapping supertwistor space, we refer to them as D5$'$ branes.

Half-BPS operators like $\Tr\,\Phi^n$ will emerge from this data through the theory of 5-5$'$ and 5$'$-5 strings. The 5-5$'$ and 5$'$-5 strings live on the brane intersection
\be\label{superline}
\P^{1|2}_{x,y,\zeta}\;:\quad\mu^\dal = x^{\dal\al}\lambda_\al\,,\quad\eta^a=\zeta^{\al a}\lambda_\al-y^{\da a}\eta_{\dot a}\,,\quad W_I = \veps_A = 0\,.
\ee
This is a superline, and it is convenient to collect its coordinates into a super-coordinate
\be
\bs{\lambda}_{\bs\al} = (\lambda_\al|\eta_{\dot a})\,.
\ee
The string fields representing 5-5, 5-$5'$, $5'$-5 and $5'$-$5'$ strings are respectively given by $\cA$ along with
\be
\begin{split}
    %&\cA \in \Pi\Omega^{0,*}(\pt,\gl_N(\C))\,,\\
    &\bs{\chi} \in \Pi\Omega^{0,*}(\P^{1|2}_{x,y,\zeta},F_N)\,,\\
    &\til{\bs\chi} \in \Pi\Omega^{0,*}(\P^{1|2}_{x,y,\zeta},F_N^\vee)\,,\\
    &\mathscr{A}\in\Pi\Omega^{0,*}(\D5',\gl_1(\C))\,,
\end{split}
\ee
where $F_N, F_N^\vee$ are the fundamental and antifundamental of $\GL_N(\C)$ respectively, and `D5$'$' is short for the locus \eqref{D5'}. The 5-$5'$ and $5'$-5 strings are governed by the action
\be
S_{55'} = \int_{\P^{1|2}_{x,y,\zeta}}\D^{1|2}\bs\lambda\;\Big(\til{\bs\chi}\,\dbar_\cA\bs\chi + \Tr(\bs\chi\mathscr{A}\til{\bs\chi}) + \cdots\Big)\,,
\ee
with the trace in the second term being over $\gl_N(\C)\simeq F_N\otimes F_N^\vee$ matrices. The ellipses denote quartic and higher-order open-open couplings that will not play any role in our analysis.

We will develop the theory of 5$'$-5$'$ strings in the coming sections, so let us decouple those for the moment. This entails setting $\mathscr{A}=0$. However, let us choose to freeze $\mathscr{A}$ to a nonzero vacuum expectation value (vev)
\be\label{12vev}
\mathscr{A} = \bar\delta^{1|2}(\bs\lambda,\bs\lambda_*)\,.
\ee
This is a weightless, $(0,1)$-form-valued delta function that fixes $\bs\lambda$ to a `reference' point $\bs\lambda_*$ on $\P^{1|2}$. For most of this work, we will choose $\bs\lambda_*$ to carry no fermionic components.

Upon partially gauge-fixing by setting the $(0,1)$-form parts of $\bs\chi,\til{\bs\chi}$ to zero, and renaming the $0$-form parts as $\chi,\til\chi$, this results in the bulk-decoupled action
\be\label{55'ac}
S_{55'} = \int_{\P^{1|2}_{x,y,\zeta}}\D^{1|2}\bs\lambda\;\,\til{\chi}\left(\dbar_\cA + \bar\delta^{1|2}(\bs\lambda,\bs\lambda_*)\right)\chi\,.
\ee
Once again, integrating out $\chi,\til{\chi}$ generates the insertion of a functional determinant in the path integral of $\cN=4$ sdYM. Since these fields are sections of $\CO_{\P^1}$, they carry one zero mode in each of their $N$ components. This would have forced a naive evaluation of their path integral to vanish. Thankfully, the $\bar\delta^{1|2}(\bs\lambda,\bs\lambda_*)\,\til\chi\chi$ mass term comes to the rescue. It generates insertions that soak up these zero modes and lead to a nonvanishing result.

The 2d CFT \eqref{55'ac} is precisely the theory that was used to construct a generating functional of half-BPS operators in \cite{Caron-Huot:2023wdh}. This generating functional is a determinant operator, and our analysis shows that its bulk dual is a D5$'$ giant with a vev \eqref{12vev} turned on. In the rest of this section, we briefly review this determinant operator. In the next section, we will develop techniques for encoding the correlators of such determinants in the configurations of giants.

\paragraph{Integrating out 5-5$'$ and 5$'$-5 strings.} Let us define a deformed dbar operator that incorporates the vev \eqref{12vev},
\be\label{dbars}
\dbar_* \equiv \dbar + \bar\delta^{1|2}(\bs\lambda,\bs\lambda_*)\,.
\ee
The free-field $\chi$-$\til\chi$ 2-point function in the presence of the vev is the Green's function of $\dbar_*$. It can be written in a gauge adapted to the reference point $\bs\lambda_*$,
\be\label{chi2}
\big\la\chi_r(\bs\lambda_i)\,\til\chi^s(\bs\lambda_j)\big\ra = \delta^s_r\big(1+[i,j,*]\big) \equiv \delta_r^s\,\Delta(\bs\lambda_i,\bs\lambda_j)\,,
\ee
with $r,s$ being the indices of $F_N$. Here, $[i,j,k]$ is a 3-point analog of the 5-point R-invariant $[i,j,k,l,m]$ commonly encountered in $\cN=4$ superamplitudes,
\be\label{Rijk}
[i,j,k] = \frac{\bar\delta^{0|2}(\la ij\ra\eta_k+\la jk\ra\eta_i + \la ki\ra\eta_j)}{\la ij\ra\la jk\ra\la ki\ra}\,.
\ee
This invariant has the useful property that it vanishes if any two of its three arguments $\bs\lambda_i,\bs\lambda_j,\bs\lambda_k$ coincide. We have also defined a color-stripped Green's function $\Delta(\bs\lambda_i,\bs\lambda_j)=1+[i,j,*]$ that will show up in the computations below.

Integrating out $\chi,\til\chi$ produces the determinant insertion
\be
\det\!\big(\dbar_\cA|_{\P^{1|2}_{x,y,\zeta}}\big)
\ee
in the path integral of holomorphic Chern-Simons on $\pt$. Unlike the case of D9 branes, this D5$'$ determinant can be checked to have no gauge anomaly, essentially owing to the ``supersymmetry'' on $\P^{1|2}$. The gauge anomaly would have been proportional to the level of the $\GL_N(\C)$ current algebra generated by the currents $J_r{}^s(\lambda) = \int\d^2\eta\,\chi_r\til\chi^s$, but this level vanishes due to a cancellation between bosonic and fermionic contributions. Hence, we do not need any gravitational dressing or anomaly inflow to make this insertion well-defined.\footnote{This is fortunate, as one can show that the closed string backreaction of our D5$'$ giant does \emph{not} couple to the D5 branes. This would have foiled our gravitational dressing argument, had we needed it.}

Having confirmed its gauge invariance, we can reduce this determinant to spacetime by working in Woodhouse gauge. To accomplish this, we evaluate its logarithm,
\begin{multline}\label{detexp1}
\log\det\!\big(\dbar_\cA|_{\P^{1|2}_{x,y,\zeta}}\big) - \log\det\!\big(\dbar|_{\P^{1|2}_{x,y,\zeta}}\big) \\
= -\sum_{n=1}^\infty\frac{1}{n}\left(\frac{i}{2\pi}\right)^n\int_{(\P^{1|2}_{x,y,\zeta})^n}\D^{1|2}\bs\lambda_1\cdots\D^{1|2}\bs\lambda_n\,\Delta_{12}\Delta_{23}\cdots\Delta_{n1}\,\Tr(\cA_1\cA_2\cdots\cA_n)\,,
\end{multline}
having abbreviated $\Delta(\bs\lambda_i,\bs\lambda_j)\equiv\Delta_{ij}$. The Woodhouse representatives of the $(0,1)$-form fields were collected in equations \eqref{wa} -- \eqref{wb}. In this gauge, the $\eta$-expansion of the $\bar e^0$-component $\cA|_{\P^1_x}$ starts at order $\eta^2$, \emph{i.e}, with the scalar $\Phi_{IJ}(x)$. Due to its manifest supersymmetry, it suffices to determine the top component of the chiral superspace expansion of our determinant. So let us set $\zeta^{\al a}=0$ for simplicity. Pulling back $\cA$ to a superline $\P^{1|2}_{x,y,\zeta=0}$ yields
\be
\cA\bigr|_{\P^{1|2}_{x,y,0}} = \frac12\,\bar e^0\,\Phi_{IJ}(x)\,\eta^I\eta^J\bigr|_{\P^{1|2}_{x,y,0}}\,,
\ee
since terms cubic or quartic in $\eta^I$ drop out upon setting $\zeta^{\al a}=0$. 

To perform the superline integrals in \eqref{detexp1}, note that when $\zeta^{\al a}=0$, the fermionic incidence relations in \eqref{superline} reduce to
\be
\eta^I = y^{\dot{b} I}\eta_{\dot b}\,,\qquad\text{where}\;\; y^{\dot{b}I}\equiv \big(\delta^{\dot b}_{\ \dot a},-y^{\dot{b}a}\big)\,.
\ee
If we define a null vector in $\C^6$ by setting
\be
y^{IJ} = \eps_{\dot a\dot b}\,y^{\da I}y^{\dot{b}J}\,,
\ee
and introduce the inner product $y\cdot\Phi\equiv \frac12\,y^{IJ}\Phi_{IJ}$, then we obtain
\be
\cA\bigr|_{\P^{1|2}_{x,y,0}} = \bar e^0\,y\cdot\Phi(x)\,\eta_{\dot0}\eta_{\dot1}\,.
\ee
Plugging this into \eqref{detexp1}, we can saturate the $\eta_{i\dot a}$ integrals purely with the $\eta_{i\dot0}\eta_{i\dot1}$ factors contributed by the $\cA_i$'s. Hence, the 2-point functions $\Delta_{i,i+1}$ each reduce to $1$ and drastically simplify the determinant.%, which additionally ensures the independence of the determinant from the reference point $\bs\lambda_*$.

Performing the sphere integrals, we are left with
\be
\begin{split}
\log\det\!\big(\dbar_\cA|_{\P^{1|2}_{x,y,0}}\big) - \log\det\!\big(\dbar|_{\P^{1|2}_{x,y,0}}\big) &= -\sum_{n=1}^\infty\frac{(-1)^n}{n}\,\Tr(y\cdot\Phi)^n\\
&= \log\det(1+y\cdot\Phi)\,.
\end{split}
\ee
Exponentiating this, we conclude that integrating out the 5-$5'$ and $5'$-5 strings $\chi,\til\chi$ inserts the determinant $\det(1+y\cdot\Phi)$ in the path integral of $\cN=4$ sdYM. It is straightforward to generalize this to nonzero $\zeta^{\al a}$ by acting with holomorphic supercharges on $\det(1+y\cdot\Phi)$ to generate its superconformal descendants.

In summary, we can encode all the half-BPS operators formed from a single scalar into a determinant operator. It is this determinant operator (and its supersymmetrization) which is found to be dual to a giant graviton D5$'$ brane in the bulk 7-fold $X$. This is markedly different from the non-chiral duality between type IIB strings on AdS$_5\times S^5$ and $\cN=4$ sYM, where these operators happen to be dual to KK modes of supergravity fields on the $S^5$ \cite{Aharony:1999ti}. 

In fact, even in the non-chiral duality, determinantal generating functions of half-BPS operators provide a useful tool for computing their correlation functions \cite{Jiang:2019xdz}. The same trick works in our case. Correlators of half-BPS operators in $\cN=4$ sdYM can be extracted from correlators of such determinant operators by picking them out as coefficients of an expansion in $y^{IJ}$. Motivated by this, we study the computation of determinant correlators and develop their bulk interpretation in the next section.

%%%%%%%%%%%%%%%%%%%%%%%%%%%%
%%%%%%%%%%%%%%%%%%%%%%%%%%%%

\section{Determinant correlators}
\label{sec:giant}

In this section, we describe the bulk dual of half-BPS correlators in $\cN=4$ sdYM. This is implemented via the holographic dictionary for determinant operators developed in the previous section. 

We begin by reviewing the calculation of correlators of determinant operators such as $\det(1+y\cdot\Phi)$ in the self-dual theory carried out in \cite{Caron-Huot:2023wdh}. This computation boils down to a matrix model governing a single matrix $\rho_{ij}$, where $i,j=1,\dots,k$ label the determinant insertions. Next, we show that saddles of this matrix model map to nontrivial `shapes' of D5$'$ giant gravitons in the bulk. The giant graviton dual to a single determinant was discussed in the previous section. For a stack of $k$ (non-coincident) giants that engineer the $k$ determinant insertions, by a `shape' we will mean an on-shell open string configuration on their worldvolume. Such a configuration will encode the backreaction of the $N$ D5 branes on the geometry of the giants. %In these configurations, we will show that the on-shell worldvolume action of the giants reproduces the saddle-point action of the $\rho$ matrix model.

Some of these computations are much cleaner in the setup of twisted holography on a 3-fold. For the convenience of the reader, as an easy warmup, we provide an appendix that describes the calculation of determinant correlators in a chiral algebra dual to the B-model on the deformed conifold $\SL_2(\C)$. Part of this is a review of the original work \cite{Budzik:2021fyh} on the subject. However, our bulk computation of the worldvolume action of the giants in the presence of the backreaction of $N$ D1 branes is new. We provide two techniques for accomplishing this, one of which readily generalizes to our higher-dimensional case.

%%%%%%%%%%%%%%%%%%%%%%%%%%%%
%%%%%%%%%%%%%%%%%%%%%%%%%%%%

\subsection{The $\rho$ matrix model}
\label{ssec:rho}

We wish to study the correlator of $k$ bosonic determinant operators of the kind
\be
\mathbb{D}(x,y) = \det(1+y\cdot\Phi(x))\,.
\ee
In this section, we review the techniques of \cite{Caron-Huot:2023wdh,Jiang:2019xdz} that reduce such a correlator to a matrix integral over a single matrix $\rho$. This computation can also be generalized to incorporate the superconformal descendants of $\mathbb{D}(x,y)$ in chiral superspace. We remark on this in section \ref{sec:outlook}, but othertwise leave the construction of giant gravitons dual to the correlators of such supersymmetrized determinants to future work.

The correlator $\la\prod_{i=1}^k\mathbb{D}(x_i,y_i)\ra$ of $k$ such determinants is given by a path integral in the holomorphic Chern-Simons theory uplifting $\cN=4$ sdYM to twistor space. If we represent each determinant as the partition function of a pair of fermions as in \eqref{55'ac}, our computation turns into a path integral over the combined action of the 6d and 2d fields:
\be
    S[\cA,\chi,\til\chi] = \int_{\pt}\D^{3|4}\cZ\;\Tr\left(\frac12\,\cA\,\dbar\cA + \frac13\,\cA^3\right)
    + \sum_{i=1}^k\int_{\P^{1|2}_i}\D^{1|2}\bs\lambda\;\big(\til{\chi}_i\dbar_*\chi_i + \til\chi_i\cA\chi_i\big)\,,
\ee
where $\P^{1|2}_i\equiv\P^{1|2}_{x_i,y_i,0}$, and $\dbar_*$ is the ``massive'' dbar operator defined in \eqref{dbars}. For simplicity, we are choosing the same reference point $\bs\lambda_*$ on each superline.

For this calculation, we will work in a gauge in which we will only need to keep the physical $(0,1)$-form fields in $\cA$. A convenient gauge to perform such path integrals is CSW gauge. This is an axial gauge in which the $\cA^3$ term in the holomorphic Chern-Simons action drops out, and the Green's function of $\dbar$ on $\pt$ takes the form
\be\label{AA}
\big\la\cA_r{}^s(\cZ_i)\,\cA_{u}{}^{v}(\cZ_j)\big\ra = \delta_r^{v}\delta_{u}^s\,\bar\delta^{2|4}(\cZ_i,\cZ_j,\cZ_*) 
\ee
for some fixed choice of reference supertwistor $\cZ^\bA_*$. The notation $\bar\delta^{2|4}(\cZ_i,\cZ_j,\cZ_k)$ represents a delta function whose support is the locus where $\cZ_i,\cZ_j,\cZ_k$ are projectively collinear (see the helpful review \cite{Adamo:2013cra}),
\be
\bar\delta^{2|4}(\cZ_i,\cZ_j,\cZ_k) = \int_{(\C^*)^2}\frac{\d s}{s}\,\frac{\d t}{t}\;\bar\delta^{4|4}(\cZ_i+s\cZ_j+t\cZ_k)\,.
\ee
2-point functions like \eqref{AA} are typically normalized with a factor of $1/N$. But we will find it convenient to instead keep the `bare' normalization displayed in \eqref{AA}. This will give rise to factors of $N$ in various expressions, which will help us keep track of terms that are generated due to backreaction.

On dropping the $\cA^3$ interaction, the integral over $\cA$ becomes Gaussian and can be performed in closed form. This results in an effective action for the 2d fields which consists of their kinetic term plus the following four-fermion interaction vertex,
\be
\frac{1}{2}\sum_{i,j=1}^k\int_{\P^{1|2}_i}\D^{1|2}\bs\lambda_i\int_{\P^{1|2}_j}\D^{1|2}\bs\lambda_j\;\bar\delta^{2|4}(\cZ_i,\cZ_j,*)\,\Tr(\chi_i\til\chi_j)\,\Tr(\chi_j\til\chi_i)\,,
\ee
with $\chi_i\equiv\chi_i(\bs\lambda_i)$, $\til\chi_i\equiv\til\chi_i(\bs\lambda_i)$, etc. The supertwistors $\cZ_i$ are being evaluated at the points $\bs\lambda_i$ on the superlines $\P^{1|2}_i$. 

Because this four-fermion vertex involves equal numbers of integrals and delta functions of either parity, it can be completely localized down to
\be\label{4f}
-\frac12\sum_{i\neq j}\frac{y_{ij}^2}{x_{ij}^2}\,(\chi\til\chi)_{ij}(\chi\til\chi)_{ji}\,,
\ee
written in terms of the products
\be
(\chi\til\chi)_{ij} \equiv \chi_{ir}(\bs\lambda_{ij})\,\til\chi_{j}^r(\bs\lambda_{ji})\,.
\ee
Here, $x_{ij}^2\equiv \eps_{\al\beta}\eps_{\dal\dot\beta}x_{ij}^{\dal\al}x_{ij}^{\dot\beta \beta}$ and $y_{ij}^2\equiv \eps_{ab}\eps_{\dot a\dot b}y_{ij}^{\da a}y_{ij}^{\dot{b} b}$ are (squares of) flat space norms of the separations $x_{ij}=x_i-x_j$ and $y_{ij}=y_i-y_j$. The points $\bs\lambda_{ij},\bs\lambda_{ji}$ are solutions of the delta function constraints. Their precise values may be found in \cite{Caron-Huot:2023wdh}, but we will not need them when dealing with purely bosonic determinants. The only fact we need is that in this case, using $\zeta_i^{\al a}=0$ we obtain the simplification
\be\label{lij}
\bs\lambda_{ij\bs\alpha} = (\lambda_{ij\al}\,|\,0)\,,
\ee
i.e., their fermionic components vanish. 

At this stage, we perform a Hubbard-Stratonovich transform. We integrate in bosonic matrix variables $\rho_{ij}$ for $i\neq j$ that allow us to replace this four-fermion vertex by
\be
\frac12\sum_{i\neq j}\frac{x_{ij}^2}{y_{ij}^2}\,\rho_{ij}\rho_{ji} + \sum_{i\neq j}\rho_{ij}(\chi\til\chi)_{ji}\,.
\ee
It is clear that integrating out $\rho_{ij}$ reverts us back to \eqref{4f}, up to an overall Gaussian integral over $\rho$ that may be accommodated as a normalization constant. The complete action for the 2d fermions $\chi,\til\chi$ is now Gaussian, so they may also be integrated out.

With hindsight, it is helpful to think of the label $i$ on the fermions as being an index of the fundamental representation $F_k$ of $\GL_k(\C)$. Integrating out the fermions inserts a functional determinant into the leftover matrix integral over $\rho$,
\be
\big(\det(\dbar_*+\cL)|_{\P^{1|2}}\big)^N\,,
\ee
where both $\dbar_*\equiv \dbar+\bar\delta^{1|2}(\bs\lambda,\bs\lambda_*)$ and $\cL$ are operators acting on quantities valued in $F_k$. The operator $\cL$ acts on the fermions $\chi_i$ as
\be
(\cL\chi)_i(\bs\lambda) = \sum_{j}\rho_{ij}\,\bar\delta^{1|2}(\bs\lambda,\bs\lambda_{ij})\,\chi_j(\bs\lambda_{ji})\,,
\ee
where we have introduced dummy diagonal entries $\rho_{ii}=0$ to streamline notation.

This determinant can be evaluated perturbatively in $\cL$, once again by computing its logarithm,
\be
\log\det(\dbar_*+\cL)) - \log\det\dbar_* = -\sum_{n=1}^\infty\frac{(-1)^n}{n}\,\tr(\dbar_*^{-1}\cL)^n
\ee
where `$\tr$' denotes a trace over $\gl_k(\C)$ matrices like $\rho_{ij}$ formed from the fusion $F_k\otimes F_k^\vee$. Keep in mind that this is different from the trace `$\Tr$' over $\gl_N(\C)$ matrices that we have been employing up to this point.

We can invert the $\dbar_*$ operator using its Green's function $\Delta(\bs\lambda,\bs\lambda')=1+[\bs\lambda,\bs\lambda',*]$ mentioned in \eqref{chi2}. This yields the trace
\be\label{trexp}
    \tr(\dbar_*^{-1}\cL)^n = \sum_{i_1,\cdots,i_n}\rho_{i_1i_2}\rho_{i_2i_3}\cdots\rho_{i_ni_1}\Delta_{i_1i_n,i_1i_2}\Delta_{i_2i_1,i_2i_3}\cdots\Delta_{i_ni_{n-1},i_ni_1}\,,
\ee
having abbreviated $\Delta(\bs\lambda_{ij},\bs\lambda_{kl})\equiv\Delta_{ij,kl}$. As mentioned in \eqref{lij}, our choice of working with bosonic determinants forced the $\bs\lambda_{ij}$ to carry no fermionic components. Without loss of generality, we can choose the reference point $\bs\lambda_*$ to also carry no fermionic components. With this choice of gauge, each propagator $\Delta_{i_mi_{m-1},i_mi_{m+1}}$ in \eqref{trexp} simply reduces to $1$, leaving us with
\be
\tr(\dbar_*^{-1}\cL)^n = \tr\,\rho^n\,.
\ee
As a result, our functional determinant evaluates to
\be
\log\det(\dbar_*+\cL)) - \log\det\dbar_* = \log\det(1+\rho)\,,
\ee
where $1$ stands for the $k\times k$ identity matrix in this determinant.

At this point, we are free to redefine the dummy components $\rho_{ii}$ by setting $\rho_{ii}=1$. This absorbs the identity matrix within $\rho$, so that we are left with a simple insertion of $(\det\rho)^N$ in the leftover integral over the off-diagonal $\rho_{ij}$. We can bring this insertion into the exponential to obtain an effective action for $\rho_{ij}$,
\be\label{rhoac}
S[\rho] = \sum_{i< j}\frac{x_{ij}^2}{y_{ij}^2}\,\rho_{ij}\rho_{ji} - N\log\det\rho\,.
\ee
We will refer to this as the ``$\rho$ matrix model''. All in all, we find that our determinant correlator evaluates to
\be\label{rhoint}
\Big\la\prod_{i=1}^k\mathbb{D}(x_i,y_i)\Big\ra = \frac{1}{Z_0}\int\prod_{i\neq j}\d\rho_{ij}\;\e^{-S[\rho]}\,,
\ee
where $Z_0$ is a free Gaussian integral over $\rho$,
\be
Z_0 = \int\prod_{i<j}\d\rho_{ij}\,\d\rho_{ji}\,\e^{-x_{ij}^2\rho_{ij}\rho_{ji}/y_{ij}^2} = \pi^{k(k-1)/2}\prod_{i<j}\frac{y_{ij}^2}{x_{ij}^2}\,.
\ee
This reduces the complicated calculation of a 6d correlation function to a 0d matrix integral! 

Key to this extreme simplification was the existence of CSW gauge that reduced the twistor action of $\cN=4$ sdYM to a purely Gaussian theory. This is a highlight of working with the self-dual theory, as the non-self-dual theory contains further interaction terms that will complicate such a reduction to zero dimensions even at small 't Hooft coupling.

At the same time, note that even though this is a 1-matrix integral over a $k\times k$ matrix, it does not possess a $\GL_k(\C)$ ``gauge'' symmetry and is not immediately amenable to the usual approaches for solving such integrals. In the next section, we will physically interpret this missing $\GL_k(\C)$ symmetry as a Higgsed phase of the worldvolume theory on a stack of $k$ giant gravitons.

%%%%%%%%%%%%%%%%%%%%%%%%%%%%
%%%%%%%%%%%%%%%%%%%%%%%%%%%%

\subsection{The worldvolume theory of giants}
\label{ssec:giant}

In the last section, following \cite{Caron-Huot:2023wdh}, we saw how to reduce the correlator of half-BPS (bosonic) determinant operators in $\cN=4$ sdYM to a 1-matrix integral over a matrix $\rho_{ij}$, where $i,j=1,\dots,k$. The diagonal entries of $\rho$ are fixed to the values $\rho_{ii}=1$. On-shell, the off-diagonal entries of $\rho$ obey the equations of motion obtained from varying the action \eqref{rhoac},
\be\label{rhoeom}
x_{ij}^2\rho_{ij} - Ny_{ij}^2\rho^{-1}_{ij} = 0\,.
\ee
Solutions of these equations can be used to evaluate the matrix integral \eqref{rhoint} in a saddle point approximation.

\begin{figure}[t]
    \centering
    \includegraphics[width=1\textwidth]{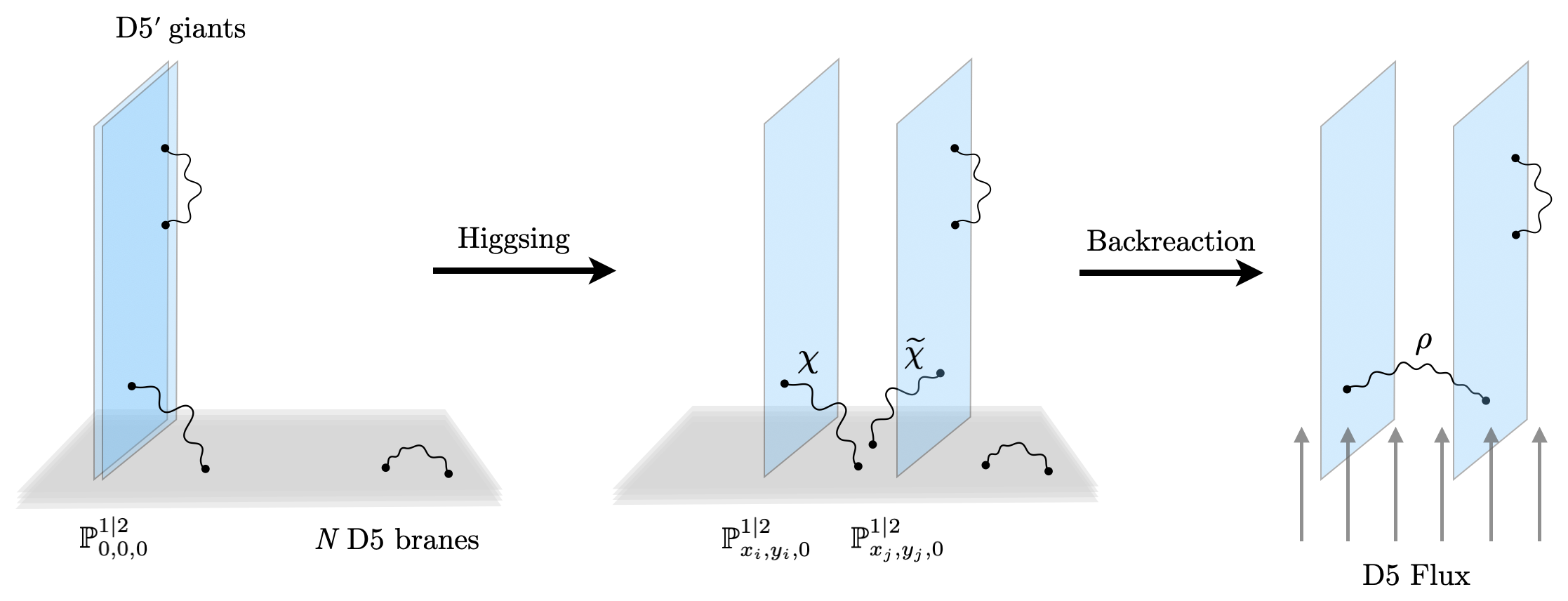}
    \caption{The D$5'$ giant gravitons start life as coincident branes. They are sent to their desired locations in the target space via Higgsing their worldvolume theory. We have displayed the 5-$5'$ and $5'$-5 strings $\chi_{ir},\til\chi_i^r$ stretching from the $N$ D5 branes to the D$5'$ giants. Post backreaction, they get replaced by the fermion bilinears $\rho_{ij}=\chi_{ir}\til\chi_j^r$ introduced during the Hubbard-Stratonovich transform. The $\rho_{ij}$ will be identified with nontrivial states of strings stretching across non-coincident giants, signaling recombination of the 5-$5'$ and $5'$-5 strings into $5'$-$5'$ strings in the presence of flux.}
    \label{fig:recombination}
\end{figure}

We propose that each saddle of the $\rho$ matrix model maps to a nontrivial configuration of giant gravitons on the gravitational side of our duality. This identification proceeds in the following steps.
\begin{itemize}
    \item We begin by constructing the theory of $k$ coincident D5$'$ giants that intersect the original stack of $N$ D5 branes in the twistor superline of the origin $x=y=\zeta=0$.
    \item  We separate these giants by Higgsing their worldvolume theory, thereby sending them to their respective locations in the fibers above the lines $\P^{1|2}_i\equiv\P^{1|2}_{x_i,y_i,0}$.
    \item Integrating out the open strings stretching between the $N$ D5 branes and the $k$ $\D5'$ giants adds an effective interaction term to the worldvolume theory of the giants. This captures the backreaction of the $N$ D5s on the giants.
    \item Solutions of the equations of motion of $\rho$ map to solutions of the equations of motion of the worldvolume theory of the giants coupled to this backreaction.
\end{itemize}
This recipe is depicted in figure \ref{fig:recombination}. The key step here is the incorporation of backreaction, which is the main result of this section. This will add a tadpole to the open string field theory on the giants that obstructs the trivial vacuum of the giants from being a solution of its equations of motion. Instead, in the next section, we will implement the last step of this list to obtain new vacua that are in one-to-one correspondence with saddles of the $\rho$ matrix model. To aid the reader, we describe the same procedure in the simpler case of giant gravitons in $\SL_2(\C)$ in appendix \ref{app:giants}.

\paragraph{Giants before backreaction.} Let us begin by setting up the worldvolume theory of our $k$ D$5'$ giants in the absence of the $N$ D5 branes wrapping supertwistor space. We first write the theory of $k$ coincident giants, and subsequently Higgs this theory to move the giants to their desired locations.

All our giants will be D$5'$ branes wrapping loci of the form \eqref{D5'}. The first step is to place all of them along the same locus
\be
\mu^\dal = \nu^{\dot a} = \veps^\al = \eta^a = 0\,.
\ee
Bosonically, these $k$ coincident branes wrap the superline $\P^1_{x=0}$ with coordinates $\lambda_\al$, and the two fibers of $\CO(-1)^{\oplus 4}\to\P^1_{x=0}$ that carry coordinates $\tau_a$. In the fermionic directions, they wrap the $\veps_\dal,\eta_{\dot a}$ dimensions. So the brane worldvolume carries the super-coordinates
\be
\bs\lambda_{\bs\alpha} \equiv (\lambda_\al|\eta_{\dot a})\,,\qquad\bs\tau_{\bs a} \equiv (\tau_a|\veps_\dal)\,.
\ee
The former are coordinates on $\P^{1|2}$, while the latter can be thought of as coordinates along the fibers of $\CO(-1)^{\oplus2}\oplus\Pi\CO(-1)^{\oplus2}\to\P^{1|2}$. We work with the natural measures $\D^{1|2}\bs\lambda = \D\lambda\,\d^2\eta$ and $\d^{2|2}\bs\tau = \d^2\tau\,\d^2\veps$ on the base and fibers of this supergeometry. The total measure $\D^{1|2}\bs\lambda\wedge\d^{2|2}\bs\tau$ is weightless, so our giants wrap a super Calabi-Yau, just like the original branes that wrapped supertwistor space.

The resemblance continues, as the worldvolume theory on the giants is again a holomorphic Chern-Simons theory,
\be\label{5'5'ac}
S_{5'5'} = \int_{\D5'}\D^{1|2}\bs\lambda\,\d^{2|2}\bs\tau\;\tr\left(\frac12\,\rA\dbar\rA + \frac13\,\rA^3\right)\,.
\ee
The crucial differences here are that the gauge group is now $\GL_k(\C)$, the trace `tr' is over matrices valued in $\gl_k(\C)$, and the open string field is given by
\be
\rA\in\Pi\Omega^{0,*}(\D5',\gl_k(\C))\,.
\ee
As before, by this we mean that $\rA$ is a $(0,*)$-polyform in the bosonic coordinates on the D5$'$, with coefficients that depend on the super-coordinates $\bs\lambda,\bs\tau$. Moreover, $\rA$ is a $k\times k$ matrix that we will label as $\rA_{ij}$, where $i,j$ are the Chan-Paton indices labeling the giants. These are the same indices that $\rho_{ij}$ carries, which proves to be extremely suggestive.

\paragraph{Higgs mechanism.} Up till now, we have described the theory of $k$ coincident giants that intersect supertwistor space in the twistor superline of the origin. However, we originally wanted the theory of \emph{non-coincident giants}, with the $i^\text{th}$ giant placed along the locus
\be\label{ithgiant}
\begin{split}
    \mu^\dal &= x_i^{\dal\al}\lambda_\al\,,\qquad \hspace{0.65em}\nu^{\dot a} = y_i^{\dot{a} a}\tau_a\,,\\
    \veps^\al &= -x_i^{\dal\al}\veps_\dal \,,\qquad\eta^a=-y_i^{\dot{a} a}\eta_{\dot a}\,.
\end{split}
\ee
This intersects $\pt$ in $\P^{1|2}_i$. 

In the absence of the shift generated by $\zeta_i^{\al a}$, such a brane's placement in the fermionic directions is completely fixed by the embedding of its bosonic part due to the constraint $\Theta|_{\D5'}=\d(\cdots)$ discussed in \eqref{ThetaD5'}. The bosonic part is a copy of $\CO(-1)^{\oplus2}\to\P^1_{x_i}$ embedded fiberwise into $X = \CO(-1)^{\oplus4}\to\PT$. The fermionic directions $(\veps_\dal,\eta_{\dot a})$ are simply coordinates on the parity-reversed conormal bundle of this bosonic submanifold. Indeed, this conormal bundle is a submanifold of $\Pi T^*_X$. It is spanned by the four $(1,0)$-forms $\d\mu^\dal-x_i^{\al\dal}\d\lambda_\al$ and $\d\nu^\da-y_i^{a\da}\d\tau_a$. A general covector in their span is 
\be
\veps_\dal(\d\mu^\dal-x_i^{\dal\al}\d\lambda_\al) + \eta_\da(\d\nu^\da-y_i^{\da a}\d\tau_a)\,,
\ee
and comparing this with $\veps_A\d Z^A+\eta^I\d W_I$ tells us that $\veps^\al=-x_i^{\dal\al}\veps_\dal$, $\eta^a=-y_i^{\da a}\tau_a$ on the $i^\text{th}$ brane. So it suffices to focus on the placement of the giants in the bosonic directions.

The giants can be separated and sent to their respective homes \eqref{ithgiant} by giving vacuum expectation values to the Higgs fields inside $\rA$ representing normal fluctuations of the $\D5'$ branes. To pick out these Higgs fields, let's expand $\rA$ as a superfield in the Grassmann coordinates $\veps_\dal,\eta_{\dot a}$ on the $\D5'$ worldvolume,
\be\label{rAexp}
\rA = \sA + \sM^\dal\veps_\dal + \sN^\da\eta_\da + \sP\veps^2 + \sQ\eta^2 + \sR^{\dal\da}\veps_\dal\eta_{\da} + \til\sM^\dal\veps_\dal\eta^2 + \til\sN^{\da}\eta_{\da}\veps^2 + \sB\veps^2\eta^2\,,
\ee
where $\veps^2\equiv\veps_{\dot0}\veps_{\dot1}$ and $\eta^2\equiv\eta_{\dot0}\eta_{\dot1}$. 

Each coefficient in this expansion is a $(0,*)$-polyform in the BV formalism. Denote by
\be
L = \CO(-1)^{\oplus2}\to\P^1_{x=0}
\ee
the total space of the bosonic vector bundle wrapped by our $\D5'$ branes. Then the coefficient fields are valued in the following bundles:
\begingroup
\allowdisplaybreaks
\begin{align}
    \sA&\in\Pi\Omega^{0,*}(L,\gl_k(\C))\nonumber\\
    \sM^\dal&\in\Omega^{0,*}(L,\CO(1)\otimes\gl_k(\C))\nonumber\\
    \sN^\da&\in\Omega^{0,*}(L,\CO(-1)\otimes\gl_k(\C))\nonumber\\
    \sP&\in\Pi\Omega^{0,*}(L,\CO(2)\otimes\gl_k(\C))\nonumber\\
    \sQ&\in\Pi\Omega^{0,*}(L,\CO(-2)\otimes\gl_k(\C))\label{wts}\\
    \sR^{\dal\da}&\in\Pi\Omega^{0,*}(L,\CO\otimes\gl_k(\C))\nonumber\\
    \til\sM^\dal&\in\Omega^{0,*}(L,\CO(-1)\otimes\gl_k(\C))\nonumber\\
    \til\sN^\da&\in\Omega^{0,*}(L,\CO(1)\otimes\gl_k(\C))\nonumber\\
    \sB&\in\Pi\Omega^{0,*}(L,\CO\otimes\gl_k(\C))\nonumber
\end{align}
\endgroup
where $\CO(n)$ denotes the pullback of $\CO(n)\to\P^1$ to the total space of $L\to\P^1$. It is worth being explicit about their weights because over half of these fields will play a nontrivial role in our story. We also collect their equations of motion for later use,
\begin{subequations}
\begingroup
\allowdisplaybreaks
\begin{align}
    &\dbar\sA+\sA^2 = 0\label{sAeq}\\
    &\dbar_\sA\sM^\dal = \dbar_\sA\sN^\da = 0\,,\label{MNeq}\\
    &\dbar_\sA\sP + \sM_\dal\sM^\dal = 0\,,\label{Peq}\\
    &\dbar_\sA\sQ + \sN_\da\sN^\da = 0\,,\label{Qeq}\\ &\dbar_\sA\sR^{\dal\da} + [\sM^\dal,\sN^\da] = 0\,,\label{Req}\\
    &\dbar_\sA\til\sM^\dal + [\sQ,\sM^\dal] + [\sR^{\dal\da},\sN_\da] = 0\,,\label{Mteq}\\
    &\dbar_\sA\til\sN^\da + [\sP,\sN^\da] - [\sR^{\dal\da},\sM_\dal] = 0\,,\label{Nteq}\\
    &\dbar_\sA\sB + [\sM_\dal,\til\sM^\dal] + [\sN_\da,\til\sN^\da] + [\sP,\sQ] - \sR_{\dal\da}\sR^{\dal\da} = 0\,,\label{sBeq}
\end{align}
\endgroup
\end{subequations}
where $\dbar_\sA\sM^\dal\equiv\dbar\sM^\dal + [\sA,\sM^\dal]$, etc. These follow from plugging the Grassmann expansion \eqref{rAexp} into the equation of motion $\dbar\rA+\rA^2=0$ obtained from varying the action \eqref{5'5'ac}.

The Higgs fields are the $0$-form parts of the coefficients $\sM^\dal,\sN^\da$ of the terms linear in the Grassmann coordinates. Indeed, by the correspondence \eqref{fermitovec}, the $0$-form parts of
\be
\sM^\dal\veps_\dal\leftrightarrow\sM^\dal\frac{\p}{\p\mu^\dal}\,,\qquad\sN^\da\eta_\da\leftrightarrow\sN^\da\frac{\p}{\p\nu^\da}
\ee
can be interpreted as a pair of vector fields normal to the branes. At the same time, they are $k\times k$ bosonic matrices $\sM^\dal_{ij},\sN^\da_{ij}$. Giving a nontrivial vev to their diagonal entries allows us to translate our branes to their desired locations.

Our branes start life at the ``origin'' $\mu^\dal=\nu^\da=0$, and we want to send the $i^\text{th}$ brane to $\mu^\dal = x_i^{\dal\al}\lambda_\al$, $\nu^\da = y_i^{\da a}\tau_a$. Since shifts in $\mu^\dal$ and $\nu^\da$ are ordinary affine translations, we can accomplish this by taking the vev for $\sM^\dal_{ii}$ to be $x_i^{\dal\al}\lambda_\al$, and that for $\sN^\da_{ii}$ to be $y_i^{\da a}\tau_a$. These vevs are allowed as they possess the requisite homogeneities. Collectively, we can package them into a pair of matrix-valued vevs
\be\label{Mvev}
\begin{split}
    \la\sM^\dal\ra &= \bX^{\dal\al}\lambda_\al\,,\\
    \la\sN^\da\ra &= \bY^{\da a}\tau_a\,,
\end{split}
\ee
where we have defined a pair of diagonal matrices containing the data of the $x_i,y_i$,
\be\label{XY}
\begin{split}
    \bX^{\dal\al}_{ij} &= x_i^{\dal\al}\delta_{ij}\,,\\
    \bY^{\da a}_{ij} &= y_i^{\da a}\delta_{ij}\,.
\end{split}
\ee
The introduction of such matrices is motivated by the use of similar diagonal matrices $\zeta,\mu$ in \cite{Budzik:2021fyh} to describe stacks of giants in twisted holography on $\SL_2(\C)$.

Let us check that these vevs move our coincident branes to the required loci in the supermanifold $\Pi T^*_X$. The total vev for the diagonal string $\rA_{ii}$ is
\be\label{diagvev}
\la\rA_{ii}\ra = x_i^{\dal\al}\lambda_\al\veps_\dal + y_i^{\da a}\tau_a\eta_\da\,.
\ee
The shift of the $i^\text{th}$ brane away from the origin $\mu^\dal=\nu^\da=\veps^\al=\eta^a=0$ is computed by the Schouten brackets of this vev with the super-coordinates normal to the brane,
\be
\begin{split}
    \delta\mu^\dal &= [\la\rA_{ii}\ra,\mu^\dal] = \frac{\p}{\p\veps_\dal}\la\rA_{ii}\ra = x_i^{\dal\al}\lambda_\al\,,\\
    \delta\nu^\da &= [\la\rA_{ii}\ra,\nu^\da] = \frac{\p}{\p\eta_\da}\la\rA_{ii}\ra = y_i^{\da a}\tau_a\,,\\
    \delta\veps^\al &= [\la\rA_{ii}\ra,\veps^\al] = -\frac{\p}{\p\lambda_\al}\la\rA_{ii}\ra = -x_i^{\dal\al}\veps_\dal\,,\\
    \delta\eta^a &= [\la\rA_{ii}\ra,\eta^a] = -\frac{\p}{\p\tau_a}\la\rA_{ii}\ra = -y_i^{\da a}\eta_\da\,.
\end{split}
\ee
This is precisely the desired brane locus \eqref{ithgiant}. It isn't hard to generalize the vev $\la\cA_{ii}\ra$ to incorporate nonzero super-coordinates $\zeta^{\al a}_i$, and we briefly remark on this in section \ref{sec:outlook}.

In this sense, the operations $[\la\rA_{ii}\ra,-]$ provide vector fields generating translations of the branes away from a chosen origin. More precisely, the Schouten bracket is a shifted holomorphic Poisson bracket derived from the symplectic potential $\Theta = \veps_A\d Z^A+\eta^I\d W_I$, and $\la\rA_{ii}\ra$ are the associated Hamiltonians generating translations of the branes. The fact that the branes satisfy $\d\Theta|_\text{brane}=0$ ensures that these translations are symplectic.

Without loss of generality, we assume that no two points $(x_i,y_i)$ coincide. If so, then our Higgsing breaks the $\GL_k(\C)$ gauge symmetry down to its maximal torus $\GL_1(\C)^k$. Furthermore, it also drastically cuts down the spectrum of open strings on the giants. 

In general, in the B-model, strings that stretch from one brane to another only carry nontrivial states\footnote{A nontrivial state of a B-model string is necessarily massless.} if the branes intersect \cite{Donagi:2003hh}. Such states are always localized to the brane intersections. In our case, since the giants never intersect after being sent to their final positions, the only open strings that can carry any states are the diagonal strings $\rA_{ii}$ that stretch from a giant to itself. The off-diagonal strings $\rA_{ij}$, $i\neq j$, become massive or get removed in unitary gauge. Once the massive strings are integrated out, one recovers an action which is simply a copy of abelian holomorphic Chern-Simons theory for every giant.

However, this naive conclusion only holds in the absence of the backreaction of the original $N$ D5 branes. This backreaction deforms the theory \eqref{5'5'ac} by the addition of extra potential terms. This potential energy can balance the tension of strings stretching across different giants, allowing them to carry nontrivial (\emph{i.e.}, massless) states even when the giants do not intersect! In the next section, we will explicitly show that it is precisely the states of such off-diagonal strings that get identified with $\rho_{ij}$.

\paragraph{Giants after backreaction.} The third and most important step in our bulk-boundary dictionary is the incorporation of backreaction. A simple example of this for giants in $\C^3$ is reviewed in the appendix. In that setup, one starts with the B-model on $\C^3$, places a large number of D1 branes along a copy of $\C\subset\C^3$, and backreacts by them. The backreaction gives rise to a complex structure deformation that turns $\C^3$ into $\SL_2(\C)\sim\AdS_3\times S^3$. Giant gravitons in this case arise as D$1'$ branes whose shapes are governed by holomorphic curves in the deformed complex structure. This is the mechanism through which they couple to the backreaction.

Our story is somewhat different in a very subtle way. Up till now, we studied the theory of $k$ $\D5'$ giants in the absence of the $N$ D5 branes wrapping $\PT$. When these D5 branes are reinstated and their backreaction taken into account, one might expect the giants to recoil in response to the  backreaction $\xi_N$ that we obtained in \eqref{xiN}. The interactions between $\xi_N$ and operators $\cO_n'$ of the gauge theory on the giants are given by open-closed couplings of the form
\be
\int_{\D5'}\D^{1|2}\bs\lambda\,\d^{2|2}\bs\tau\;\p^n\xi_N\,\cO_n'
\ee
where $\p^n$ are some set of bosonic or fermionic derivatives normal to the giants. Since $\xi_N$ is a $(0,3)$-form, the operators $\cO_n'$ must be $0$-forms. However, $\xi_N$ points purely along $\D^3\br W$ in the antiholomorphic-form directions. As a result, the pullback of $\xi_N$ to the loci \eqref{ithgiant} that the giants wrap vanishes. So all such couplings appear to vanish. Open-closed couplings higher-order in $\xi_N$ vanish for the same reasons.

One might also suspect that the factor of $\|W\|^8$ in the denominator of $\xi_N$ obstructs this vanishing as one gets too close to $W_I=0$, leading to a delta function behavior instead of a strict zero. This is indeed what happens in the case of D$1'$ giants in $\C^3$, as we prove in the appendix. Similarly, this would also happen if our $\D5'$ giants intersected the D5 branes in a point instead of a $\P^1$, which may have been a more typical setup. However, in our case, this does not help us as $\xi_N$ does not carry any antiholomorphic-form dependence pointing along $\D\hat\lambda\equiv\hat\lambda^\al\d\hat\lambda_\al$. And integration over the $\D5'$s necessarily requires the $(0,3)$-form dependence to assemble into $\D\hat\lambda\,\d^2\br\tau$.

Nonetheless, in all of these cases, there is a morally equivalent, uniform approach for obtaining the theory on the giants in the presence of the backreaction: open-closed duality. Just as closed strings in the bulk feel the presence of the $N$ D5 branes through their coupling to 5-5 strings, open strings on the $\D5'$ giants feel the D5 branes through their coupling to the 5-$5'$ and $5'$-5 strings. Integrating out the 5-5 strings deforms the closed string BCOV theory by the backreaction $\xi_N$. Similarly, integrating out the 5-$5'$ and $5'$-5 strings will deform the open string theory on the giants. In the absence of a better understanding of the coupling of $\xi_N$ to our giants, we will take this as our proxy for the backreaction of the D5 branes on the $\D5'$ giants. It is useful to know that this alternative approach indeed reproduces the correct backreaction in the case of D$1'$ giants in $\SL_2(\C)$, as checked in the appendix.

The basic idea is very simple and has partially appeared in a similar setting in \cite{Bittleston:2024efo} where it was applied to study flavor instanton backgrounds in celestial holography. Suppose we have integrated out the 5-5 strings and replaced them with their backreaction. By our arguments above, this does not appear to affect the giants to leading order in $1/N$, so let us forget about it. This leaves us with the worldvolume theory of the $k$ giants -- which we take to be coincident to start with -- coupled to 5-$5'$ and $5'$-5 strings governed by the action
\be\label{5'5'ac1}
S_{55'} = \int_{\P^{1|2}}\D^{1|2}\bs\lambda\,\bigg(\sum_i\til\chi_i^r\dbar_*\chi_{ir} + \sum_{i,j}\til\chi_i^r\rA_{ij}\chi_{jr}\bigg)\,.
\ee
Along the lines of \eqref{dbars}, we have shifted $\rA$ by the additional vev $\la\rA_{ij}\ra = \delta_{ij}\bar\delta^{1|2}(\bs\lambda,\bs\lambda_*)$ so as to obtain an invertible Dolbeault operator $\dbar_*\equiv\dbar + \bar\delta^{1|2}(\bs\lambda,\bs\lambda_*)$. This action no longer contains the 5-5 strings $\cA$ that we integrated out, and the $\GL_N(\C)$ gauge symmetry acting on the $r$ indices has been relegated to the role of a flavor symmetry.

The 2-point function of $\chi_i$ with $\til\chi_j$ is obtained by dressing the result quoted in \eqref{chi2} with $\delta_{ij}$. The R-invariant occurring in it vanishes when its arguments coincide. Using this fact, we observe that the $\chi\til\chi$ fermion bilinear acquires the vev
\be
\la\chi_{ir}\til\chi_j^r\ra = \delta_r^r\delta_{ij} = N\delta_{ij}
\ee
in this vacuum. As a result, the cubic interaction in \eqref{5'5'ac1} generates a tadpole for $\rA$,
\be\label{1Nterm}
N\int_{\P^{1|2}}\D^{1|2}\bs\lambda\;\tr\,\rA\,.
\ee
The equation of motion of $\rA$ obtained from the combined action $S_{5'5'}+S_{55'}$ sees this tadpole as a source term which cannot be ignored in the large $N$ limit. The trivial vacuum $\rA=0$ is no longer an allowed state of the giants, as it fails to solve the equations of motion. Instead, this tadpole sources a nontrivial background for $\rA$, thereby deforming the ``shape'' of the D5$'$ branes in a generalized sense.

Replacing $\chi\til\chi$ by its vev is a linearization of the operation of integrating out the 5-$5'$ and $5'$-5 strings. To obtain the full effective action of $\rA$, let us integrate out $\chi,\til\chi$ nonlinearly. This is possible because their path integral is Gaussian. Just as integrating out these fields inserted a set of determinant operators in the path integral of 5-5 strings, integrating them out here inserts a determinant operator in the path integral of $5'$-$5'$ strings,
\be
\big(\det(\dbar_\rA|_{\P^{1|2}})\big)^N\,,
\ee
where $\dbar_\rA = \dbar+\rA$ and the vev $\la\rA\ra = \bar\delta^{1|2}(\bs\lambda,\bs\lambda_*)$ is now understood as being included within $\rA$. This contributes a potential energy term to the effective action of $\rA$,
\be\label{D5'ac}
S_{5'5'}^\text{eff}[\rA] = S_{5'5'}[\rA] - N\log\det(\dbar_\rA|_{\P^{1|2}})\,,
\ee
where $S_{5'5'}$ was the holomorphic Chern-Simons action given in \eqref{5'5'ac}. The tadpole \eqref{1Nterm} is just the leading order term in the expansion of this potential.

At this stage, we can Higgs the giants by turning on the vevs \eqref{Mvev}. The resulting $S_{5'5'}^\text{eff}$ will be our ansatz for the action of the giants in the presence of the backreaction of $N$ D5 branes wrapping $\pt$. It would be very interesting to see if this can be reproduced from the flux $\xi_N$ in a more direct way, but we leave this possibility to the future.

%%%%%%%%%%%%%%%%%%%%%%%%%%%%
%%%%%%%%%%%%%%%%%%%%%%%%%%%%

\subsection{The map from saddles to strings} 
\label{ssec:map}

This brings us to the final step in the procedure outlined at the start of the previous section. We have two theories that we claim are dual to each other at leading order in the large $N$ limit. On the boundary side of our duality lives the $\rho$ matrix model with action \eqref{rhoac}. In the bulk, we have obtained the worldvolume theory \eqref{D5'ac} of giant gravitons coupled to the backreaction. In this section, we show how our duality relates the solutions of the equations of motion of the two theories.

We begin by reexpressing the saddle equations of the $\rho$ matrix model in matrix notation. Using the diagonal matrices introduced in \eqref{XY}, we observe that
\be
\begin{split}
    [\bX^{\dal\al},[\bX_{\dal\al},\rho]]_{ij} &= x_{ij}^{\dal\al}\,[\bX_{\dal\al},\rho]_{ij} = x_{ij}^2\rho_{ij}\,,\\
    [\bY^{\da a},[\bY_{\da a},\rho^{-1}]]_{ij} &= y_{ij}^{\da a}\,[\bY_{\da a},\rho^{-1}]_{ij} = y_{ij}^2\rho_{ij}^{-1}\,.
\end{split}
\ee
Therefore, the saddle equations \eqref{rhoeom} can be framed as the matrix commutator equation
\be\label{sadeq}
[\bX^{\dal\al},[\bX_{\dal\al},\rho]] = N[\bY^{\da a},[\bY_{\da a},\rho^{-1}]]\,.
\ee
This rewriting will help us connect with the equations obeyed by the fields on the D$5'$ worldvolume. A similar rewriting was the basis of the spectral curve approach to giant gravitons in $\SL_2(\C)$ pioneered in \cite{Budzik:2021fyh}.

Let us compare this with the equations of motion of the open strings on our giants. The variation of \eqref{D5'ac} results in the equation
\be\label{rAeq}
\dbar\rA + \rA^2 = N\mathscr{W}\bar\delta^{2|2}(\bs\tau)\,.
\ee
The quantity $\mathscr{W}(\bs\lambda)$ on the right is a Wilson loop that arises in the variation of the functional determinant
\be
\delta\log\det(\dbar_\rA|_{\P^{1|2}}) = \int_{\P^{1|2}}\D^{1|2}\bs\lambda\;\tr\,\delta\rA\,\mathscr{W}\,.
\ee
% The source is described by a Wilson loop $\mathscr{W}$ formed out of $\rA$,
% \begin{multline}
%     \mathscr{W}(\bs\lambda) = \sum_{n=0}^\infty(-1)^n\int_{(\P^{1|2})^n}\prod_{m=1}^n\D^{1|2}\bs\lambda_m\;\rA_1\rA_2\cdots\rA_n\\
%     \times \Delta(\bs\lambda,\bs\lambda_1)\Delta(\bs\lambda_1,\bs\lambda_2)\cdots\Delta(\bs\lambda_{n-1},\bs\lambda_n)\Delta(\bs\lambda_n,\bs\lambda)
% \end{multline}
% To construct it, let $\mathrm{W}(\bs\lambda,\bs\lambda')$ be a holomorphic Wilson line obeying the equation
% \be
% \dbar\mathrm{W} + \rA\mathrm{W} = 0\,.
% \ee
% Then $\mathscr{W}$ is defined by the coincident limit
% \be
% \mathscr{W}(\bs\lambda) = \lim_{\bs\lambda'\to\bs\lambda}\mathrm{W}(\bs\lambda,\bs\lambda')\,.
% \ee
After shifting $\dbar\mapsto\dbar_*$ by giving the requisite vev to $\rA$, we can obtain an explicit expression for $\mathscr{W}$ by varying the series expansion of the determinant. But this tends to be hard to work with in general. For our purposes, it will suffice to simplify the equation of motion using a judicious ansatz for $\rA$.

As explained previously, we are implementing the separation of giants indirectly through Higgsing. So by the superline $\P^{1|2}$ here, we still mean the superline of the origin in supertwistor space. On this superline, one sets $\bs\tau_{\bs a} = (\tau_a|\veps_\dal) = 0$. The pullback of the Grassmann expansion of $\rA$ given in \eqref{rAexp} to $\P^{1|2}$ is found to be
\be
\rA|_{\P^{1|2}} = \sA + \sN^\da\eta_\da + \sQ\eta^2\,,
\ee
with the right-hand side being implicitly restricted to $\tau_a=0$. This tells us that the only fields that enter $\log\det(\dbar_\rA|_{\P^{1|2}})$ are the $(0,1)$-parts of $\sA,\sN^\da,\sQ$.

The $(0,1)$-part of $\sA$ is a $\GL_k(\C)$ partial connection. It makes intuitive sense to look for vacua in which this gauge field has not been excited. By the parity assignments made in \eqref{wts}, the $(0,1)$-part of $\sN^\da$ is seen to be a fermion.\footnote{Keep in mind that the generators of differential forms like $\d\hat\lambda_\al$, etc.\ are treated as Grassmann odd objects in topological-holomorphic theories such as ours.} A state with a nontrivial profile for $\sN^\da$ might well be a necessary requirement for describing giant graviton configurations dual to correlators of supersymmetrized determinants. But for our purposes, we will not need to excite these fermions either. 

So we will make the ansatz that the $(0,1)$-parts of $\sA,\sN^\da$ vanish. Then the only field that contributes to the determinant is $\sQ$. At this stage, define a constant $k\times k$ matrix $\rho$ as the sphere zero mode of the $(0,1)$-part of $\sQ$,
\be\label{rhodef}
\rho \vcentcolon= \int_{\P^1}\D\lambda\;\sQ\bigr|_{\tau_a=0}\,.
\ee
This is not a gauge-invariant definition. It only makes sense here because we are looking for solutions in which the gauge field is zero. If this were nonzero, we would have to choose a frame for the gauge bundle on the giants that is covariantly constant (if possible) over $\P^{1|2}$, and evaluate the components $\sQ_{ij}$ of $\sQ$ in that frame.

Next recall that, in total, we would like to give $\rA$ the vev
\be
\la\rA\ra = \bX^{\dal\al}\lambda_\al\veps_\dal + \bY^{\da a}\tau_a\eta_\da + \bar\delta^{1|2}(\bs\lambda,\bs\lambda_*)\mathbbm{1}_{k}\,,
\ee
where $\mathbbm{1}_{k}$ is the $k\times k$ identity matrix. The first two terms here signify the vevs discussed in \eqref{Mvev} that we gave to the Higgs fields, \emph{i.e.}, the $0$-form parts of $\sM^\dal,\sN^\da$. The third term can be viewed as a vev for $\sQ$ if we work in a gauge in which the reference point $\bs\lambda_*\in\P^{1|2}$ only carries bosonic components: $\bs\lambda_*\equiv(\lambda_*|0)$. Since this is the choice we used to obtain the $\rho$ matrix model, it is the choice that we will continue to employ in this section. For this choice, the delta function $\bar\delta^{1|2}(\bs\lambda,\bs\lambda_*)$ is a product of the weight $-2$ bosonic delta function $\bar\delta_{-2}(\la\lambda\lambda_*\ra)$ and the fermionic dependence $\eta^2$. Therefore, the third term can be interpreted as a vev
\be
\la\sQ_{ij}\ra = \bar\delta_{-2}(\la\lambda\lambda_*\ra)\,\delta_{ij}
\ee
for the diagonal entries of $\sQ$. 

Plugging this vev into \eqref{rhodef} tells us that the diagonal entries of the zero mode $\rho$ carry the vevs $\la\rho_{ij}\ra=\delta_{ij}$. We would like to identify this giant graviton zero mode $\rho$ with the matrix $\rho$ that governs the determinant correlators. Since the diagonal entries of the latter are fixed to be $1$ (in our conventions), we will choose an ansatz in which the diagonal fluctuations of the giant graviton mode $\rho$ around this vev vanish as well. All the dynamical data will then be contained in its off-diagonal entries $\rho_{ij}$, $i\neq j$.

With these motivations, the simplest choice of ansatz to seed the equations of motion would be the profiles
\be
\begin{split}
    \sA &= 0\,,\\
    \sM^\dal &= \bX^{\dal\al}\lambda_\al\,,\\
    \sN^\da &= \bY^{\da a}\tau_a\,,\\
    \sQ &= \rho\,\bar\delta_{-2}(\la\lambda\lambda_*\ra)\,,
\end{split}
\ee
with $\rho_{ii}=1$ held fixed. In this ansatz, we have dropped fluctuations of $\sM^\dal,\sN^\da$. Remarkably, this will suffice for our applications and allow us to conclude that the bosonic geometry of the giants remains undeformed by the backreaction! 

We are also dropping higher harmonics of $\sQ$ on $\P^1$, as well as any $\bs\tau$-dependence. Notwithstanding these extreme simplifying assumptions, this clean ansatz allows us to identify the matrix model degree of freedom $\rho_{ij}$ with the open string that stretches from the $i^\text{th}$ giant to the $j^\text{th}$ giant. This bestows a precise bulk interpretation upon these matrix variables, concretely realizing some of the aspirations of \cite{Jiang:2019xdz}. In particular, when $N=0$, the only solution of the saddle equations \eqref{rhoeom} is $\rho_{ij}=0$, $i\neq j$. This is consistent with the fact that in the absence of backreaction, strings between non-intersecting giants cannot carry any states.

Our task now is to solve the equations of motion of $\rA$ to determine the remaining fields in the multiplet \eqref{rAexp}. The equations of motion before backreaction were given in equations \eqref{sAeq} -- \eqref{sBeq}. For our ansatz, a by now standard calculation shows that the determinant generated by the backreaction reduces to
\be
\log\det(\dbar_\rA|_{\P^{1|2}}) = \log\det\rho\,.
\ee
Since $\rho$ was a mode of $\sQ$, we find that the effective interaction $N\log\det\rho$ only deforms the equation of motion \eqref{Peq} that is obtained from varying the worldvolume action with respect to $\sQ$. The deformed equation reads
\be
\dbar_\sA\sP + \sM_\dal\sM^\dal = -N\rho^{-1}\bar\delta^2(\tau)\,.
\ee
Setting $\sA=0$ and $\sM^\dal=\bX^{\dal\al}\lambda_\al$ reduces the left-hand side to $\dbar\sP$. Hence, this equation can be solved using the Bochner-Martinelli kernel on $\C^2$,
\be
\sP = -N\rho^{-1}\frac{\D\hat\tau}{\|\tau\|^4}\,,
\ee
where we set $\hat\tau_a \equiv \eps_{ab}\bar\tau^b$.

The remaining equations of motion are undeformed. Equations \eqref{sAeq}, \eqref{MNeq} and \eqref{Qeq} are trivially solved by our ansatz, and \eqref{Req} is thereafter solved by
\be
\sR^{\dal\da} = 0\,.
\ee
The equations \eqref{Mteq} and \eqref{Nteq} governing $\til\sM^\dal$, $\til\sN^\da$ are more nontrivial. For our ansatz and the value of $\sP$ obtained above, they simplify to
\be
\begin{split}
    \dbar\til\sM^\dal &= [\sM^\dal,\sQ] = [\bX^{\dal\al}\lambda_\al,\rho]\,\bar\delta_{-2}(\la\lambda\lambda_*\ra)\,,\\
    \dbar\til\sN^\da &= [\sN^\da,\sP] = -N[\bY^{\da a}\hat\tau_a,\rho^{-1}]\,\frac{\D\hat\tau}{\|\tau\|^4}\,.
\end{split}
\ee
In turn, these are satisfied by the profiles
\be
\til\sM^\dal = \frac{[\bX^{\dal\al}\lambda_{*\al},\rho]}{\la\lambda\lambda_*\ra}\,,\qquad
\til\sN^\da = -\frac{N[\bY^{\da a}\hat\tau_a,\rho^{-1}]}{\|\tau\|^2}\,.
\ee
Lastly, we can show that the equation \eqref{sBeq} governing $\sB$ reduces to $\dbar\sB=0$:
\begin{align}
    0&\overset{!}{=} \dbar_\sA\sB + [\sM_\dal,\til\sM^\dal] + [\sN_\da,\til\sN^\da] + [\sP,\sQ] - \sR_{\dal\da}\sR^{\dal\da}\nonumber\\
    &= \dbar\sB - \frac{\lambda_\al\lambda_{*\beta}}{\la\lambda\lambda_*\ra}\,[\bX^{\dal\al},[\bX_{\dal}^{\ \beta},\rho]] + \frac{N\tau_a\hat\tau_b}{\|\tau\|^2}\,[\bY^{\da a},[\bY_{\da}^{\ b},\rho^{-1}]] - N[\rho^{-1},\rho]\,\frac{\D\hat\tau}{\|\tau\|^4}\,\bar\delta_{-2}(\la\lambda\lambda_*\ra)\nonumber\\
    &= \dbar\sB + [\bX^{\dal\al},[\bX_{\dal\al},\rho]]-N[\bY^{\da a},[\bY_{\da a},\rho^{-1}]]\nonumber\\
    &= \dbar\sB\,.
\end{align}
In getting from the third to the fourth line, we finally invoked the saddle equations \eqref{sadeq} of the $\rho$ matrix model. The simplest solution of $\dbar\sB=0$ is clearly
\be
\sB = 0\,.
\ee
With this, we have obtained a solution of the backreaction-coupled open string field equations on the giants.

To summarize, solutions of the saddle equations \eqref{sadeq} translate into the following open string configurations on the giants:
\begin{multline}\label{rhotoA}
\rA = \bX^{\dal\al}\lambda_\al\veps_\dal + \bY^{\da a}\tau_a\eta_\da + \rho\,\eta^2\,\bar\delta_{-2}(\la\lambda\lambda_*\ra)\\
-N\rho^{-1}\veps^2\,\frac{\D\hat\tau}{\|\tau\|^4} + \frac{[\bX^{\dal\al}\lambda_{*\al},\rho]}{\la\lambda\lambda_*\ra}\,\veps_\dal\eta^2 -\frac{N[\bY^{\da a}\hat\tau_a,\rho^{-1}]}{\|\tau\|^2}\,\eta_\da\veps^2
\end{multline}
in the gauge where the reference point $\bs\lambda_*$ had zero fermionic components. In this way, we have provided a holographic map from saddles of the $\rho$ matrix model to open string vacua of the D$5'$ worldvolume theory. 

As mentioned before, the bosonic embedding of the giants remains undeformed. All the terms that contain $\rho$ are second- or higher-order polyvectors. This indicates a deformation of the locus wrapped by the giants in the supergeometry $\Pi T^*_X\oplus\Pi\br T_X$. It would be interesting to explore such super-deformations in more detail. This should be possible by using the gauge symmetries of holomorphic super Chern-Simons theory to diagonalize the entries of \eqref{rhotoA}. In the $\SL_2(\C)$ case of \cite{Budzik:2021fyh}, the giants frequently experience brane recombination, and it is such a diagonalization that gives rise to the spectral curves that they wrap in the presence of backreaction.

Although we lack a deeper geometric understanding of the open string vacua \eqref{rhotoA}, it is nevertheless remarkable that we can explicitly find giant gravitons in $\cN=4$ sdYM for determinant correlators at arbitrary multiplicity $k$. This is a dramatic improvement upon the situation in the non-self-dual theory, where giant graviton shapes dual to determinant correlators are only known at very low multiplicity, see e.g., \cite{Jiang:2019xdz,McGreevy:2000cw,Grisaru:2000zn,Hashimoto:2000zp,Balasubramanian:2001nh,Bak:2011yy,Bissi:2011dc}.

We leave the study of the path integral of the D$5'$ worldvolume theory in these vacua to future work. Plugging our solution \eqref{rhotoA} into the potential energy $N\log\det(\dbar_\rA|_{\P^{1|2}})$ was already seen to reproduce the interaction term $N\log\det\rho$ of the matrix model action \eqref{rhoac}. Similarly, we expect that the Gaussian term in this matrix model will also emerge upon systematically integrating out the remaining open string fluctuations on the giants.

%%%%%%%%%%%%%%%%%%%%%%%%%%%%
%%%%%%%%%%%%%%%%%%%%%%%%%%%%

\section{Outlook}
\label{sec:outlook}

We close by highlighting some salient aspects of our duality. We also discuss many open questions and interesting applications that remain to be explored.

\paragraph{Chiral holography.} In this work, we have proposed a gravitational holographic dual of the self-dual sector of $\cN=4$ sYM theory. The magic ingredient behind this construction was twistor string theory, which engineers $\cN=4$ sdYM as the worldvolume theory of B-model branes wrapping supertwistor space.

Along the way, we brought up many subtleties in the construction of twistor strings that had been poorly understood in the past. Our reformulation of twistor strings involved a B-model on the Calabi-Yau 7-fold $\CO(-1)^{\oplus4}\to\PT$. We wrapped a stack of $N$ D5 branes along the zero section of $\CO(-1)^{\oplus4}$ to engineer the $\GL_N(\C)$ holomorphic super Chern-Simons twistor action uplifting $\U(N)$ $\cN=4$ sdYM to twistor space. 

In the large $N$ limit, these D5 branes backreact to source a closed string flux $\xi_N$ that we obtained in \eqref{xiN}. We identified the bulk dual of $\cN=4$ sdYM as being the closed string B-model on $\CO(-1)^{\oplus4}$ minus its zero section, in the presence of $N$ units of this flux. The closed string theory was described in terms of a Bershadsky-Cecotti-Ooguri-Vafa (BCOV) theory, which is an adaptation of Kodaira-Spencer gravity \cite{Bershadsky:1993cx} to Calabi-Yau manifolds of higher dimensions \cite{Costello:2012cy}.

Our holographic bulk theory has the upshot of not suffering from any uncontrollable towers of massless higher-spin particles, unlike the anticipated dual to free sYM \cite{Gaberdiel:2021jrv}. Moreover, naively one would have worried that in any zero coupling limit of sYM, the bulk would have become highly stringy and no easy target space effective action description would exist. Contrary to this expectation, we found a relatively well-understood target space description by means of BCOV theory. This is perhaps one of the most surprising features of our duality and leads to our first open question.

\paragraph{The search for AdS$_5\times {\bf\emph{S}}^5$.} One of the most important problems our discussion leaves open is to understand the relation between the closed string B-model on $\CO(-1)^{\oplus4}$ and some form of string theory on AdS$_5\times S^5$. The regular free limit of sYM is expected to be dual to the tensionless limit of type IIB strings in AdS$_5\times S^5$. So one may have expected our chiral free limit to also give rise to a bulk dual that is a string theory with an AdS$_5\times S^5$ or related target, even if it cannot be described through a supergravity limit. This was the idea implemented by Gaberdiel and Gopakumar in \cite{Gaberdiel:2021qbb}, whose worldsheet dual to free sYM was a sigma model into a supersymmetrized twistor space associated to AdS$_5$.

The most obvious way to obtain AdS$_5\times S^5$ from our space is as follows. First, recall that $\PT$ fibres (non-holomorphically) over Euclidean $\R^4$. Specifically, a given twistor $Z^A = (\mu^\dal,\lambda_\al)\in\PT$ lies on the $\P^1$ fibre above the point
\be
    x^{\dal\al} = \frac{\mu^\dal\hat\lambda^\al - \hat\mu^\dal \lambda^\al}{\la\hat\lambda\lambda\ra}\in \R^4\,,
\ee
constructed with the help of Euclidean spinor conjugation. Equivalently, one can write this as 
\be
    x^\mu = \frac{ Z^A \sigma^\mu_{AB}\hat{Z}^B}{I(\hat{Z},Z)}
\ee
where $I_{AB}$ is the usual infinity twistor of flat $\R^4$ and $\sigma^\mu_{AB}$ are four $4\times4$ matrices with the usual Pauli matrices on their block off-diagonals. Similarly, using the SU(4)-invariant inner product on the fibres of $\CO(-1)^{\oplus4}\to\PT$ we define
\be
n^m = \frac{W_I(\Gamma^m)^I_{\ J}\br{W}^J}{\|W\|^2}
\ee
where  $\Gamma^m$ are the Dirac matrices for Weyl spinors on $\R^6$. The $n^m$ are six, weightless, real coordinates. As a consequence of Pauli-Fierz identities, they obey the constraint $\delta_{mn}n^m n^n=1$ and so define a point on $S^5$. Finally, the scale $\|W\|$ may be identified with the radial coordinate of AdS$_5$. More precisely, any such radial coordinate requires a choice of point either within AdS$_5$ and acting as the `origin' or else on the boundary. For example, the obvious scale-invariant combination representing the radial coordinate of the Poincar{\'e} patch is $r = \sqrt{\|W\|^2 \,I(\hat{Z},Z)}$. It would be very interesting to understand whether this identification helps reduce the B-model on $\CO(-1)^{\oplus4}$ to a theory on AdS$_5\times S^5$.

%The (mini)twistor space of AdS$_5$ is a quadric \cite{Adamo:2016rtr,Bailey:1998zif}. To give an explicit presentation of this quadric, one coordinatizes $\P^3\times\P^3$ by a pair $(Z^A,W_B)$ consisting of a twistor and a dual twistor. The relevant quadric is cut out by the equation $Z\cdot W\equiv Z^AW_A = 0$ . Now, even though our coordinates $(Z^A,W_I)$ on the base and fibers of $\CO(-1)^{\oplus4}\to\PT$ resemble the coordinates $(Z^A,W_B)$ on $\P^3\times\P^3$, we cannot identify the R-symmetry index $I$ on $W_I$ with a twistor index $A$ without breaking the conformal and R-symmetries of the theory down to a diagonal subgroup. And $(Z^A,W_I)$ do not satisfy any analog of $Z\cdot W=0$, except for the super-quadric constraint \eqref{euler} which looks like a potential relative of $Z\cdot W=0$. Due to these reasons, we have been unable to recast the target space of our closed twistor string theory in terms of the minitwistor space of $\AdS_5$.

There may be other potential relationships between our target space and $\AdS_5\times S^5$.\footnote{However, there does not seem to be any clear relationship between $\CO(-1)^{\oplus4}$ and the minitwistor space of AdS$_5$ (see \emph{e.g.} \cite{Adamo:2016rtr,Bailey:1998zif}). This is because the minitwistor space can be identified with the quadric $Z^A W_A=0$ inside $\P^3\times (\P^3)^*$. However, in our case there is no natural way to identify the R-symmetry index $I$ on $W_I$ with a twistor (conformal) index $A$.} 
One tantalizing hint seems to be implicit in the calculation of our backreacted target superspace. As discussed in section \ref{ssec:deformed}, the backreaction $\xi_N$ deforms the target superspace of the B-model into a nontrivial determinantal variety whose bosonic and fermionic coordinates naturally assemble into a parity-reversed element of $\psl(4|4,\C)$ as in equation \eqref{GG}. Exponentiating this gives an element of $\mathrm{PSL}(4|4,\C)$, and it is well-known that $\AdS_5\times S^5$ can be obtained as a coset space from a real form of this supergroup. For instance, this led to the construction of a Green-Schwarz action for strings in $\AdS_5\times S^5$ \cite{Metsaev:1998it}. 

A second possibility is that our deformed bosonic coordinates $G^A{}_I$ found in equation \eqref{gAI} may be related to the variables $Z^A{}_I$ utilized in the pure spinor formulation of strings in $\AdS_5\times S^5$ \cite{Berkovits:2007rj,Berkovits:2007zk}. This interpretation is bolstered by another recent attempt by Berkovits at finding the bulk dual of free sYM based on twistor-like variables \cite{Berkovits:2025xok}. Moreover, in lower dimensional twisted holographic setups such as \cite{Costello:2018zrm}, it is indeed the deformed bosonic coordinates that get interpreted as coordinates on $\SL_2(\C)\sim\AdS_3\times S^3$. We expect that our deformed supercoordinates will allow for a similar interpretation as coordinates on an appropriate superspace associated to $\AdS_5\times S^5$.

A third possibility is that our duality may secretly be equivalent to that of Gaberdiel and Gopakumar \cite{Gaberdiel:2021qbb}.\footnote{We thank Matthias Gaberdiel and Edward Mazenc for discussions on this point.} Since the latter duality also operates via a worldsheet theory with a twistorial target, it could plausibly be reinterpreted as a string dual to \emph{self-dual} instead of free sYM. If so, then it might be closely related to the worldsheet theory of our B-model (in the presence of the backreaction) through a worldsheet analog of graph duality. In the terminology of \cite{Gopakumar:2022djw,Gopakumar:2024jfq}, the proposal of \cite{Gaberdiel:2021qbb} may be thought of as being a V-type duality, and our proposal may be its F-type cousin. This interpretation is supported by the fact that while the holographic dictionary of \cite{Gaberdiel:2021qbb} operates at the level of closed string vertex operators inserted on the worldsheet, our dictionary is phrased in terms of D-branes and operates at the level of the faces (boundaries) of the worldsheet.

\paragraph{Extending the holographic dictionary.} In this work, we have only uncovered a small portion of a larger holographic dictionary that must exist between $\cN=4$ sdYM and closed twistor strings. Because local operators on $\R^4$ uplift to operators on $\PT$ that are nonlocal integrals over twistor lines, the standard holographic dictionary between local operators on $\PT$ and closed string states does not translate into a dictionary relating general gauge invariant local operators on $\R^4$ to closed string states. 

The exceptions to this were the conserved currents of global symmetries such as the $\cN=4$ superconformal symmetries. As summarized in table \ref{tab:dict}, the stress tensor and R-symmetry currents of $\cN=4$ sdYM were respectively found to be dual to bulk complex structure deformations longitudinal and transverse to $\PT$ inside $\CO(-1)^{\oplus4}\to\PT$, restricted to the brane locus $W_I=0$. These complex structure deformations were interpreted in \cite{Berkovits:2004jj} as 4d conformal supergravity modes, but we believe they are better seen as the leading terms in an expansion of bulk gravitational modes around the brane locus. It is certainly also possible to construct couplings between the open strings and arbitrary \emph{normal derivatives} ($\p/\p W_I)$ of the Beltrami differential in the bulk 7-fold. It would be very interesting to understand whether these relate to higher KK modes of some field in the supergravity multiplet on AdS$_5\times S^5$. 

For general local operators we had to work harder to find their bulk duals, even in the half-BPS case. Pleasingly, we found that all the half-BPS operators of $\cN=4$ sdYM, when restricted to chiral superspace, can be packaged into compact determinant operators that are dual to giant graviton D5$'$ branes. Such determinant operators were first proposed in \cite{Caron-Huot:2023wdh} as a helpful trick for computing a large class of half-BPS correlators in the self-dual theory in one go. We discovered in section \ref{sec:giant} that far from being a trick, this is precisely a mirror to the bulk dynamics of the splitting and joining of strings on D$5'$ giant gravitons in the presence of $N$ units of D5 flux. In this direction, it would be very interesting to explore the role of the emergent 10d symmetry of half-BPS correlators found in \cite{Caron-Huot:2021usw,Caron-Huot:2023wdh} as a symmetry of the worldvolume theory of our giants.

In summary, we have constructed a holographic dictionary for half-BPS operators by means of D$5'$ giants wrapping a twistor line in $\PT$ and two out of the four fibers of $\CO(-1)^{\oplus4}$. We have also interpreted Witten's D-instantons as D9 branes wrapping all four fibres of $\CO(-1)^{\oplus4}\to\PT$ and a twistor line in the base. Suitable superspace integrals of the associated determinant operators give rise to insertions of the non-self-dual interaction $L_\text{int}$ of equation \eqref{S-8} that deforms the self-dual theory into non-self-dual $\cN=4$ sYM. 

The D9 branes discussed in this work intersect supertwistor space in twistor lines $\P^1$. Similarly, the D$5'$ giants intersect supertwistor space in twistor superlines that are copies of $\P^{1|2}$. One can also consider other branes. For instance, a D1 brane that wraps none of the fiber directions would intersect supertwistor space in a $\P^{1|4}$. It could be interesting to explore what spacetime operators the associated determinants give rise to, and which BPS sectors they belong to. One could also consider D3 and D7 branes that wrap one or three of the fiber directions. These intersect supertwistor space in $\P^{1|3}$ and $\P^{1|1}$ respectively, so they could possibly give rise to fermionic operators on $\R^4$. A generalization to branes intersecting $\pt$ in higher degree rational curves \cite{Witten:2003nn} and supercurves would also be desirable, as it could lead to novel holographic reformulations of the RSVW formula \cite{Roiban:2004yf}. In the same vein, one could also explore determinant modifications like sub-determinant operators, as well as heavier operators.

There is, of course, a whole universe of operators beyond the half-BPS ones. Perhaps the most interesting from the viewpoint of twisted holography are operators in the $\tfrac{1}{8}$-BPS sector. These include the chiral algebra sector of $\cN=4$ sYM \cite{Beem:2013sza,Costello:2018zrm}, which is dual to the B-model on $\SL_2(\C)$.\footnote{The relation between the twistor string and such chiral algebras will be studied in work to appear by Seraphim Jarov.} More generally, it would be interesting to explore the correlators of $\frac{1}{16}$-BPS or even non-BPS operators in the self-dual sector. By virtue of the absence of a 't Hooft coupling, these should be computable perturbatively. It will be very instructive to understand which bulk gravitational states or D-branes encode such weakly or completely non-protected operators and correlators.

\paragraph{Supersymmetrized determinants.} In section \ref{sec:giant}, we built the giant graviton configurations dual to the correlators of $k$ bosonic determinants $\det(1+y_i\cdot\Phi(x_i))$. In the original work \cite{Caron-Huot:2023wdh} on the $\rho$ matrix model, the authors also derive an extended matrix model for supersymmetrized determinants. Such determinants fit in a chiral supermultiplet that starts on $\det(1+y_i\cdot\Phi(x_i))$ and includes its superconformal descendants.

It would be interesting to generalize our giant graviton geometries to these cases. The Higgsing step is quite simple. Once again, we start with $k$ coincident giants that intersect supertwistor space in the twistor superline of the origin $x^{\dal\al}=y^{\da a}=\zeta^{\al a}=0$. We would like to move the $i^\text{th}$ giant to the locus
\be
\begin{split}
    \mu^\dal &= x_i^{\dal\al}\lambda_\al\,,\qquad \veps^\al = \zeta_i^{\al a}\tau_a - x_i^{\dal\al}\veps_\dal\,,\\
    \nu^{\dot a} &= y_i^{\da a}\tau_a\,,\qquad\hspace{0.5em}\eta^a = \zeta_i^{\al a}\lambda_\al - y_i^{\da a}\eta_{\dot a}\,,
\end{split}
\ee
which intersects supertwistor space in the superline $\P^{1|2}_{x_i,y_i,\zeta_i}$. To accomplish this, we simply modify the vev of the diagonal string $\rA_{ii}$ given in \eqref{diagvev} to
\be\label{diagvev1}
\la\rA_{ii}\ra = -\zeta_i^{\al a}\lambda_\al\tau_a + x_i^{\dal\al}\lambda_\al\veps_\dal + y_i^{\da a}\tau_a\eta_\da\,.
\ee
The leading term $\zeta_i^{\al a}\lambda_\al\tau_a$ is curiously novel. It is fermionic and is a vev for the $ii$ diagonal component of the $\GL_k(\C)$ \emph{ghost}. As discussed in our observations below \eqref{diagvev}, the vev for $\rA_{ii}$ acts as a Hamiltonian -- with respect to the Schouten bracket -- for a vector field that translates the $i^\text{th}$ brane to its desired location. Using this fact, we may compute the shift in the location of the $i^\text{th}$ giant generated by the Hamiltonian \eqref{diagvev1},
\be
\begin{split}
    \delta\mu^\dal &= [\la\rA_{ii}\ra,\mu^\dal] = \frac{\p}{\p\veps_\dal}\la\rA_{ii}\ra = x_i^{\al\dal}\lambda_\al\,,\\
    \delta\nu^\da &= [\la\rA_{ii}\ra,\nu^\da] = \frac{\p}{\p\eta_\da}\la\rA_{ii}\ra = y_i^{a\da}\tau_a\,,\\
    \delta\veps^\al &= [\la\rA_{ii}\ra,\veps^\al] = -\frac{\p}{\p\lambda_\al}\la\rA_{ii}\ra = \zeta_i^{\al a}\tau_a -x_i^{\dal\al}\veps_\dal\,,\\
    \delta\eta^a &= [\la\rA_{ii}\ra,\eta^a] = -\frac{\p}{\p\tau_a}\la\rA_{ii}\ra = \zeta_i^{\al a}\lambda_\al -y_i^{\da a}\eta_\da\,.
\end{split}
\ee
So our ghostly vev does the job.

Plugging these vevs into the worldvolume action of the giants obtained in \eqref{D5'ac} should produce a theory that is dual to the supersymmetrized $\rho$ matrix model of \cite{Caron-Huot:2023wdh} at large $N$. We leave the study of the map from saddles of this supersymmetrized matrix model to open string states on the giants to the future. The corresponding saddle equations depend on the detailed data of the points $\bs{\lambda}_{ij}$ and $\bs{\lambda}_{ji}$ that the $5'$-$5'$ string $\rho_{ij}$ is suspended between. This data drops out from the bosonic $\rho$ matrix model, so it would be interesting to see how it enters the giant graviton states in the supersymmetrized setting.

\paragraph{The Amplitude/Wilson loop duality.} The most natural next target for the application of our duality is the Amplitude/Wilson loop duality of $\cN=4$ sYM. In its original form \cite{Alday:2007hr}, this correspondence operates at strong 't Hooft coupling and is motivated from standard AdS$_5$/CFT$_4$. But one can also find weak coupling analogs of this duality \cite{Drummond:2007cf} by studying holomorphic Wilson loops in twistor space \cite{Bullimore:2011ni,Mason:2010yk,Caron-Huot:2010ryg}. The complete tree-level S-matrix of $\cN=4$ sYM can be computed as the expectation value of a Wilson loop in the twistor uplift of the self-dual theory, and planar loop corrections can be obtained by turning on the non-self-dual interaction $L_\text{int}$. To date, a holographic understanding of this weak coupling Amplitude/Wilson loop duality has not been achieved.

For an $n$-particle amplitude, such a Wilson loop is comprised of $n$ holomorphic Wilson lines wrapping the twistor lines $\P^1_i$ associated to $n$ points $(x_i,\theta_i)$ in region super-momentum space. The consecutive lines $\P^1_i$ and $\P^1_{i+1}$ intersect, so the set of lines closes to form a loop. The bulk dual of this setup is a set of $n$ D9 branes that wrap the fibers of $\CO(-1)^{\oplus4}$ and intersect $\PT$ in the lines $\P^1_i$. Each Wilson line operator wrapping $\P^1_i$ can be engineered as a fermion 2-point function in the theory of 5-9 and 9-5 strings living on $\P^1_i$.

Let $C=\cup_{i=1}^n\P^1_i$, and let $\W[C]$ denote the Wilson loop formed from the ordered product of such Wilson lines. Just as the correlator of half-BPS determinants in the self-dual sector was reduced to a $\rho$ matrix model, we can reduce the evaluation of the expectation value of $\W[C]$ to a simple 2-matrix integral,
\be
\la\W[C]\ra \propto \prod_{k=1}^n\frac{\p}{\p\rho_{k+1,k}}\int\prod_{i\neq j,j\pm1}\d\rho_{ij}\,\d\sigma_{ij}\;\e^{-S[\rho,\sigma]}\biggr|_{\rho_{k+1,k}=0}\,.
\ee
The matrix variables here are denoted $\rho_{ij},\sigma_{ij}$. We set $\rho_{ii}=\rho_{i,i+1}=\sigma_{ii}=\sigma_{i,i\pm1}=0$, and the matrix integral is over the entries $\rho_{ij},\sigma_{ij}$ for $i\neq j,j\pm1$. Finally, we also set $\rho_{i+1,i}=0$ after computing the derivatives. 

Again, this 2-matrix integral is not totally trivial, as it does not possess a $\GL_n(\C)$ ``gauge'' symmetry. The action for this matrix model reads
\be
S[\rho,\sigma] = \sum_{i\neq j}\left(\frac12\,R_{ij}\sigma_{ij}\sigma_{ji} - \rho_{ij}\sigma_{ji}\right) + N\sum_{p=3}^\infty\frac{(-1)^p}{p}\frac{\rho_{i_1i_2}}{\la\lambda_{i_1i_p}\lambda_{i_1i_2}\ra}\cdots\frac{\rho_{i_pi_1}}{\la\lambda_{i_pi_{p-1}}\lambda_{i_pi_1}\ra}\,,
\ee
where $R_{ij}=[i,i+1,j,j+1,*]$ is an R-invariant evaluated in a CSW gauge with CSW reference supertwistor $\cZ_*^\bA$. The infinite sum on the right is generated by a 1-loop determinant when integrating out the 5-5, 5-9 and 9-5 strings in CSW gauge. Apart from a standard choice of the intersections $\P^1_i\cap\P^1_{i+1}$ given by $\lambda_{i,i+1\,\al}=(1,0)$ and $\lambda_{i+1,i\,\al}=(0,1)$, the points $\lambda_{ij\al}\in\P^1_i$ are given by solutions of the four linear equations
\be
\lambda_{ij0}Z_i^A + \lambda_{ij1}Z_{i+1}^A + \lambda_{ji0}Z_j^A + \lambda_{ji1}Z_{j+1}^A + Z_*^A = 0\,.
\ee
They also occur in the supersymmetrized $\rho$ matrix model for the half-BPS determinants.

Just as we saw in the case of D$5'$ giants, we expect the saddles of this matrix model to map to open string vacua on the D9 branes in the presence of backreaction. One can again start with coincident D9 branes, in which case the open string field on their worldvolume takes the form
\be
\rA = \sA + \sM^\dal\veps_\dal + \sB\veps^2\,,
\ee
where $\rA$ is overall graded odd, and $\sA,\sM^\dal,\sB$ are $(0,*)$-polyforms on $\CO(-1)^{\oplus4}\to\P^1$ of alternating parity. In the absence of flux, $\rA$ obeys the field equation $\dbar\rA+\rA^2=0$. One can move the D9 branes to the fibers above the lines $\P^1_i$ by giving the diagonal entries $\rA_{ii}$ the vevs
\be
\la\rA_{ii}\ra = -\theta^{I\al}_i\lambda_\al W_I + x_i^{\dal\al}\lambda_\al\veps_\dal\,.
\ee
Our task would then be to turn on backreaction, and find a map from saddles $\rho,\sigma$ to solutions of the flux-coupled field equations of 9-9 strings.

We hope to return to this problem and provide more details of this 2-matrix integral in future work. It would also be interesting to study the interplay between D9 and D$5'$ branes by turning them on simultaneously. The D$5'$ branes can also be used to engineer insertions of $L_\text{int}$ through the superconformal descendants of $\det(1+y\cdot\Phi)$ \cite{Caron-Huot:2023wdh}, which may help in finding the holographic dual of loop integrands in $\cN=4$ sYM.

\paragraph{Chiral ABJM and higher spins.} The tensionless string dual to free $\text{Sym}^N(T^4)$, and our twistor string dual to $\cN=4$ sdYM, naturally fill the gap between twisted holography and the more ``physical'' AdS/CFT dualities of Maldacena.\footnote{Recently, the tensionless dual of free $\text{Sym}^N(T^4)$ has also been reformulated as a sigma model into the minitwistor space of $\AdS_3$ \cite{McStay:2024dtk}. This puts these two dualities on a similar footing.} In a more ambitious direction for future work, it might also be possible to find a similar duality for a free or chiral limit of ABJM theory. This might entail looking for a free limit of the duality relating ABJM to M-theory on AdS$_4\times S^7/\Z_k$ \cite{Aharony:2008ug}.

Similarly, there has been recent work \cite{Aharony:2024nqs,Jain:2024bza} on finding closed chiral subsectors of Chern-Simons-matter theories, with an eye toward finding the boundary dual to chiral higher-spin gravity in AdS$_4$ \cite{Metsaev:2018xip,Skvortsov:2018uru,Sharapov:2022wpz}. Seeing as how the twistor uplift of sdYM played a key role in our discovery of its B-model dual, we anticipate that the twistor uplifts of Chern-Simons-matter theories and chiral higher-spin gravity could play an important role in the top-down construction of such a duality. For chiral theories of higher-spin Yang-Mills and gravity, many variations of twistor actions have now been proposed \cite{Tran:2022tft,Herfray:2022prf,Adamo:2022lah,Mason:2025pbz,Tran:2025uad}. It would be interesting to see if any of them admit a topological string or (twisted) M-theory uplift.

\paragraph{Lessons for the physical string.} This work was primarily occupied with the study of giant gravitons in the topological string. But many of the general lessons acquired in section \ref{sec:giant} may extend in one form or another to the RNS string. Giant gravitons in type IIB string theory have a rich history. For the most part, they have been studied as classical geometries that branes can wrap in the presence of flux \cite{McGreevy:2000cw,Grisaru:2000zn,Hashimoto:2000zp}. In this paper, we have presented a slightly more general perspective. 

The giant gravitons in our work are branes carrying a nontrivial open string configuration on their worldvolume. Our key insight was to study configurations of multiple giants through a non-abelian gauge theory enriched with Higgs vevs, as opposed to attaching a separate abelian gauge theory to every giant. This allowed us to find the open string vacua on their worldvolume in a much easier way than looking for clever ans\"atze for curved geometries that the giants can wrap. 

It might be worth importing this trick to the physical string in the setup of \cite{Jiang:2019xdz} (albeit this could get complicated due to our poor understanding of a non-abelian DBI action). In large $N$ QCD or $\cN=4$ sYM, single-trace type operators correspond to mesons and glueballs, whereas determinant operators correspond to baryons. This makes them a physically interesting target for holographic applications. Due to the stringy exclusion principle \cite{McGreevy:2000cw}, determinants have also emerged as a tool to study holography at finite $N$, where they account for trace relations and give rise to giant graviton expansions \cite{Arai:2019xmp,Arai:2019wgv,Arai:2019aou,Imamura:2021ytr,Gaiotto:2021xce,Murthy:2022ien}.

A key feature of our duality is that even at large $N$, it handles determinant operators in a much more natural manner than single-trace operators. This is highly promising. In principle, one can compute the correlator of determinants in the non-self-dual theory perturbatively around the self-dual sector. Our duality provides a bulk interpretation of this computation within the self-dual sector, i.e., at zero 't Hooft coupling. But we expect that it can be extended to nonzero 't Hooft coupling via Witten's D-instanton expansion.

Similarly, the bulk interpretation of giant graviton expansions of indices of 4d $\cN=4$ sYM is often clearest in the $\al'\to\infty$ limit or perturbatively in the 't Hooft coupling \cite{Lee:2023iil,Lee:2024hef}. Our duality is a closed subsector of the non-self-dual duality which operates precisely in this limit. So we hope that it can provide a clean tablet for exploring the bulk origin of such expansions.

%%%%%%%%%%%%%%%%%%%%%%%%%%%%
%%%%%%%%%%%%%%%%%%%%%%%%%%%%

\acknowledgments

We are grateful to Kasia Budzik for sharing her insights over extensive discussions. We thank Tim Adamo for a careful reading of the draft and helpful comments. We also thank Roland Bittleston, Charlotte Christensen, Andrea Ferrari, Simon Heuveline, Matthew Heydeman, Max H\"ubner, Claude LeBrun, Lionel Mason, Walker Melton, Surya Raghavendran, Debmalya Sarkar, Sean Seet and Andrew Strominger for useful conversations, and Kevin Costello and Seraphim Jarov for sharing details of their parallel work. 

We would like to thank the Isaac Newton Institute (INI) for Mathematical Sciences, Cambridge, for support and hospitality during the programme ``Twistor theory'', where the ideas of this paper originated. The INI is supported by the EPSRC grant EP/Z000580/1. We also thank the organizers of the workshops ``From Asymptotic Symmetries to Flat Holography: Theoretical Aspects and Observable Consequences'' at the Galileo Galilei Institute, and ``From Good Cuts to Celestial Holography'' at St Antony's College, University of Oxford, for their hospitality, where parts of this work were completed. A key component of this work was performed at the Aspen Center for Physics, which is supported by the National Science Foundation grant PHY-2210452. We would like to thank the participants of the Aspen workshop ``Recent Developments in String Theory'', in particular Matthias Gaberdiel, Shota Komatsu, and Edward Mazenc, for many important suggestions and clarifications. 

A.S.\ is supported by the Gordon and Betty Moore Foundation and the John Templeton Foundation via the Black Hole Initiative. D.S. is supported by the STFC (UK) HEP Theory Consolidated grant ST/X000664/1. This work is supported in part by the Simons Foundation through the Simons Collaboration on Celestial Holography.
%%%%%%%%%%%%%%%%%%%%%%%%%%%%
%%%%%%%%%%%%%%%%%%%%%%%%%%%%

\begin{appendix}

\section{Giant gravitons in $\SL_2(\C)$}
\label{app:giants}

In this appendix, we review the appearance of giant gravitons in twisted holography on $\C^3$. They have been explored in \cite{Budzik:2021fyh} and furnish a B-brane interpretation of determinant correlators in the chiral algebra subsector \cite{Beem:2013sza} of 4d $\cN=4$ $\U(N)$ sYM at large $N$. We also provide two novel derivations of the coupling of these giants to the backreaction of the stack of branes hosting the large $N$ chiral algebra, one of which generalizes easily to our higher-dimensional context.

%%%%%%%%%%%%%%%%%%%%%%%%%%%%
%%%%%%%%%%%%%%%%%%%%%%%%%%%%

\subsection{Chiral algebra determinants}

Let us begin by reviewing the setup of twisted holography on $\C^3$ as presented in \cite{Costello:2018zrm,Budzik:2021fyh}. These works study the B-model on a copy of $\C^3$, which may be thought of as a patch of the resolved conifold $\CO(-1)^{\oplus2}\to\P^1$. Let $x,y,z$ be complex coordinates on $\C^3$, and wrap a stack of $N$ D1 branes along the locus $x=y=0$. These branes backreact to generate a complex structure deformation $\dbar\mapsto\dbar+\beta$, where the Beltrami differential $\beta$ is given by
\be\label{betaback}
\beta = \frac{N(\by\,\d\bx-\bx\,\d\by)}{(|x|^2+|y|^2)^2}\,\p_z\,.
\ee
The subset $\C^3-\{x=y=0\}$ equipped with this deformed complex structure can be identified with a patch of the complex manifold $\SL_2(\C)$, the deformed conifold.

The conjecture of \cite{Costello:2018zrm} is that the B-model in the presence of this backreaction is dual to the worldvolume theory of the D1 branes in the large $N$ limit. The latter is a 2d chiral gauge theory with gauge group $\GL_N(\C)$. It is described by the action of 1-1 strings,
\be
S_{11} = \int_{\C}\d z\;\Tr(Y\dbar_a X)\,,
\ee
where $X,Y$ are a pair of adjoint-valued Higgs fields describing normal fluctuations of the branes, and $a\equiv a_\bz\d\bz$ is a $\GL_N(\C)$ $(0,1)$-form gauge field appearing in the covariant derivative $\dbar_a=\dbar+a$. This may be written in the BV formalism, but we leave the details to the original references. One typically gauge fixes by setting $a=0$, in which case this theory reduces to a chiral algebra with generators $X,Y$ along with a pair of adjoint-valued ghosts $b,c$.

In \cite{Budzik:2021fyh}, the authors identify giant gravitons in the bulk with determinant operators in the chiral algebra,
\be
\det(m_i+X+u_iY)(z_i)\,,
\ee
for some complex parameters $u_i,m_i$ and insertion points $z=z_i$. These determinants are engineered by giant graviton D$1'$ branes wrapping the loci
\be\label{zxi}
z=z_i\,,\qquad x+u_iy+m_i=0\,.
\ee
For multiple giants labeled by indices $i,j=1,\dots,k$, the theory of 1-$1'$ and $1'$-1 strings living on the intersection of these branes is a 0-dimensional fermionic integral with ``action''
\be
S_{11'} = \sum_{i=1}^k\til\chi_i(m_i+X+u_iY)\bigr|_{z_i}\chi_i\,,
\ee
where $\chi_i$ and $\til\chi_i$ are valued in the fundamental and antifundamental of $\GL_N(\C)$ respectively. The determinant insertions are engineered through their ``path integral''
\be
\int\prod_{i=1}^k\d^N\chi_i\,\d^N\til\chi_i\;\e^{-S_{11'}} = \prod_{i=1}^k\det(m_i+X+u_iY)(z_i)\,,
\ee
obtained after decoupling the $1'1'$ strings that are to be thought of as being part of the bulk theory.

The determinant correlators in the chiral algebra are thus computed by the combined path integral over 1-1, 1-$1'$ and $1'$-1 strings. Following the same kind of steps that we used to reduce our $\cN=4$ sdYM determinant correlators to a 1-matrix model, the authors of \cite{Budzik:2021fyh} reduce this computation to the matrix integral
\be
\Big\la\prod_{i=1}^k\det(m_i+X+u_iY)(z_i)\Big\ra \propto \int\prod_{i\neq j}\d\rho_{ij}\;\e^{-S[\rho]}\,,
\ee
where we are being cavalier about a normalization factor independent of the external data $z_i,u_i,m_i$. The $k\times k$ matrix $\rho$ in this integral has diagonal entries fixed to the values $\rho_{ii}=m_i$, while its off-diagonal entries are governed by the action
\be
S[\rho] = \sum_{i<j}\frac{z_{ij}}{u_{ij}}\,\rho_{ij}\rho_{ji} - N\log\det\rho\,.
\ee
This is a precise lower-dimensional analog of the matrix action \eqref{rhoac}.

The saddle-point equations following from this action read
\be
z_{ij}\rho_{ij} - Nu_{ij}\rho^{-1}_{ij} = 0\,.
\ee
Introducing the $k\times k$ diagonal matrices
\be\label{zetamu}
\zeta_{ij} = z_i\delta_{ij}\,,\qquad\mu_{ij} = u_i\delta_{ij}\,,
\ee
these equations may be recast as a matrix commutator equation
\be\label{D1saddle}
[\zeta,\rho] = N[\mu,\rho^{-1}]\,.
\ee
When $N=0$, the only solution for $\rho_{ij}$, $i\neq j$, is $\rho_{ij}=0$ (assuming the $z_i$'s are distinct). This is consistent with the absence of open string states stretching across non-intersecting branes in the B-model.

When $N\neq0$, we would like to interpret the solutions of \eqref{D1saddle} as being in 1:1 correspondence with nontrivial shapes of the $k$ giants. These shapes are meant to arise from the giants recoiling in the presence of the D1 backreaction as one sends $N\to\infty$. To see this, we need to explicitly construct the theory on the giants in the presence of the backreaction. 

%%%%%%%%%%%%%%%%%%%%%%%%%%%%
%%%%%%%%%%%%%%%%%%%%%%%%%%%%

\subsection{Incorporating backreaction}

Before accounting for backreaction, the theory on the D$1'$ giants is easy to construct. To begin with, take the $k$ giants to be coincident and wrap them along the $y$-plane $\C\subset\C^3$ given by $x=z=0$. Then the open strings on the giants consist of a pair of $\gl_k(\C)$-valued Higgs fields $\sX,\sZ$ and a $\GL_k(\C)$ $(0,1)$-form gauge field $\msf{a}$. Their action is the $\GL_k(\C)$ gauge theory
\be
S_{1'1'} = \int_\C\d y\;\tr(\sZ\,\dbar_{\msf{a}}\sX)\,,
\ee
where `$\tr$' denotes the trace of $k\times k$ matrices.\footnote{The idea of representing open strings on the giants in terms of $k\times k$ fields $\sX,\sZ$ was suggested to us by Edward Mazenc based on work to appear with Debmalya Sarkar.}

Since the giants are not coincident but actually wrap the loci \eqref{zxi}, we need to endow the Higgs fields $\sX(y),\sZ(y)$ with the appropriate vevs. These vevs can be written in terms of the diagonal matrices $\zeta,\mu$ in \eqref{zetamu} along with a matrix
\be
\msf{m}_{ij} = m_i\delta_{ij}
\ee
that encodes the parameters $m_i$. In terms of these, we set
\be
\begin{split}
    \la\sX\ra &= -y\mu-\msf{m}\,,\\
    \la\sZ\ra &= \zeta\,.
\end{split}
\ee
This completes the definition of the theory of $1'$-$1'$ strings in the absence of backreaction.

In this lower dimensional context, we have found two methods for incorporating the backreaction. We discuss each of them in turn.

\paragraph{Backreaction via 1-1$'$ and 1$'$-1 strings.} The simplest trick is to use the idea of open-closed duality that also applies to our case of determinants in $\cN=4$ sdYM. Just as we employed the integral over $\chi_i,\til\chi_i$ to engineer the determinant insertions in the chiral algebra path integral, we can complementarily use them to engineer an effective interaction term in the worldvolume theory of the giants.

To do this, decouple the 1-1 strings and focus purely on the coupled system of the $1'$-$1'$, 1-$1'$ and $1'$-1 strings. By a standard calculation of open-open couplings in the B-model, the total action of this system is found to be
\be
S_{1'1'+11'} = S_{1'1'} + \sum_{i,j=1}^k\til\chi_{i}^r\,\sX_{ij}|_{y=0}\chi_{jr}\,,
\ee
where $r=1,\dots,N$ is an index of the fundamental of $\GL_N(\C)$. When the fermions $\chi_i,\til\chi_i$ are integrated out, one obtains an effective action
\be\label{D1'ac}
S_{1'1'}^\text{eff} = S_{1'1'} - N\log\det(\sX)\bigr|_{y=0}\,.
\ee
This is a direct analog of our higher-dimensional worldvolume theory \eqref{D5'ac} on $\D5'$ giants. The $\log\det\sX$ term is invariant under the adjoint action of $\GL_k(\C)$ on $\sX$, so this action continues to be a $\GL_k(\C)$ gauge theory before Higgsing.

\paragraph{Backreaction via Beltrami couplings.} An intuitive alternative for deriving \eqref{D1'ac} that exists in this setup is to directly couple the giants to the Beltrami background \eqref{betaback}. We know from \cite{Budzik:2021fyh} that this must work, because these giants should have to deform from holomorphic curves in $\C^3$ to holomorphic curves in $\SL_2(\C)$. Indeed, we are able to show that the Beltrami backreaction knows about both the 1-1 and the 1-$1'$, $1'$-1 strings at the same time, thereby reproducing \eqref{D1'ac}. This gives an a posteriori justification for the previous trick of using open-closed duality.

The structure our backreaction takes is that of a Beltrami pointing purely along the D1 branes, $\beta = \beta^z\p_z$, where the coefficient
\be\label{bmk}
\beta^z = \frac{N(\by\,\d\bx-\bx\,\d\by)}{(|x|^2+|y|^2)^2}
\ee
is proportional to the Bochner-Martinelli kernel along the normal $\C^2$ directions. The coupling of a such background to the fields of the worldvolume theory on the giants is an infinite sum given by
\be\label{belcoup}
\sum_{n=-1}^\infty\frac{1}{(n+1)!}\int_\C\d y\;\tr(\sX^{n+1})\;\p_x^n\beta^z\bigr|_{x=z=0}\,.
\ee
In this expression, the Beltrami has been pulled back to the coincident stack of giants at $x=z=0$, as we are implementing non-coincident giants via Higgsing around this ``origin''. 

The first term in this sum arises from a well-known tadpole $\p^{-1}\beta$ for the Beltrami that the giants contribute by virtue of being D1 branes. The remaining terms are supposed to evoke the Taylor expansion of $\p^{-1}\beta|_{x=\sX}$ around $x=0$, at least when $\sX$ is a number instead of a matrix. They can be derived systematically by computing open-closed worldsheet amplitudes using the techniques of \cite{Hofman:2002cw}. Open-closed couplings that are higher order in $\beta^z$ do not appear due to the nilpotence $\beta^z\wedge\beta^z=0$.

Naively, the pullback of the Bochner-Martinelli kernel \eqref{bmk} to $x=0$ appears to vanish. However, it has the structure of a Poisson kernel or an AdS bulk-to-boundary propagator: it vanishes at $x=0$ away from $y=0$, but encounters a singularity at $x=y=0$. So its restriction to $x=0$ is not zero but a delta function,
\be
\beta^z \overset{x\to0}{\sim} -\frac{2\pi i N}{x}\;\bar\delta(y) + \rO(\bx)\,.
\ee
We will neglect factors of $2\pi i$ in what follows. In this asymptotic expansion, the leading term has a pole at $x=0$, but the residue at this pole vanishes away from $y=0$. The subleading terms are non-singular and always come with positive powers of $\bx$, so they vanish in the limit $x\to0$. As their $x$-derivatives will also vanish due to the positive powers of $\bx$, they indeed drop out of \eqref{belcoup}.

Plugging the asymptotics of $\beta^z$ into the interaction vertices \eqref{belcoup} produces
\begingroup
\allowdisplaybreaks
\begin{align}
&\lim_{x\to0}\sum_{n=-1}^\infty\frac{N}{(n+1)!}\int_\C\d y\;\tr(\sX^{n+1})\;\p_x^n\bigg(\frac{1}{x}\bigg)\,\bar\delta(y)\nonumber\\
&\hspace{3cm}= -\lim_{x\to0}\bigg\{Nk\log x + N\sum_{n=0}^\infty\frac{(-1)^{n}}{n+1}\,\tr\bigg(\frac{\sX^{n+1}}{x^{n+1}}\bigg)\bigg\}\biggr|_{y=0}\nonumber\\
&\hspace{3cm}= -\lim_{x\to0}\bigg\{N\log x^k + N\,\tr\log\bigg(1+\frac{\sX}{x}\bigg)\bigg\}\biggr|_{y=0}\nonumber\\
&\hspace{3cm}= -\lim_{x\to0}N\log\det(x+\sX)\bigr|_{y=0}\nonumber\\
&\hspace{3cm}= -N\log\det(\sX)\bigr|_{y=0}\,,
\end{align}
\endgroup
having used $\p^{-1}(1/x) = \log x$ and $\tr(1)=k$ to evaluate the $n=-1$ summand. We also employed the identity $\tr\log=\log\det$ for $k\times k$ matrices to simplify the third line. In the end, we find exactly the same interaction term that we obtained in \eqref{D1'ac} by integrating out the 1-$1'$ and $1'$-1 strings. 

This puts the open-closed duality interpretation on a stronger footing and provides indirect evidence for our application of it to $\cN=4$ sdYM. It would be highly desirable to repeat a version of this calculation with our D5 backreaction \eqref{xiN}, but we have been unable to do so due to reasons discussed in section \ref{ssec:giant}.

%%%%%%%%%%%%%%%%%%%%%%%%%%%%
%%%%%%%%%%%%%%%%%%%%%%%%%%%%

\subsection{The shapes of giants} 

Let us take another look at the action of the giants in the presence of this Beltrami coupling,
\be
S_{1'1'}^\text{eff} = \int\tr(\sZ\,\dbar_\sa\sX)\;\d y - N\log\det(\sX)\bigr|_{y=0}\,.
\ee
The equations of motion following from this action read
\be
\begin{split}
    \dbar_\sa\sZ &= -N\,\sX^{-1}\,\bar\delta(y)\,,\\
    \dbar_\sa\sX &= 0\,,\\
    [\sZ,\sX] &= 0\,,
\end{split}
\ee
with the final equation arising as a constraint from integrating out $\sa$.

Our Higgsed vacuum $\sX = -y\mu-\msf{m}$, $\sZ=\zeta$, $\sa=0$, or for that matter even the trivial vacuum $\sX=\sZ=\sa=0$, are no longer solutions of these equations at nonzero $N$. This is because when the giants come too close to the source locus $x=y=0$, they experience an infinite potential barrier and receive a delta function kick. Nevertheless, it is easy to solve these modified equations. 

True vacua of the giants are obtained as follows. Define
\be
\rho \vcentcolon= -\sX|_{y=0}\,.
\ee
We will look for vacua in which $\sa=0$. %\footnote{Due to the Higgsing, we can no longer use the $\GL_k(\C)$ symmetry to gauge fix $\sa$ to zero without simultaneously also deforming $\sX,\sZ$. So we simply take $\sa=0$ to be our ansatz for this calculation.} 
In such states, the solutions for $\sX$ are holomorphic curves on the annulus $\{y\neq0,\infty\}$. So $\sX(y)$ admits a Laurent expansion in $y$. We impose the boundary conditions that $\sX$ be regular at $y=0$, as we want $\rho$ to not blow up, and asymptote at $y=\infty$ to
\be\label{Xsol}
\sX \overset{y\to\infty}{\sim} -y\mu-\msf{m}\,,
\ee
up to off-diagonal entries regular at $y=0,\infty$. Then the solution for $\sX$ is found to be
\be
\sX = -y\mu - \rho\,,
\ee
with the diagonal entries of $\rho$ being fixed to the entries of $\msf{m}$ due to these boundary conditions: $\rho_{ii}=m_i$.

Away from $y=0$, the equation for $\sZ$ also reduces to $\dbar\sZ=0$ upon setting $\sa=0$. So $\sZ(y)$ is again given by a Laurent expansion in $y$. We impose the boundary condition
\be
\sZ \overset{y\to\infty}{\sim} \zeta\,.
\ee
With this boundary condition, the sourced equation for $\sZ$ has a unique solution
\be\label{Zsol}
\sZ = \zeta + \frac{N\rho^{-1}}{y}\,,
\ee
having used $\sX^{-1}|_{y=0}=-\rho^{-1}$ and assumed the invertibility of $\rho$ (which follows from the invertibility of $\msf{m}$ in perturbation theory). There cannot be any terms with higher order poles at $y=0$ as they would have contributed further delta function sources to $\dbar\sZ$.

Finally, even with the ansatz $\sa=0$, we still need to impose the equation of motion of $\sa$. This gives rise to an ADHM-type constraint:
\be
[\sZ,\sX] = -[\zeta,\rho] + N[\mu,\rho^{-1}]=0\,.
\ee
This reproduces precisely the saddle equation \eqref{D1saddle} of the $\rho$ matrix model governing chiral algebra determinant correlators. We conclude that saddles of the $\rho$ matrix model are in one-to-one correspondence with vacua of the giant worldvolume theory obeying $\sa=0$ and our holographic boundary conditions.

The spectral curve interpretation of giant graviton shapes arises as follows. The backreaction \eqref{betaback} is expressed in a harmonic gauge. We can easily transform it to an axial gauge,
\be
\beta = -\frac{N}{x}\,\bar\delta(y)\,\p_z\,.
\ee
Away from its singularities at $x=0$ and $y=0$, holomorphic coordinates in the presence of this backreaction are given by
\be\label{wdef}
\begin{split}
    a &= y\,,\\
    b &= x\,,\\
    c &= yz\,,\\
    d &= xz + \frac{N}{y}\,.
\end{split}
\ee
It is easily tested that $d$ is annihilated by $\dbar+\beta$ (again, neglecting factors of $2\pi i$). And the product $c=yz$ is trivially holomorphic because $\beta\propto\bar\delta(y)$. These quantities may be arranged into a $2\times 2$ unimodular matrix
\be
\frac{1}{\sqrt{N}}\begin{pmatrix}
    a&&b\\c&&d
\end{pmatrix}\,,
\ee
confirming that the deformed complex structure describes a patch of $\SL_2(\C)$.

After backreaction, the Higgs field $\sX$ retains its interpretation as describing fluctuations in the $b$ direction, but the Higgs field $\sZ$ must now be replaced by the product $y\sZ = a\sZ$ so as to be interpreted as capturing the holomorphic fluctuations of $c$. This motivates us to introduce the matrices
\be
\begin{split}
    B &= \sX = -(a\mu+\rho)\,,\\
    C &= y\sZ = a\zeta+N\rho^{-1}\,,\\
    D &= -(a\mu\zeta+\zeta\rho+N\rho^{-1}\mu)\,,
\end{split}
\ee
having plugged in our solutions for $\sX,\sZ$. They satisfy the matrix equation
\be
aD-BC = N
\ee
on the support of the saddle equation \eqref{D1saddle}. With this new interpretation, we can conclude that our solutions for $\sX,\sZ$ describe branes wrapping holomorphic curves in the deformed complex geometry $\SL_2(\C)$.

To make these holomorphic curves manifest, one simply diagonalizes our expressions for $B,C,D$. They are simultaneously diagonalizable. Indeed, $B,C$ commute with each other because $[B,C]=y[\sX,\sZ]=0$, and thereafter they both commute with $D=a^{-1}(N+BC)$. The resulting eigenvalues spread out across a holomorphic curve in $a$ which is reducible in general. This was referred to as a \emph{spectral curve} in \cite{Budzik:2021fyh}. The number of irreducible components of this curve can be identified with the effective number of giant gravitons. If this is smaller than our original number of giants $k$, it signals brane recombination in the presence of flux. This is a common occurrence and happens when two or more triplets of eigenvalues $(b,c,d)$ coincide for some lucky value of $a$.

\end{appendix}

%%%%%%%%%%%%%%%%%%%%%%%%%%%%
%%%%%%%%%%%%%%%%%%%%%%%%%%%%

\bibliographystyle{JHEP}
\bibliography{sdym}

@article{Penrose:1967wn,
    author = "Penrose, R.",
    title = "{Twistor algebra}",
    doi = "10.1063/1.1705200",
    journal = "J. Math. Phys.",
    volume = "8",
    pages = "345",
    year = "1967"
}

@article{Ferber:1977qx,
    author = "Ferber, Alan",
    title = "{Supertwistors and Conformal Supersymmetry}",
    reportNumber = "EFI 77/50-CHICAGO",
    doi = "10.1016/0550-3213(78)90257-2",
    journal = "Nucl. Phys. B",
    volume = "132",
    pages = "55--64",
    year = "1978"
}

@article{Skinner:2013xp,
    author = "Skinner, David",
    title = "{Twistor strings for $ \mathcal{N} $ = 8 supergravity}",
    eprint = "1301.0868",
    archivePrefix = "arXiv",
    primaryClass = "hep-th",
    doi = "10.1007/JHEP04(2020)047",
    journal = "JHEP",
    volume = "04",
    pages = "047",
    year = "2020"
}

@article{Nair:1988bq,
    author = "Nair, V. P.",
    title = "{A Current Algebra for Some Gauge Theory Amplitudes}",
    reportNumber = "CU-TP-408",
    doi = "10.1016/0370-2693(88)91471-2",
    journal = "Phys. Lett. B",
    volume = "214",
    pages = "215--218",
    year = "1988"
}

@article{Budzik:2021fyh,
    author = "Budzik, Kasia and Gaiotto, Davide",
    title = "{Giant gravitons in twisted holography}",
    eprint = "2106.14859",
    archivePrefix = "arXiv",
    primaryClass = "hep-th",
    doi = "10.1007/JHEP10(2023)131",
    journal = "JHEP",
    volume = "10",
    pages = "131",
    year = "2023"
}

@article{Roiban:2004yf,
    author = "Roiban, Radu and Spradlin, Marcus and Volovich, Anastasia",
    title = "{On the tree level S matrix of Yang-Mills theory}",
    eprint = "hep-th/0403190",
    archivePrefix = "arXiv",
    reportNumber = "NSF-KITP-04-35",
    doi = "10.1103/PhysRevD.70.026009",
    journal = "Phys. Rev. D",
    volume = "70",
    pages = "026009",
    year = "2004"
}

@article{Adamo:2013tsa,
    author = "Adamo, Tim and Casali, Eduardo and Skinner, David",
    title = "{Ambitwistor strings and the scattering equations at one loop}",
    eprint = "1312.3828",
    archivePrefix = "arXiv",
    primaryClass = "hep-th",
    reportNumber = "DAMTP-2013-74",
    doi = "10.1007/JHEP04(2014)104",
    journal = "JHEP",
    volume = "04",
    pages = "104",
    year = "2014"
}

@article{Geyer:2016wjx,
    author = "Geyer, Yvonne and Mason, Lionel and Monteiro, Ricardo and Tourkine, Piotr",
    title = "{Two-Loop Scattering Amplitudes from the Riemann Sphere}",
    eprint = "1607.08887",
    archivePrefix = "arXiv",
    primaryClass = "hep-th",
    reportNumber = "CERN-TH-2016-172, DAMTP-2016-54",
    doi = "10.1103/PhysRevD.94.125029",
    journal = "Phys. Rev. D",
    volume = "94",
    number = "12",
    pages = "125029",
    year = "2016"
}

@article{Ikeda:2013wh,
    author = "Ikeda, Noriaki and Xu, Xiaomeng",
    title = "{Canonical functions, differential graded symplectic pairs in supergeometry, and Alexandrov-Kontsevich-Schwartz-Zaboronsky sigma models with boundaries}",
    eprint = "1301.4805",
    archivePrefix = "arXiv",
    primaryClass = "math.SG",
    doi = "10.1063/1.4900834",
    journal = "J. Math. Phys.",
    volume = "55",
    pages = "113505",
    year = "2014"
}

@article{Mason:2007zv,
    author = "Mason, L. J. and Skinner, David",
    title = "{Heterotic twistor-string theory}",
    eprint = "0708.2276",
    archivePrefix = "arXiv",
    primaryClass = "hep-th",
    doi = "10.1016/j.nuclphysb.2007.11.010",
    journal = "Nucl. Phys. B",
    volume = "795",
    pages = "105--137",
    year = "2008"
}

@article{Geyer:2015bja,
    author = "Geyer, Yvonne and Mason, Lionel and Monteiro, Ricardo and Tourkine, Piotr",
    title = "{Loop Integrands for Scattering Amplitudes from the Riemann Sphere}",
    eprint = "1507.00321",
    archivePrefix = "arXiv",
    primaryClass = "hep-th",
    doi = "10.1103/PhysRevLett.115.121603",
    journal = "Phys. Rev. Lett.",
    volume = "115",
    number = "12",
    pages = "121603",
    year = "2015"
}

@article{Seet:2025mes,
    author = "Seet, Sean",
    title = "{Single-trace current correlators for 2d models of 4d gluon scattering}",
    eprint = "2509.12200",
    archivePrefix = "arXiv",
    primaryClass = "hep-th",
    month = "9",
    year = "2025"
}

@article{Costello:2022wso,
    author = "Costello, Kevin and Paquette, Natalie M.",
    title = "{Celestial holography meets twisted holography: 4d amplitudes from chiral correlators}",
    eprint = "2201.02595",
    archivePrefix = "arXiv",
    primaryClass = "hep-th",
    doi = "10.1007/JHEP10(2022)193",
    journal = "JHEP",
    volume = "10",
    pages = "193",
    year = "2022"
}

@article{Adamo:2017qyl,
    author = "Adamo, Tim",
    title = "{Lectures on twistor theory}",
    eprint = "1712.02196",
    archivePrefix = "arXiv",
    primaryClass = "hep-th",
    doi = "10.22323/1.323.0003",
    journal = "PoS",
    volume = "Modave2017",
    pages = "003",
    year = "2018"
}

@article{Woodhouse:1985id,
    author = "Woodhouse, N. M. J.",
    title = "{Real methods in twistor theory}",
    doi = "10.1088/0264-9381/2/3/006",
    journal = "Class. Quant. Grav.",
    volume = "2",
    pages = "257--291",
    year = "1985"
}

@article{Ward:1977ta,
    author = "Ward, R. S.",
    title = "{On self-dual gauge fields}",
    doi = "10.1016/0375-9601(77)90842-8",
    journal = "Phys. Lett. A",
    volume = "61",
    pages = "81--82",
    year = "1977"
}

@article{Costello:2018zrm,
    author = "Costello, Kevin and Gaiotto, Davide",
    title = "{Twisted Holography}",
    eprint = "1812.09257",
    archivePrefix = "arXiv",
    primaryClass = "hep-th",
    month = "12",
    year = "2018"
}

@article{Mason:2005zm,
    author = "Mason, L. J.",
    title = "{Twistor actions for non-self-dual fields: A Derivation of twistor-string theory}",
    eprint = "hep-th/0507269",
    archivePrefix = "arXiv",
    doi = "10.1088/1126-6708/2005/10/009",
    journal = "JHEP",
    volume = "10",
    pages = "009",
    year = "2005"
}

@article{Witten:2003nn,
    author = "Witten, Edward",
    title = "{Perturbative gauge theory as a string theory in twistor space}",
    eprint = "hep-th/0312171",
    archivePrefix = "arXiv",
    doi = "10.1007/s00220-004-1187-3",
    journal = "Commun. Math. Phys.",
    volume = "252",
    pages = "189--258",
    year = "2004"
}

@article{Berkovits:2004jj,
    author = "Berkovits, Nathan and Witten, Edward",
    title = "{Conformal supergravity in twistor-string theory}",
    eprint = "hep-th/0406051",
    archivePrefix = "arXiv",
    reportNumber = "IFT-P-019-2004",
    doi = "10.1088/1126-6708/2004/08/009",
    journal = "JHEP",
    volume = "08",
    pages = "009",
    year = "2004"
}

@article{Costello:2021bah,
    author = "Costello, Kevin J.",
    title = "{Quantizing local holomorphic field theories on twistor space}",
    eprint = "2111.08879",
    archivePrefix = "arXiv",
    primaryClass = "hep-th",
    month = "11",
    year = "2021"
}

@article{Caron-Huot:2023wdh,
    author = {Caron-Huot, Simon and Coronado, Frank and M\"uhlmann, Beatrix},
    title = "{Determinants in self-dual $ \mathcal{N} $ = 4 SYM and twistor space}",
    eprint = "2304.12341",
    archivePrefix = "arXiv",
    primaryClass = "hep-th",
    doi = "10.1007/JHEP08(2023)008",
    journal = "JHEP",
    volume = "08",
    pages = "008",
    year = "2023"
}

@article{Bak:2011yy,
    author = "Bak, Dongsu and Chen, Bin and Wu, Jun-Bao",
    title = "{Holographic Correlation Functions for Open Strings and Branes}",
    eprint = "1103.2024",
    archivePrefix = "arXiv",
    primaryClass = "hep-th",
    reportNumber = "UOSTP-110301",
    doi = "10.1007/JHEP06(2011)014",
    journal = "JHEP",
    volume = "06",
    pages = "014",
    year = "2011"
}

@article{Bissi:2011dc,
    author = "Bissi, A. and Kristjansen, C. and Young, D. and Zoubos, K.",
    title = "{Holographic three-point functions of giant gravitons}",
    eprint = "1103.4079",
    archivePrefix = "arXiv",
    primaryClass = "hep-th",
    doi = "10.1007/JHEP06(2011)085",
    journal = "JHEP",
    volume = "06",
    pages = "085",
    year = "2011"
}

@article{Caron-Huot:2010ryg,
    author = "Caron-Huot, Simon",
    title = "{Notes on the scattering amplitude / Wilson loop duality}",
    eprint = "1010.1167",
    archivePrefix = "arXiv",
    primaryClass = "hep-th",
    doi = "10.1007/JHEP07(2011)058",
    journal = "JHEP",
    volume = "07",
    pages = "058",
    year = "2011"
}

@article{Drummond:2007cf,
    author = "Drummond, J. M. and Henn, J. and Korchemsky, G. P. and Sokatchev, E.",
    title = "{On planar gluon amplitudes/Wilson loops duality}",
    eprint = "0709.2368",
    archivePrefix = "arXiv",
    primaryClass = "hep-th",
    reportNumber = "LAPTH-1206-07, LPT-ORSAY-07-82",
    doi = "10.1016/j.nuclphysb.2007.11.007",
    journal = "Nucl. Phys. B",
    volume = "795",
    pages = "52--68",
    year = "2008"
}

@article{Bailey:1998zif,
    author = "Bailey, Toby N. and Dunne, Edward G.",
    title = "{A twistor correspondence and Penrose transform for odd-dimensional hyperbolic space}",
    doi = "10.1090/S0002-9939-98-04215-4",
    journal = "Proc. Am. Math. Soc.",
    volume = "126",
    number = "04",
    pages = "1245--1253",
    year = "1998"
}

@article{Ferrara:1998ej,
    author = "Ferrara, Sergio and Fronsdal, Christian and Zaffaroni, Alberto",
    title = "{On N=8 supergravity on AdS(5) and N=4 superconformal Yang-Mills theory}",
    eprint = "hep-th/9802203",
    archivePrefix = "arXiv",
    reportNumber = "CERN-TH-98-62, UCLA-98-TEP-6",
    doi = "10.1016/S0550-3213(98)00444-1",
    journal = "Nucl. Phys. B",
    volume = "532",
    pages = "153--162",
    year = "1998"
}

@article{Liu:1998bu,
    author = "Liu, Hong and Tseytlin, Arkady A.",
    title = "{D = 4 superYang-Mills, D = 5 gauged supergravity, and D = 4 conformal supergravity}",
    eprint = "hep-th/9804083",
    archivePrefix = "arXiv",
    reportNumber = "IMPERIAL-TP-97-98-39, NSF-ITP-98-049",
    doi = "10.1016/S0550-3213(98)00443-X",
    journal = "Nucl. Phys. B",
    volume = "533",
    pages = "88--108",
    year = "1998"
}

@article{Hofman:2002cw,
    author = "Hofman, Christiaan",
    title = "{On the open closed B model}",
    eprint = "hep-th/0204157",
    archivePrefix = "arXiv",
    reportNumber = "RUNHETC-2002-04",
    doi = "10.1088/1126-6708/2003/11/069",
    journal = "JHEP",
    volume = "11",
    pages = "069",
    year = "2003"
}

@article{Costello:2019jsy,
    author = "Costello, Kevin and Li, Si",
    title = "{Anomaly cancellation in the topological string}",
    eprint = "1905.09269",
    archivePrefix = "arXiv",
    primaryClass = "hep-th",
    doi = "10.4310/ATMP.2020.v24.n7.a2",
    journal = "Adv. Theor. Math. Phys.",
    volume = "24",
    number = "7",
    pages = "1723--1771",
    year = "2020"
}

@article{Budzik:2023xbr,
    author = "Budzik, Kasia and Gaiotto, Davide and Kulp, Justin and Williams, Brian R. and Wu, Jingxiang and Yu, Matthew",
    title = "{Semi-chiral operators in 4d $ \mathcal{N} $ = 1 gauge theories}",
    eprint = "2306.01039",
    archivePrefix = "arXiv",
    primaryClass = "hep-th",
    doi = "10.1007/JHEP05(2024)245",
    journal = "JHEP",
    volume = "05",
    pages = "245",
    year = "2024"
}

@article{Costello:2020jbh,
    author = "Costello, Kevin and Paquette, Natalie M.",
    title = "{Twisted Supergravity and Koszul Duality: A case study in AdS$_3$}",
    eprint = "2001.02177",
    archivePrefix = "arXiv",
    primaryClass = "hep-th",
    doi = "10.1007/s00220-021-04065-3",
    journal = "Commun. Math. Phys.",
    volume = "384",
    number = "1",
    pages = "279--339",
    year = "2021"
}

@article{Dijkgraaf:2002fc,
    author = "Dijkgraaf, Robbert and Vafa, Cumrun",
    title = "{Matrix models, topological strings, and supersymmetric gauge theories}",
    eprint = "hep-th/0206255",
    archivePrefix = "arXiv",
    reportNumber = "HUTP-02-A028, ITFA-2002-22",
    doi = "10.1016/S0550-3213(02)00766-6",
    journal = "Nucl. Phys. B",
    volume = "644",
    pages = "3--20",
    year = "2002"
}

@article{Gaberdiel:2021qbb,
    author = "Gaberdiel, Matthias R. and Gopakumar, Rajesh",
    title = "{String Dual to Free N=4 Supersymmetric Yang-Mills Theory}",
    eprint = "2104.08263",
    archivePrefix = "arXiv",
    primaryClass = "hep-th",
    doi = "10.1103/PhysRevLett.127.131601",
    journal = "Phys. Rev. Lett.",
    volume = "127",
    number = "13",
    pages = "131601",
    year = "2021"
}

@article{Gaberdiel:2021jrv,
    author = "Gaberdiel, Matthias R. and Gopakumar, Rajesh",
    title = "{The worldsheet dual of free super Yang-Mills in 4D}",
    eprint = "2105.10496",
    archivePrefix = "arXiv",
    primaryClass = "hep-th",
    doi = "10.1007/JHEP11(2021)129",
    journal = "JHEP",
    volume = "11",
    pages = "129",
    year = "2021"
}

@article{Mason:2013sva,
    author = "Mason, Lionel and Skinner, David",
    title = "{Ambitwistor strings and the scattering equations}",
    eprint = "1311.2564",
    archivePrefix = "arXiv",
    primaryClass = "hep-th",
    doi = "10.1007/JHEP07(2014)048",
    journal = "JHEP",
    volume = "07",
    pages = "048",
    year = "2014"
}

@article{Geyer:2014fka,
    author = "Geyer, Yvonne and Lipstein, Arthur E. and Mason, Lionel J.",
    title = "{Ambitwistor Strings in Four Dimensions}",
    eprint = "1404.6219",
    archivePrefix = "arXiv",
    primaryClass = "hep-th",
    doi = "10.1103/PhysRevLett.113.081602",
    journal = "Phys. Rev. Lett.",
    volume = "113",
    number = "8",
    pages = "081602",
    year = "2014"
}

@article{Bittleston:2024efo,
    author = "Bittleston, Roland and Costello, Kevin and Zeng, Keyou",
    title = "{Self-Dual Gauge Theory from the Top Down}",
    eprint = "2412.02680",
    archivePrefix = "arXiv",
    primaryClass = "hep-th",
    month = "12",
    year = "2024"
}

@article{Geyer:2022cey,
    author = "Geyer, Yvonne and Mason, Lionel",
    title = "{The SAGEX review on scattering amplitudes Chapter 6: Ambitwistor Strings and Amplitudes from the Worldsheet}",
    eprint = "2203.13017",
    archivePrefix = "arXiv",
    primaryClass = "hep-th",
    reportNumber = "SAGEX-22-07",
    doi = "10.1088/1751-8121/ac8190",
    journal = "J. Phys. A",
    volume = "55",
    number = "44",
    pages = "443007",
    year = "2022"
}

@article{Berkovits:2004hg,
    author = "Berkovits, Nathan",
    title = "{An Alternative string theory in twistor space for N=4 superYang-Mills}",
    eprint = "hep-th/0402045",
    archivePrefix = "arXiv",
    doi = "10.1103/PhysRevLett.93.011601",
    journal = "Phys. Rev. Lett.",
    volume = "93",
    pages = "011601",
    year = "2004"
}

@article{Alexandrov:1995kv,
    author = "Alexandrov, M. and Schwarz, A. and Zaboronsky, O. and Kontsevich, M.",
    title = "{The Geometry of the master equation and topological quantum field theory}",
    eprint = "hep-th/9502010",
    archivePrefix = "arXiv",
    reportNumber = "UCD-94-01, UCD{\&}B-94-01",
    doi = "10.1142/S0217751X97001031",
    journal = "Int. J. Mod. Phys. A",
    volume = "12",
    pages = "1405--1429",
    year = "1997"
}

@article{Boels:2006ir,
    author = "Boels, Rutger and Mason, L. J. and Skinner, David",
    title = "{Supersymmetric Gauge Theories in Twistor Space}",
    eprint = "hep-th/0604040",
    archivePrefix = "arXiv",
    doi = "10.1088/1126-6708/2007/02/014",
    journal = "JHEP",
    volume = "02",
    pages = "014",
    year = "2007"
}

@article{Costello:2012cy,
    author = "Costello, Kevin J. and Li, Si",
    title = "{Quantum BCOV theory on Calabi-Yau manifolds and the higher genus B-model}",
    eprint = "1201.4501",
    archivePrefix = "arXiv",
    primaryClass = "math.QA",
    month = "1",
    year = "2012"
}

@mastersthesis{Adamo:2013cra,
    author = "Adamo, Tim",
    title = "{Twistor actions for gauge theory and gravity}",
    eprint = "1308.2820",
    archivePrefix = "arXiv",
    primaryClass = "hep-th",
    type = "Other thesis",
    month = "8",
    year = "2013"
}

@book{Hori:2003ic,
    author = "Hori, K. and Katz, S. and Klemm, A. and Pandharipande, R. and Thomas, R. and Vafa, C. and Vakil, R. and Zaslow, E.",
    title = "{Mirror symmetry}",
    publisher = "AMS",
    address = "Providence, USA",
    series = "Clay mathematics monographs",
    volume = "1",
    year = "2003"
}

@article{Bershadsky:1993cx,
    author = "Bershadsky, M. and Cecotti, S. and Ooguri, H. and Vafa, C.",
    title = "{Kodaira-Spencer theory of gravity and exact results for quantum string amplitudes}",
    eprint = "hep-th/9309140",
    archivePrefix = "arXiv",
    reportNumber = "HUTP-93-A025, RIMS-946, SISSA-142-93-EP",
    doi = "10.1007/BF02099774",
    journal = "Commun. Math. Phys.",
    volume = "165",
    pages = "311--428",
    year = "1994"
}

@article{Maldacena:1997re,
    author = "Maldacena, Juan Martin",
    title = "{The Large $N$ limit of superconformal field theories and supergravity}",
    eprint = "hep-th/9711200",
    archivePrefix = "arXiv",
    reportNumber = "HUTP-97-A097, HUTP-98-A097",
    doi = "10.4310/ATMP.1998.v2.n2.a1",
    journal = "Adv. Theor. Math. Phys.",
    volume = "2",
    pages = "231--252",
    year = "1998"
}

@article{Fernandez:2024tue,
    author = "Fern{\'a}ndez, V{\'\i}ctor E. and Paquette, Natalie M. and Williams, Brian R.",
    title = "{Twisted holography on AdS$_3 \times S^3 \times$ K3 {\&} the planar chiral algebra}",
    eprint = "2404.14318",
    archivePrefix = "arXiv",
    primaryClass = "hep-th",
    doi = "10.21468/SciPostPhys.17.4.109",
    journal = "SciPost Phys.",
    volume = "17",
    number = "4",
    pages = "109",
    year = "2024"
}

@article{Witten:1998qj,
    author = "Witten, Edward",
    title = "{Anti de Sitter space and holography}",
    eprint = "hep-th/9802150",
    archivePrefix = "arXiv",
    reportNumber = "IASSNS-HEP-98-15",
    doi = "10.4310/ATMP.1998.v2.n2.a2",
    journal = "Adv. Theor. Math. Phys.",
    volume = "2",
    pages = "253--291",
    year = "1998"
}

@article{Bogna:2023bbd,
    author = "Bogna, Giuseppe and Mason, Lionel",
    title = "{Yang-Mills form factors on self-dual backgrounds}",
    eprint = "2305.07542",
    archivePrefix = "arXiv",
    primaryClass = "hep-th",
    doi = "10.1007/JHEP08(2023)165",
    journal = "JHEP",
    volume = "08",
    pages = "165",
    year = "2023"
}

@article{Mason:2010yk,
    author = "Mason, L. J. and Skinner, David",
    title = "{The Complete Planar S-matrix of N=4 SYM as a Wilson Loop in Twistor Space}",
    eprint = "1009.2225",
    archivePrefix = "arXiv",
    primaryClass = "hep-th",
    doi = "10.1007/JHEP12(2010)018",
    journal = "JHEP",
    volume = "12",
    pages = "018",
    year = "2010"
}

@article{Adamo:2011dq,
    author = "Adamo, Tim and Bullimore, Mathew and Mason, Lionel and Skinner, David",
    title = "{A Proof of the Supersymmetric Correlation Function / Wilson Loop Correspondence}",
    eprint = "1103.4119",
    archivePrefix = "arXiv",
    primaryClass = "hep-th",
    doi = "10.1007/JHEP08(2011)076",
    journal = "JHEP",
    volume = "08",
    pages = "076",
    year = "2011"
}

@article{Chicherin:2014uca,
    author = "Chicherin, Dmitry and Doobary, Reza and Eden, Burkhard and Heslop, Paul and Korchemsky, Gregory P. and Mason, Lionel and Sokatchev, Emery",
    title = "{Correlation functions of the chiral stress-tensor multiplet in $ \mathcal{N}=4 $ SYM}",
    eprint = "1412.8718",
    archivePrefix = "arXiv",
    primaryClass = "hep-th",
    reportNumber = "CERN-PH-TH-2014-270, DCPT-14-79, HU-EP-14-66, IPHT-T14-242, LAPTH-239-14",
    doi = "10.1007/JHEP06(2015)198",
    journal = "JHEP",
    volume = "06",
    pages = "198",
    year = "2015"
}

@article{Koster:2016fna,
    author = "Koster, Laura and Mitev, Vladimir and Staudacher, Matthias and Wilhelm, Matthias",
    title = "{On Form Factors and Correlation Functions in Twistor Space}",
    eprint = "1611.08599",
    archivePrefix = "arXiv",
    primaryClass = "hep-th",
    reportNumber = "HU-MATHEMATIK-2016-20, HU-EP-16-40, MITP-16-123",
    doi = "10.1007/JHEP03(2017)131",
    journal = "JHEP",
    volume = "03",
    pages = "131",
    year = "2017"
}

@article{Jiang:2019xdz,
    author = "Jiang, Yunfeng and Komatsu, Shota and Vescovi, Edoardo",
    title = "{Structure constants in $ \mathcal{N} $ = 4 SYM at finite coupling as worldsheet g-function}",
    eprint = "1906.07733",
    archivePrefix = "arXiv",
    primaryClass = "hep-th",
    reportNumber = "CERN-TH-2019-093, Imperial-TP-EV-2019-01",
    doi = "10.1007/JHEP07(2020)037",
    journal = "JHEP",
    volume = "07",
    number = "07",
    pages = "037",
    year = "2020"
}

@article{Lee:2023iil,
    author = "Lee, Ji Hoon",
    title = "{Trace relations and open string vacua}",
    eprint = "2312.00242",
    archivePrefix = "arXiv",
    primaryClass = "hep-th",
    doi = "10.1007/JHEP02(2024)224",
    journal = "JHEP",
    volume = "02",
    pages = "224",
    year = "2024"
}

@article{Aharony:1999ti,
    author = "Aharony, Ofer and Gubser, Steven S. and Maldacena, Juan Martin and Ooguri, Hirosi and Oz, Yaron",
    title = "{Large N field theories, string theory and gravity}",
    eprint = "hep-th/9905111",
    archivePrefix = "arXiv",
    reportNumber = "CERN-TH-99-122, HUTP-99-A027, LBNL-43113, RU-99-18, UCB-PTH-99-16, LBL-43113",
    doi = "10.1016/S0370-1573(99)00083-6",
    journal = "Phys. Rept.",
    volume = "323",
    pages = "183--386",
    year = "2000"
}

@article{Donagi:2003hh,
    author = "Donagi, Ron and Katz, S. and Sharpe, E.",
    title = "{Spectra of D-branes with higgs vevs}",
    eprint = "hep-th/0309270",
    archivePrefix = "arXiv",
    reportNumber = "ILL-TH-03-09, ILL-(TH)-03-09",
    doi = "10.4310/ATMP.2004.v8.n5.a3",
    journal = "Adv. Theor. Math. Phys.",
    volume = "8",
    number = "5",
    pages = "813--859",
    year = "2004"
}

@article{Beem:2013sza,
    author = "Beem, Christopher and Lemos, Madalena and Liendo, Pedro and Peelaers, Wolfger and Rastelli, Leonardo and van Rees, Balt C.",
    title = "{Infinite Chiral Symmetry in Four Dimensions}",
    eprint = "1312.5344",
    archivePrefix = "arXiv",
    primaryClass = "hep-th",
    reportNumber = "YITP-SB-13-45, CERN-PH-TH-2013-311, HU-EP-13-78",
    doi = "10.1007/s00220-014-2272-x",
    journal = "Commun. Math. Phys.",
    volume = "336",
    number = "3",
    pages = "1359--1433",
    year = "2015"
}

@article{Koster:2016ebi,
    author = "Koster, Laura and Mitev, Vladimir and Staudacher, Matthias and Wilhelm, Matthias",
    title = "{Composite Operators in the Twistor Formulation of N=4 Supersymmetric Yang-Mills Theory}",
    eprint = "1603.04471",
    archivePrefix = "arXiv",
    primaryClass = "hep-th",
    reportNumber = "HU-MATHEMATIK-16-05, HU-EP-16-09, MITP-16-024",
    doi = "10.1103/PhysRevLett.117.011601",
    journal = "Phys. Rev. Lett.",
    volume = "117",
    number = "1",
    pages = "011601",
    year = "2016"
}

@article{Balasubramanian:2001nh,
    author = "Balasubramanian, Vijay and Berkooz, Micha and Naqvi, Asad and Strassler, Matthew J.",
    title = "{Giant gravitons in conformal field theory}",
    eprint = "hep-th/0107119",
    archivePrefix = "arXiv",
    reportNumber = "UPR-T-943, WIS-15-01-DPP",
    doi = "10.1088/1126-6708/2002/04/034",
    journal = "JHEP",
    volume = "04",
    pages = "034",
    year = "2002"
}

@article{McGreevy:2000cw,
    author = "McGreevy, John and Susskind, Leonard and Toumbas, Nicolaos",
    title = "{Invasion of the giant gravitons from Anti-de Sitter space}",
    eprint = "hep-th/0003075",
    archivePrefix = "arXiv",
    reportNumber = "SU-ITP-00-09",
    doi = "10.1088/1126-6708/2000/06/008",
    journal = "JHEP",
    volume = "06",
    pages = "008",
    year = "2000"
}

@article{Siegel:1992xp,
    author = "Siegel, W.",
    title = "{N=2 (4) string theory is self-dual N=4 Yang-Mills theory}",
    eprint = "hep-th/9205075",
    archivePrefix = "arXiv",
    doi = "10.1103/PhysRevD.46.R3235",
    journal = "Phys. Rev. D",
    volume = "46",
    number = "8",
    pages = "R3235",
    year = "1992"
}

@article{Siegel:1992wd,
    author = "Siegel, W.",
    title = "{Selfdual N=8 supergravity as closed N=2 (N=4) strings}",
    eprint = "hep-th/9207043",
    archivePrefix = "arXiv",
    reportNumber = "ITP-SB-92-31",
    doi = "10.1103/PhysRevD.47.2504",
    journal = "Phys. Rev. D",
    volume = "47",
    pages = "2504--2511",
    year = "1993"
}

@article{Chalmers:1996rq,
    author = "Chalmers, G. and Siegel, W.",
    title = "{The Selfdual sector of QCD amplitudes}",
    eprint = "hep-th/9606061",
    archivePrefix = "arXiv",
    reportNumber = "ITP-SB-96-29",
    doi = "10.1103/PhysRevD.54.7628",
    journal = "Phys. Rev. D",
    volume = "54",
    pages = "7628--7633",
    year = "1996"
}

@article{Cachazo:2004kj,
    author = "Cachazo, Freddy and Svrcek, Peter and Witten, Edward",
    title = "{MHV vertices and tree amplitudes in gauge theory}",
    eprint = "hep-th/0403047",
    archivePrefix = "arXiv",
    doi = "10.1088/1126-6708/2004/09/006",
    journal = "JHEP",
    volume = "09",
    pages = "006",
    year = "2004"
}

@article{Cachazo:2004zb,
    author = "Cachazo, Freddy and Svrcek, Peter and Witten, Edward",
    title = "{Twistor space structure of one-loop amplitudes in gauge theory}",
    eprint = "hep-th/0406177",
    archivePrefix = "arXiv",
    doi = "10.1088/1126-6708/2004/10/074",
    journal = "JHEP",
    volume = "10",
    pages = "074",
    year = "2004"
}

@article{Cachazo:2004by,
    author = "Cachazo, Freddy and Svrcek, Peter and Witten, Edward",
    title = "{Gauge theory amplitudes in twistor space and holomorphic anomaly}",
    eprint = "hep-th/0409245",
    archivePrefix = "arXiv",
    doi = "10.1088/1126-6708/2004/10/077",
    journal = "JHEP",
    volume = "10",
    pages = "077",
    year = "2004"
}

@article{Adamo:2011pv,
    author = "Adamo, Tim and Bullimore, Mathew and Mason, Lionel and Skinner, David",
    title = "{Scattering Amplitudes and Wilson Loops in Twistor Space}",
    eprint = "1104.2890",
    archivePrefix = "arXiv",
    primaryClass = "hep-th",
    doi = "10.1088/1751-8113/44/45/454008",
    journal = "J. Phys. A",
    volume = "44",
    pages = "454008",
    year = "2011"
}

@article{Fioresi:2017cax,
    author = "Fioresi, R. and Latini, E. and Lledo, M. A. and Nadal, F. A.",
    title = "{The Segre embedding of the quantum conformal superspace}",
    eprint = "1709.03075",
    archivePrefix = "arXiv",
    primaryClass = "hep-th",
    reportNumber = "IFIC-17-39",
    doi = "10.4310/ATMP.2018.v22.n8.a4",
    journal = "Adv. Theor. Math. Phys.",
    volume = "22",
    pages = "1939--2000",
    year = "2018"
}

@article{lebrun1990projective,
  title={Projective embeddings of complex supermanifolds},
  author={LeBrun, Claude and Poon, Yat-Sun and Wells Jr, R. O.},
  journal={Commun. Math. Phys.},
  volume={126},
  number={3},
  pages={433--452},
  year={1990},
  publisher={Springer}
}

@article{Witten:1991zz,
    author = "Witten, Edward",
    editor = "Yau, Shing-Tung",
    title = "{Mirror manifolds and topological field theory}",
    eprint = "hep-th/9112056",
    archivePrefix = "arXiv",
    reportNumber = "IASSNS-HEP-91-83",
    journal = "AMS/IP Stud. Adv. Math.",
    volume = "9",
    pages = "121--160",
    year = "1998"
}

@article{Beisert:2010jr,
    author = "Beisert, Niklas and others",
    title = "{Review of AdS/CFT Integrability: An Overview}",
    eprint = "1012.3982",
    archivePrefix = "arXiv",
    primaryClass = "hep-th",
    reportNumber = "AEI-2010-175, CERN-PH-TH-2010-306, HU-EP-10-87, HU-MATH-2010-22, KCL-MTH-10-10, UMTG-270, UUITP-41-10",
    doi = "10.1007/s11005-011-0529-2",
    journal = "Lett. Math. Phys.",
    volume = "99",
    pages = "3--32",
    year = "2012"
}

@book{Ammon:2015wua,
    author = "Ammon, Martin and Erdmenger, Johanna",
    title = "{Gauge/gravity duality}: {Foundations and applications}",
    doi = "10.1017/CBO9780511846373",
    isbn = "978-1-107-01034-5, 978-1-316-23594-2",
    publisher = "Cambridge University Press",
    address = "Cambridge",
    month = "4",
    year = "2015"
}

@article{Ishtiaque:2018str,
    author = "Ishtiaque, Nafiz and Faroogh Moosavian, Seyed and Zhou, Yehao",
    title = "{Topological holography: The example of the D2-D4 brane system}",
    eprint = "1809.00372",
    archivePrefix = "arXiv",
    primaryClass = "hep-th",
    doi = "10.21468/SciPostPhys.9.2.017",
    journal = "SciPost Phys.",
    volume = "9",
    number = "2",
    pages = "017",
    year = "2020"
}

@article{Bonetti:2016nma,
    author = "Bonetti, Federico and Rastelli, Leonardo",
    title = "{Supersymmetric localization in AdS$_{5}$ and the protected chiral algebra}",
    eprint = "1612.06514",
    archivePrefix = "arXiv",
    primaryClass = "hep-th",
    doi = "10.1007/JHEP08(2018)098",
    journal = "JHEP",
    volume = "08",
    pages = "098",
    year = "2018"
}

@article{Eberhardt:2019ywk,
    author = "Eberhardt, Lorenz and Gaberdiel, Matthias R. and Gopakumar, Rajesh",
    title = "{Deriving the AdS$_{3}$/CFT$_{2}$ correspondence}",
    eprint = "1911.00378",
    archivePrefix = "arXiv",
    primaryClass = "hep-th",
    doi = "10.1007/JHEP02(2020)136",
    journal = "JHEP",
    volume = "02",
    pages = "136",
    year = "2020"
}

@article{Haggi-Mani:2000dxu,
    author = "Haggi-Mani, Parviz and Sundborg, Bo",
    title = "{Free large N supersymmetric Yang-Mills theory as a string theory}",
    eprint = "hep-th/0002189",
    archivePrefix = "arXiv",
    reportNumber = "USITP-00-01",
    doi = "10.1088/1126-6708/2000/04/031",
    journal = "JHEP",
    volume = "04",
    pages = "031",
    year = "2000"
}

@article{Sundborg:2000wp,
    author = "Sundborg, Bo",
    editor = "Sorokin, Dmitri P.",
    title = "{Stringy gravity, interacting tensionless strings and massless higher spins}",
    eprint = "hep-th/0103247",
    archivePrefix = "arXiv",
    doi = "10.1016/S0920-5632(01)01545-6",
    journal = "Nucl. Phys. B Proc. Suppl.",
    volume = "102",
    pages = "113--119",
    year = "2001"
}

@article{Sezgin:2001zs,
    author = "Sezgin, E. and Sundell, P.",
    title = "{Doubletons and 5-D higher spin gauge theory}",
    eprint = "hep-th/0105001",
    archivePrefix = "arXiv",
    reportNumber = "CTP-TAMU-13-01, UG-01-27",
    doi = "10.1088/1126-6708/2001/09/036",
    journal = "JHEP",
    volume = "09",
    pages = "036",
    year = "2001"
}

@article{Mikhailov:2002bp,
    author = "Mikhailov, Andrei",
    title = "{Notes on higher spin symmetries}",
    eprint = "hep-th/0201019",
    archivePrefix = "arXiv",
    reportNumber = "NSF-ITP-01-181, ITEP-TH-66-01",
    month = "1",
    year = "2002"
}

@article{Sezgin:2002rt,
    author = "Sezgin, E. and Sundell, P.",
    title = "{Massless higher spins and holography}",
    eprint = "hep-th/0205131",
    archivePrefix = "arXiv",
    reportNumber = "CTP-TAMU-08-02, UU-01-10",
    doi = "10.1016/S0550-3213(02)00739-3",
    journal = "Nucl. Phys. B",
    volume = "644",
    pages = "303--370",
    year = "2002",
    note = "[Erratum: Nucl.Phys.B 660, 403--403 (2003)]"
}

@article{Gaberdiel:2018rqv,
    author = "Gaberdiel, Matthias R. and Gopakumar, Rajesh",
    title = "{Tensionless string spectra on AdS$_{3}$}",
    eprint = "1803.04423",
    archivePrefix = "arXiv",
    primaryClass = "hep-th",
    doi = "10.1007/JHEP05(2018)085",
    journal = "JHEP",
    volume = "05",
    pages = "085",
    year = "2018"
}

@article{Strominger:1996sh,
    author = "Strominger, Andrew and Vafa, Cumrun",
    title = "{Microscopic origin of the Bekenstein-Hawking entropy}",
    eprint = "hep-th/9601029",
    archivePrefix = "arXiv",
    reportNumber = "HUTP-96-A002, RU-96-01",
    doi = "10.1016/0370-2693(96)00345-0",
    journal = "Phys. Lett. B",
    volume = "379",
    pages = "99--104",
    year = "1996"
}

@article{Gubser:1998bc,
    author = "Gubser, S. S. and Klebanov, Igor R. and Polyakov, Alexander M.",
    title = "{Gauge theory correlators from noncritical string theory}",
    eprint = "hep-th/9802109",
    archivePrefix = "arXiv",
    reportNumber = "PUPT-1767",
    doi = "10.1016/S0370-2693(98)00377-3",
    journal = "Phys. Lett. B",
    volume = "428",
    pages = "105--114",
    year = "1998"
}

@article{Eberhardt:2018ouy,
    author = "Eberhardt, Lorenz and Gaberdiel, Matthias R. and Gopakumar, Rajesh",
    title = "{The Worldsheet Dual of the Symmetric Product CFT}",
    eprint = "1812.01007",
    archivePrefix = "arXiv",
    primaryClass = "hep-th",
    doi = "10.1007/JHEP04(2019)103",
    journal = "JHEP",
    volume = "04",
    pages = "103",
    year = "2019"
}

@article{Gopakumar:1998ki,
    author = "Gopakumar, Rajesh and Vafa, Cumrun",
    editor = "Vafa, Cumrun and Yau, S. -T.",
    title = "{On the gauge theory / geometry correspondence}",
    eprint = "hep-th/9811131",
    archivePrefix = "arXiv",
    reportNumber = "HUTP-98-A078",
    doi = "10.4310/ATMP.1999.v3.n5.a5",
    journal = "Adv. Theor. Math. Phys.",
    volume = "3",
    pages = "1415--1443",
    year = "1999"
}

@article{Adamo:2016rtr,
    author = "Adamo, Tim and Skinner, David and Williams, Jack",
    title = "{Twistor methods for AdS$_{5}$}",
    eprint = "1607.03763",
    archivePrefix = "arXiv",
    primaryClass = "hep-th",
    reportNumber = "DAMTP-2016-49",
    doi = "10.1007/JHEP08(2016)167",
    journal = "JHEP",
    volume = "08",
    pages = "167",
    year = "2016"
}

@article{Metsaev:1998it,
    author = "Metsaev, R. R. and Tseytlin, Arkady A.",
    title = "{Type IIB superstring action in AdS$_5\times S^5$ background}",
    eprint = "hep-th/9805028",
    archivePrefix = "arXiv",
    reportNumber = "FIAN-TD-98-21, IMPERIAL-TP-97-98-44, NSF-ITP-98-055",
    doi = "10.1016/S0550-3213(98)00570-7",
    journal = "Nucl. Phys. B",
    volume = "533",
    pages = "109--126",
    year = "1998"
}

@article{Berkovits:2007rj,
    author = "Berkovits, Nathan and Vafa, Cumrun",
    editor = "Edelstein, Jose D. and Grandi, Nicolas and Nunez, Carmen A. and Schvellinger, Martin",
    title = "{Towards a Worldsheet Derivation of the Maldacena Conjecture}",
    eprint = "0711.1799",
    archivePrefix = "arXiv",
    primaryClass = "hep-th",
    reportNumber = "IFT-P-018-2007",
    doi = "10.1088/1126-6708/2008/03/031",
    journal = "JHEP",
    volume = "03",
    pages = "031",
    year = "2008"
}

@article{Berkovits:2007zk,
    author = "Berkovits, Nathan",
    title = "{A New Limit of the AdS$_5\times S^5$ Sigma Model}",
    eprint = "hep-th/0703282",
    archivePrefix = "arXiv",
    reportNumber = "IFT-P-06-2007",
    doi = "10.1088/1126-6708/2007/08/011",
    journal = "JHEP",
    volume = "08",
    pages = "011",
    year = "2007"
}

@article{Berkovits:2025xok,
    author = "Berkovits, Nathan",
    title = "{Topological A-Model for $AdS_5\times S^5$ Superstring and the Maldacena Conjecture}",
    eprint = "2506.10907",
    archivePrefix = "arXiv",
    primaryClass = "hep-th",
    month = "6",
    year = "2025"
}

@article{Gopakumar:2022djw,
    author = "Gopakumar, Rajesh and Mazenc, Edward A.",
    title = "{Deriving the Simplest Gauge-String Duality -- I: Open-Closed-Open Triality}",
    eprint = "2212.05999",
    archivePrefix = "arXiv",
    primaryClass = "hep-th",
    month = "12",
    year = "2022"
}

@article{Alday:2007hr,
    author = "Alday, Luis F. and Maldacena, Juan Martin",
    title = "{Gluon scattering amplitudes at strong coupling}",
    eprint = "0705.0303",
    archivePrefix = "arXiv",
    primaryClass = "hep-th",
    reportNumber = "SPIN-07-16, ITP-UU-07-24",
    doi = "10.1088/1126-6708/2007/06/064",
    journal = "JHEP",
    volume = "06",
    pages = "064",
    year = "2007"
}

@article{Caron-Huot:2021usw,
    author = "Caron-Huot, Simon and Coronado, Frank",
    title = "{Ten dimensional symmetry of $ \mathcal{N} $ = 4 SYM correlators}",
    eprint = "2106.03892",
    archivePrefix = "arXiv",
    primaryClass = "hep-th",
    doi = "10.1007/JHEP03(2022)151",
    journal = "JHEP",
    volume = "03",
    pages = "151",
    year = "2022"
}

@article{Aharony:2008ug,
    author = "Aharony, Ofer and Bergman, Oren and Jafferis, Daniel Louis and Maldacena, Juan",
    title = "{N=6 superconformal Chern-Simons-matter theories, M2-branes and their gravity duals}",
    eprint = "0806.1218",
    archivePrefix = "arXiv",
    primaryClass = "hep-th",
    reportNumber = "WIS-12-08-JUN-DPP",
    doi = "10.1088/1126-6708/2008/10/091",
    journal = "JHEP",
    volume = "10",
    pages = "091",
    year = "2008"
}

@article{Aharony:2024nqs,
    author = "Aharony, Ofer and Kalloor, Rohit R. and Kukolj, Trivko",
    title = "{A chiral limit for Chern-Simons-matter theories}",
    eprint = "2405.01647",
    archivePrefix = "arXiv",
    primaryClass = "hep-th",
    doi = "10.1007/JHEP10(2024)051",
    journal = "JHEP",
    volume = "10",
    pages = "051",
    year = "2024"
}

@article{Jain:2024bza,
    author = "Jain, Sachin and S, Dhruva K. and Skvortsov, Evgeny",
    title = "{Hidden sectors of Chern-Simons matter theories and exact holography}",
    eprint = "2405.00773",
    archivePrefix = "arXiv",
    primaryClass = "hep-th",
    doi = "10.1103/PhysRevD.111.106017",
    journal = "Phys. Rev. D",
    volume = "111",
    number = "10",
    pages = "106017",
    year = "2025"
}

@article{Metsaev:2018xip,
    author = "Metsaev, R. R.",
    title = "{Light-cone gauge cubic interaction vertices for massless fields in AdS(4)}",
    eprint = "1807.07542",
    archivePrefix = "arXiv",
    primaryClass = "hep-th",
    reportNumber = "FIAN-TD-2018-15",
    doi = "10.1016/j.nuclphysb.2018.09.021",
    journal = "Nucl. Phys. B",
    volume = "936",
    pages = "320--351",
    year = "2018"
}

@article{Skvortsov:2018uru,
    author = "Skvortsov, Evgeny",
    title = "{Light-Front Bootstrap for Chern-Simons Matter Theories}",
    eprint = "1811.12333",
    archivePrefix = "arXiv",
    primaryClass = "hep-th",
    doi = "10.1007/JHEP06(2019)058",
    journal = "JHEP",
    volume = "06",
    pages = "058",
    year = "2019"
}

@article{Sharapov:2022wpz,
    author = "Sharapov, Alexey and Skvortsov, Evgeny and Van Dongen, Richard",
    title = "{Chiral higher spin gravity and convex geometry}",
    eprint = "2209.01796",
    archivePrefix = "arXiv",
    primaryClass = "hep-th",
    doi = "10.21468/SciPostPhys.14.6.162",
    journal = "SciPost Phys.",
    volume = "14",
    number = "6",
    pages = "162",
    year = "2023"
}

@article{Mason:2025pbz,
    author = "Mason, Lionel and Sharma, Atul",
    title = "{Chiral higher-spin theories from twistor space}",
    eprint = "2505.09419",
    archivePrefix = "arXiv",
    primaryClass = "hep-th",
    month = "5",
    year = "2025"
}

@article{Adamo:2022lah,
    author = "Adamo, Tim and Tran, Tung",
    title = "{Higher-spin Yang{\textendash}Mills, amplitudes and self-duality}",
    eprint = "2210.07130",
    archivePrefix = "arXiv",
    primaryClass = "hep-th",
    doi = "10.1007/s11005-023-01673-z",
    journal = "Lett. Math. Phys.",
    volume = "113",
    number = "3",
    pages = "50",
    year = "2023"
}

@article{Herfray:2022prf,
    author = "Herfray, Yannick and Krasnov, Kirill and Skvortsov, Evgeny",
    title = "{Higher-spin self-dual Yang-Mills and gravity from the twistor space}",
    eprint = "2210.06209",
    archivePrefix = "arXiv",
    primaryClass = "hep-th",
    doi = "10.1007/JHEP01(2023)158",
    journal = "JHEP",
    volume = "01",
    pages = "158",
    year = "2023"
}

@article{Tran:2025uad,
    author = "Tran, Tung",
    title = "{Anomaly-free twistorial higher-spin theories}",
    eprint = "2505.13785",
    archivePrefix = "arXiv",
    primaryClass = "hep-th",
    month = "5",
    year = "2025"
}

@article{Tran:2022tft,
    author = "Tran, Tung",
    title = "{Toward a twistor action for chiral higher-spin gravity}",
    eprint = "2209.00925",
    archivePrefix = "arXiv",
    primaryClass = "hep-th",
    doi = "10.1103/PhysRevD.107.046015",
    journal = "Phys. Rev. D",
    volume = "107",
    number = "4",
    pages = "046015",
    year = "2023"
}

@article{Grisaru:2000zn,
    author = "Grisaru, Marcus T. and Myers, Robert C. and Tafjord, Oyvind",
    title = "{SUSY and goliath}",
    eprint = "hep-th/0008015",
    archivePrefix = "arXiv",
    reportNumber = "MCGILL-00-21, BRX-TH-472",
    doi = "10.1088/1126-6708/2000/08/040",
    journal = "JHEP",
    volume = "08",
    pages = "040",
    year = "2000"
}

@article{Hashimoto:2000zp,
    author = "Hashimoto, Akikazu and Hirano, Shinji and Itzhaki, N.",
    title = "{Large branes in AdS and their field theory dual}",
    eprint = "hep-th/0008016",
    archivePrefix = "arXiv",
    reportNumber = "NSF-ITP-00-063",
    doi = "10.1088/1126-6708/2000/08/051",
    journal = "JHEP",
    volume = "08",
    pages = "051",
    year = "2000"
}

@article{McStay:2024dtk,
    author = "McStay, N. M. and Reid-Edwards, R. A.",
    title = "{String theory in twistor space and minimal tension holography}",
    eprint = "2411.08836",
    archivePrefix = "arXiv",
    primaryClass = "hep-th",
    doi = "10.1007/JHEP10(2025)073",
    journal = "JHEP",
    volume = "10",
    pages = "073",
    year = "2025"
}

@article{Drummond:2008vq,
    author = "Drummond, J. M. and Henn, J. and Korchemsky, G. P. and Sokatchev, E.",
    title = "{Dual superconformal symmetry of scattering amplitudes in N=4 super-Yang-Mills theory}",
    eprint = "0807.1095",
    archivePrefix = "arXiv",
    primaryClass = "hep-th",
    reportNumber = "LAPTH-1257-08, LPT-ORSAY-08-60",
    doi = "10.1016/j.nuclphysb.2009.11.022",
    journal = "Nucl. Phys. B",
    volume = "828",
    pages = "317--374",
    year = "2010"
}

@article{Drummond:2009fd,
    author = "Drummond, James M. and Henn, Johannes M. and Plefka, Jan",
    editor = "Liu, Feng and Xiao, Zhigang and Zhuang, Pengfei",
    title = "{Yangian symmetry of scattering amplitudes in N=4 super Yang-Mills theory}",
    eprint = "0902.2987",
    archivePrefix = "arXiv",
    primaryClass = "hep-th",
    reportNumber = "HU-EP-09-06, LAPTH-1308-09",
    doi = "10.1088/1126-6708/2009/05/046",
    journal = "JHEP",
    volume = "05",
    pages = "046",
    year = "2009"
}

@article{Dirac:1955uv,
    author = "Dirac, Paul A. M.",
    title = "{Gauge invariant formulation of quantum electrodynamics}",
    doi = "10.1139/p55-081",
    journal = "Can. J. Phys.",
    volume = "33",
    pages = "650",
    year = "1955"
}

@article{Donnelly:2015hta,
    author = "Donnelly, William and Giddings, Steven B.",
    title = "{Diffeomorphism-invariant observables and their nonlocal algebra}",
    eprint = "1507.07921",
    archivePrefix = "arXiv",
    primaryClass = "hep-th",
    reportNumber = "NSF-KITP-15-133",
    doi = "10.1103/PhysRevD.93.024030",
    journal = "Phys. Rev. D",
    volume = "93",
    number = "2",
    pages = "024030",
    year = "2016",
    note = "[Erratum: Phys.Rev.D 94, 029903 (2016)]"
}

@article{Donnelly:2016rvo,
    author = "Donnelly, William and Giddings, Steven B.",
    title = "{Observables, gravitational dressing, and obstructions to locality and subsystems}",
    eprint = "1607.01025",
    archivePrefix = "arXiv",
    primaryClass = "hep-th",
    doi = "10.1103/PhysRevD.94.104038",
    journal = "Phys. Rev. D",
    volume = "94",
    number = "10",
    pages = "104038",
    year = "2016"
}

@article{Arai:2019xmp,
    author = "Arai, Reona and Imamura, Yosuke",
    title = "{Finite $N$ Corrections to the Superconformal Index of S-fold Theories}",
    eprint = "1904.09776",
    archivePrefix = "arXiv",
    primaryClass = "hep-th",
    reportNumber = "TIT-HEP-671",
    doi = "10.1093/ptep/ptz088",
    journal = "PTEP",
    volume = "2019",
    number = "8",
    pages = "083B04",
    year = "2019"
}

@article{Arai:2019wgv,
    author = "Arai, Reona and Fujiwara, Shota and Imamura, Yosuke and Mori, Tatsuya",
    title = "{Finite $N$ corrections to the superconformal index of orbifold quiver gauge theories}",
    eprint = "1907.05660",
    archivePrefix = "arXiv",
    primaryClass = "hep-th",
    reportNumber = "TIT/HEP-674",
    doi = "10.1007/JHEP10(2019)243",
    journal = "JHEP",
    volume = "10",
    pages = "243",
    year = "2019"
}

@article{Arai:2019aou,
    author = "Arai, Reona and Fujiwara, Shota and Imamura, Yosuke and Mori, Tatsuya",
    title = "{Finite $N$ corrections to the superconformal index of toric quiver gauge theories}",
    eprint = "1911.10794",
    archivePrefix = "arXiv",
    primaryClass = "hep-th",
    reportNumber = "TIT/HEP-675",
    doi = "10.1093/ptep/ptaa023",
    journal = "PTEP",
    volume = "2020",
    number = "4",
    pages = "043B09",
    year = "2020"
}

@article{Gaiotto:2021xce,
    author = "Gaiotto, Davide and Lee, Ji Hoon",
    title = "{The giant graviton expansion}",
    eprint = "2109.02545",
    archivePrefix = "arXiv",
    primaryClass = "hep-th",
    doi = "10.1007/JHEP08(2024)025",
    journal = "JHEP",
    volume = "08",
    pages = "025",
    year = "2024"
}

@article{Imamura:2021ytr,
    author = "Imamura, Yosuke",
    title = "{Finite-N superconformal index via the AdS/CFT correspondence}",
    eprint = "2108.12090",
    archivePrefix = "arXiv",
    primaryClass = "hep-th",
    reportNumber = "TIT/HEP-686",
    doi = "10.1093/ptep/ptab141",
    journal = "PTEP",
    volume = "2021",
    number = "12",
    pages = "123B05",
    year = "2021"
}

@article{Lee:2024hef,
    author = "Lee, Ji Hoon and Stanford, Douglas",
    title = "{Bulk thimbles dual to trace relations}",
    eprint = "2412.20769",
    archivePrefix = "arXiv",
    primaryClass = "hep-th",
    month = "12",
    year = "2024"
}

@article{Murthy:2022ien,
    author = "Murthy, Sameer",
    title = "{Unitary matrix models, free fermions, and the giant graviton expansion}",
    eprint = "2202.06897",
    archivePrefix = "arXiv",
    primaryClass = "hep-th",
    doi = "10.4310/PAMQ.2023.v19.n1.a12",
    journal = "Pure Appl. Math. Quart.",
    volume = "19",
    number = "1",
    pages = "299--340",
    year = "2023"
}

@article{Adamo:2011cd,
    author = "Adamo, Tim",
    title = "{Correlation functions, null polygonal Wilson loops, and local operators}",
    eprint = "1110.3925",
    archivePrefix = "arXiv",
    primaryClass = "hep-th",
    doi = "10.1007/JHEP12(2011)006",
    journal = "JHEP",
    volume = "12",
    pages = "006",
    year = "2011"
}

@article{Roiban:2004vt,
    author = "Roiban, Radu and Spradlin, Marcus and Volovich, Anastasia",
    title = "{A Googly amplitude from the B model in twistor space}",
    eprint = "hep-th/0402016",
    archivePrefix = "arXiv",
    reportNumber = "NSF-KITP-04-15",
    doi = "10.1088/1126-6708/2004/04/012",
    journal = "JHEP",
    volume = "04",
    pages = "012",
    year = "2004"
}

@article{Gopakumar:2024jfq,
    author = "Gopakumar, Rajesh and Kaushik, Rishabh and Komatsu, Shota and Mazenc, Edward A. and Sarkar, Debmalya",
    title = "{Strings from Feynman Diagrams}",
    eprint = "2412.13397",
    archivePrefix = "arXiv",
    primaryClass = "hep-th",
    month = "12",
    year = "2024"
}

@article{Bullimore:2011ni,
    author = "Bullimore, Mathew and Skinner, David",
    title = "{Holomorphic Linking, Loop Equations and Scattering Amplitudes in Twistor Space}",
    eprint = "1101.1329",
    archivePrefix = "arXiv",
    primaryClass = "hep-th",
    month = "1",
    year = "2011"
}

\end{document}